\documentclass[aps,prd,twocolumn,nofootinbib,superscriptaddress,preprintnumbers,floatfix]{revtex4-2}
\usepackage{float,dcolumn}
\usepackage{amsmath,amssymb}
\usepackage{mathtools,slashed}
\usepackage{xcolor}
\usepackage{verbatim}
\usepackage{array}
\usepackage{multirow}
\usepackage{booktabs}
\usepackage{url}
\usepackage{color}
\usepackage[T1]{fontenc}
\usepackage[utf8]{inputenc}
\usepackage{graphicx}
\usepackage{subcaption}
\usepackage{caption}
\usepackage{ragged2e}
\usepackage{esint}
\usepackage[hypertexnames=false]{hyperref}
\usepackage{atbegshi}
\usepackage{placeins}
\usepackage{tikz}
\usepackage[capitalize,english]{cleveref}
\usepackage{comment}

\usepackage{physics}

\usepackage[normalem]{ulem}

\makeatletter
\g@addto@macro\bfseries{\boldmath}
\makeatother

\crefname{equation}{Eq.}{Eqs.}
\crefname{section}{Sec.}{Secs.}
\Crefname{section}{Sec.}{Secs.}
\Crefname{appendix}{Appendix}{Appendices}
\newcommand{\rcite}[1]{Ref.~\cite{#1}}
\newcommand{\rrcite}[1]{Refs.~\cite{#1}}

\newcommand{\cC}[0]{\mathcal C}

\newcommand{\cE}[0]{\mathcal E}

\newcommand{\cK}[0]{\mathcal K}
\newcommand{\cL}[0]{\mathcal L}
\newcommand{\cM}[0]{\mathcal M}

\newcommand{\cR}[0]{\mathcal R}

\newcommand{\cY}[0]{\mathcal Y}

\newcommand{\pprime}{{}^{(}{'}{}^{)}}

\newcommand\tmin{t_{\rm min}}
\newcommand\nstates{N_{\rm st}}
\newcommand\tminnstates{\left(\tmin,\,\nstates\right)}

\newcommand\Ni{\left(N_i\right)}

\DeclareMathOperator{\argmin}{argmin}

\newcommand{\mathbbm}[1]{\text{\usefont{U}{bbm}{m}{n}#1}}

\hypersetup{ 
pdfnewwindow=true,
colorlinks=true,    
linkcolor=blue,
citecolor=blue,
filecolor=blue,
urlcolor=blue
} 

\begin{document}

\preprint{DESY-26-136}
\title{Two-nucleon scattering amplitudes at light quark masses from lattice QCD }

\author{Jorge~Baeza-Ballesteros}
\email{jorge.baeza.ballesteros@desy.de}
\affiliation{John von Neumann-Institut f\"ur Computing NIC, Deutsches Elektronen-Synchrotron DESY, Platanenallee 6, 15738 Zeuthen, Germany}

\author{Andr\'e~Bai\~{a}o-Raposo}
\email{andre.baiaoraposo@ruhr-uni-bochum.de}
\affiliation{Fakult\"at f\"ur Physik und Astronomie, Institut f\"ur Theoretische Physik II, Ruhr-Universit\"at Bochum, 44780 Bochum, Germany}

\author{Davide~Laudicina}
\email{davide.laudicina@ruhr-uni-bochum.de}
\affiliation{Fakult\"at f\"ur Physik und Astronomie, Institut f\"ur Theoretische Physik II, Ruhr-Universit\"at Bochum, 44780 Bochum, Germany}

\author{Malcolm~Lazarow}
\email{mlazarow@berkeley.edu}
\affiliation{Department of Physics, University of California, Berkeley, California 94720, USA}
\affiliation{Nuclear Science Division, Lawrence Berkeley National Laboratory, Berkeley, California 94720, USA}

\author{Miguel~Salg}
\email{miguel.salg@unibe.ch}
\affiliation{Institute for Theoretical Physics, Albert Einstein Center for Fundamental Physics, University of Bern, 3012 Bern, Switzerland}

\author{John~Bulava}
\affiliation{Fakult\"at f\"ur Physik und Astronomie, Institut f\"ur Theoretische Physik II, Ruhr-Universit\"at Bochum, 44780 Bochum, Germany}

\author{M.A.~Clark}
\affiliation{NVIDIA Corporation, Santa Clara, California 95051, USA}

\author{Jeremy~R.~Green}
\affiliation{John von Neumann-Institut f\"ur Computing NIC, Deutsches Elektronen-Synchrotron DESY, Platanenallee 6, 15738 Zeuthen, Germany}

\author{Andrew~Hanlon}
\affiliation{Center for Nuclear Research, Department of Physics, Kent State University, Kent, Ohio 44242, USA}

\author{Renwick~J.~Hudspith}
\affiliation{Department of Physics, Carnegie Mellon University, Pittsburgh, Pennsylvania 15213, USA}

\author{Daniel~Mohler}
\affiliation{GSI Helmholtzzentrum f\"ur Schwerionenforschung GmbH, Planckstra{\ss}e 1, 64291 Darmstadt, Germany}
\affiliation{Institut f\"ur Kernphysik, Technische Universit\"at Darmstadt, Schlossgartenstra{\ss}e 2, 64289 Darmstadt, Germany}

\author{Colin~Morningstar}
\affiliation{Department of Physics, Carnegie Mellon University, Pittsburgh, Pennsylvania 15213, USA}

\author{Amy~Nicholson}
\affiliation{Department of Physics and Astronomy, University of North Carolina, Chapel Hill, North Carolina 27516-3255, USA}
\affiliation{Nuclear Science Division, Lawrence Berkeley National Laboratory, Berkeley, California 94720, USA}

\author{Fernando~Romero-L\'opez}
\affiliation{Institute for Theoretical Physics, Albert Einstein Center for Fundamental Physics, University of Bern, 3012 Bern, Switzerland}

\author{Thomas R.~Richardson}
\affiliation{Department of Physics, University of California, Berkeley, California 94720, USA}
\affiliation{Nuclear Science Division, Lawrence Berkeley National Laboratory, Berkeley, California 94720, USA}

\author{Andr\'e~Walker-Loud}
\affiliation{Nuclear Science Division, Lawrence Berkeley National Laboratory, Berkeley, California 94720, USA}
\affiliation{Department of Physics, University of California, Berkeley, California 94720, USA}
\affiliation{NERSC, Lawrence Berkeley National Laboratory, Berkeley, CA 94720, USA}

\author{Hartmut~Wittig}
\affiliation{PRISMA$^{++}$ Cluster of Excellence and Institut f\"ur Kernphysik, Johannes Gutenberg-Universit\"at Mainz, 55099 Mainz, Germany}
\affiliation{Helmholtz-Institut Mainz, Johannes Gutenberg-Universit\"at Mainz, 55099 Mainz, Germany}
\affiliation{GSI Helmholtzzentrum f\"ur Schwerionenforschung GmbH, Planckstra{\ss}e 1, 64291 Darmstadt, Germany}

\collaboration{for the Baryon Scattering (BaSc) Collaboration}

\date{\today}

\begin{abstract}
We present a lattice-QCD determination of nucleon-nucleon scattering amplitudes at pion masses of \(M_\pi\approx200\) MeV and \(M_\pi\approx280\) MeV at a single lattice spacing $a\approx 0.064$ fm on two CLS ensembles. The finite-volume energy spectra are extracted from large Hermitian correlation matrices constructed within the distillation framework using an extensive basis of two-nucleon interpolating operators. To reliably relate the finite-volume spectra to the infinite-volume scattering amplitudes, we employ two recently developed formalisms that explicitly account for the long-range interaction generated by one-pion exchange: the plane-wave quantization condition and the relativistic left-hand-cut formalism. 
We discuss the quark-mass dependence of the nucleon-nucleon scattering lengths, effective ranges, shape parameters and pole positions.
At all pion masses considered, our analysis prefers virtual states in both isospin channels by more than $1\sigma$.
Comparing also results from our previous calculation at $M_\pi\approx710$~MeV, we observe that the inverse scattering lengths and binding momenta approach their physical-point phenomenological values monotonically as the pion mass is varied.

\begin{tikzpicture}[remember picture,overlay,shift={(current page.north west)}]
  \node[anchor=north west,xshift=0cm,yshift=0cm]
  {\href{https://basc-collaboration.github.io/}{\includegraphics[width=1.8cm]{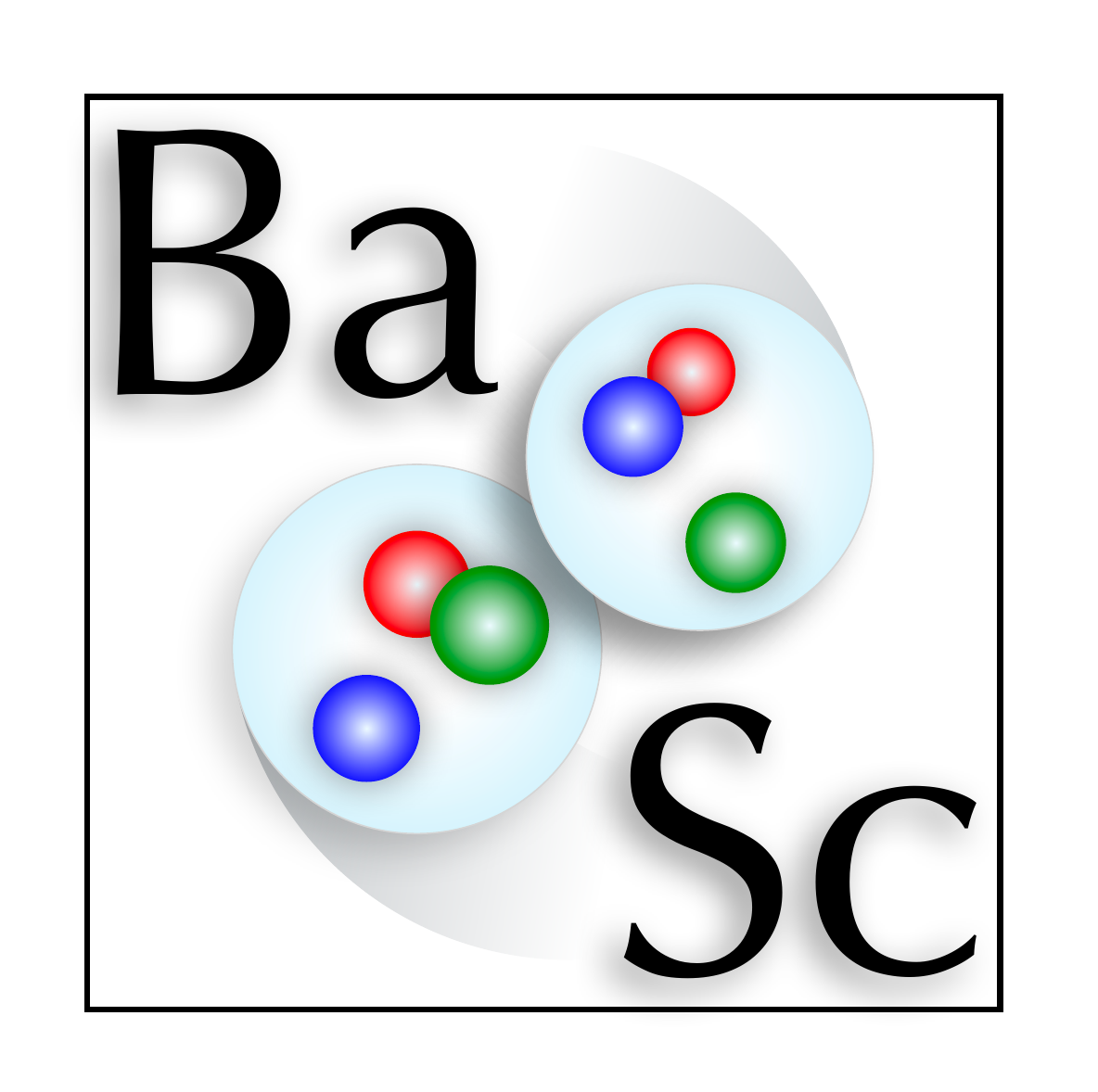}}};
  \end{tikzpicture}
\end{abstract}

\keywords{lattice QCD, scattering amplitudes, two-nucleon scattering}
\maketitle

\section{Introduction}
\label{sec:intro}

Computing nuclear forces and nuclear properties directly from quantum chromodynamics (QCD) is a major goal of hadronic physics~\cite{Detmold:2019ghl,Drischler:2019xuo}. In addition to rigorously rooting nuclear physics in the Standard Model of particle physics, new insights can potentially be obtained into two-body currents, neutrinoless double beta decay~\cite{Cirigliano:2020yhp}, and interactions of three neutrons and of hyperons. The two-nucleon system plays an important role in big bang nucleosynthesis via the deuterium bottleneck, and the sensitivity of the deuteron binding energy to quark masses is thus a key input for constraints on variations of Standard Model parameters between about a minute after the big bang and the present~\cite{Beane:2002vs,Epelbaum:2002gb,Bedaque:2010hr,Berengut:2013nh,Hall:2014dfa,Burns:2024ods, Meyer:2024auq}.

This major goal is only now becoming feasible~\cite{Bulava:2022ovd,Tews:2022yfb}.
After several decades of advances in finite-volume formalism~\cite{Luscher:1986pf}, numerical techniques~\cite{HadronSpectrum:2009krc} and controlling systematic uncertainties~\cite{Luscher:1990ck}, paired with an enormous increase in 
computational resources, lattice-QCD calculations have begun to yield consistent and reliable determinations of nucleon-nucleon ($NN$) scattering phase shifts. To date, however, calculations employing these methods have been restricted to unphysical quark masses~\cite{Beane:2006mx,Ishii:2006ec,Ishii:2012ssm,NPLQCD:2011naw,NPLQCD:2012mex,Yamazaki:2012hi,Yamazaki:2015asa,Berkowitz:2015eaa,Beane:2017edf,Wagman:2017tmp,Iritani:2016jie,Iritani:2017rlk,Iritani:2018vfn,Francis:2018qch,Horz:2020zvv,Amarasinghe:2021lqa,Detmold:2026vjx,Detmold:2026shh,Detmold:2024iwz,Green:2022rjj,BaryonScattering:2025ziz,Zhang:2026cco,Chakraborty:2026zeq,Aoki:2020bew}, typically corresponding to pion masses substantially heavier than those found in nature.

Lattice-QCD studies of nuclear systems have generally followed two strategies: finite-volume approaches that use the discrete energy spectrum~\cite{Luscher:1990ck} and potential-based methods~\cite{Ishii:2012ssm,Aoki:2020bew}. These methods have different systematics, and therefore their usage has been complementary. Early discrepancies between results obtained using these approaches have since been traced, at least in part, to systematics in the determination of lattice energies that are then fed into the finite-volume formalism. Specifically, the usage of different interpolating operators at the sink and source (off-diagonal correlators with hexaquark operators at source and bilocal operators at sink) leads to the misidentification of the finite-volume energy~\cite{Iritani:2016jie, Green:2025rel, Green:2026jgv}.

More recent studies of two-baryon systems have therefore employed large Hermitian correlation matrices constructed from extensive operator bases~\cite{Horz:2020zvv,Amarasinghe:2021lqa,Detmold:2026vjx,Detmold:2026shh,Detmold:2024iwz,Green:2022rjj,BaryonScattering:2025ziz,Zhang:2026cco,Green:2021qol}. This enables a systematic isolation of the finite-volume energy levels and brings excited-state contamination under control, with residual effects exponentially suppressed by the energy gap to the lowest state omitted from the analysis. Within this framework, nucleon-nucleon interactions at heavy quark masses are found to be attractive, but not sufficiently strong to produce bound states in either isospin channel. These claims have been recently supported by our collaboration's study that utilized both the finite-volume and potential methods to determine the $NN$ phase shifts~\cite{BaryonScattering:2025ziz}. A consensus is thus emerging that, at the heavy quark masses studied so far, the two-nucleon amplitudes most likely do not contain physical-sheet bound-state poles. Instead, the extracted poles lie on the unphysical sheet within the quoted uncertainties.

Having established that controlled lattice calculations of two-nucleon systems are possible at heavy quark masses, the next step is to approach the physical point and determine the quark-mass dependence of the interactions. This introduces several new challenges. First, the deterioration of the signal-to-noise ratio~\cite{Parisi:1983ae,Lepage:1989hd} makes the calculations computationally much more demanding. Second, as the pion mass is lowered, the long-range interaction generated by one-pion exchange becomes increasingly important and must be incorporated explicitly when constraining nucleon-nucleon scattering amplitudes from the finite-volume spectrum~\cite{Green:2021qol,Meng:2021uhz,Raposo:2023oru,Raposo:2025dkb,Dawid:2024oey,Bubna:2024izx}.

In this work, we determine nucleon-nucleon scattering amplitudes at near-physical quark masses using state-of-the-art numerical and theoretical techniques. On the numerical side, we employ the distillation framework~\cite{HadronSpectrum:2009krc}. Using in part the new QUDA-LapH software suite~\cite{Clark:2025cuz}, we compute smeared quark propagators, i.e.~perambulators. This provides the flexibility required to construct large correlation matrices for $NN$ systems with a broad basis of single- and two-hadron interpolating operators.

On the formal side, we employ two recently developed finite-volume frameworks that explicitly account for the one-pion-exchange left-hand cut: the plane-wave quantization condition~\cite{Meng:2021uhz} and the relativistic left-hand-cut formalism~\cite{Raposo:2023oru,Raposo:2025dkb}. These approaches allow the finite-volume spectrum to be related consistently to the infinite-volume nucleon-nucleon amplitudes in the regime where the long-range pion-exchange interaction can no longer be neglected. See Refs.~\cite{Meng:2023bmz,Dawid:2024dgy,Dawid:2025wsn, Yu:2025gzg, Prelovsek:2025vbr, Alharazin:2026lno, Abolnikov:2026sik,Rodas:2026zmm,PitangaLachini:2026lyd,Baiao-Raposo:2026uyo,Baiao-Raposo:2026gte, Baeza_inprep} for recent applications and developments of the left-hand cut methods to lattice-QCD data. 

Using the strategy described above, we compute the pole positions at $M_\pi\approx 200$ MeV, $M_\pi\approx 280$ MeV and compare them against the previous results at $M_\pi\approx 710$ MeV. We discuss the pion-mass dependence of scattering amplitudes and threshold observables. 
This paper is released in parallel with a letter with the highlights of this work and an effective field theory analysis of the results~\cite{Baeza-Ballesteros:2026qxo}.

This paper is organized as follows. In \Cref{sec:setup}, we describe the lattice setup, the operators and the extraction of the spectrum.
In \Cref{sec:FV}, we describe the finite-volume formalism used for the extraction of scattering amplitudes from the lattice energies, while in \Cref{sec:methodology} the procedure of our corresponding analysis is presented. The main results of this work are presented in \Cref{sec:results}. We conclude in \Cref{sec:conclusions}.
This work has several appendices.
\Cref{app:disp} discusses the single-nucleon dispersion relation.
\Cref{app:stability} describes additional details about the energy determination, \Cref{app:standardFV} compares the results from the standard finite-volume formalism to the left-hand cut case, and \Cref{app:tables} collects more information regarding the fits.

\section{Finite-volume spectrum from lattice QCD}
\label{sec:setup}

\subsection{Lattice QCD setup}

This work is based on $N_f = 2+1$ CLS ensembles~\cite{Bruno:2014jqa}, which use nonperturbatively $O(a)$-improved Wilson fermions and the tree-level $O(a^2)$-improved Lüscher-Weisz gauge action.

We use three ensembles, listed in \Cref{tab:ensems}, with pion masses ranging from $\approx 710$ MeV to $\approx 200$ MeV. The heaviest ensemble, C103\footnote{In the CLS convention, this ensemble would be C150.}, is SU(3) symmetric and was generated in Ref.~\cite{Horz:2020zvv}, with the current dataset presented in Ref.~\cite{BaryonScattering:2025ziz}. The intermediate ensemble, D251~\cite{Kuberski:2024pms}, with a pion mass of roughly $280$ MeV, lies on the chiral trajectory with $2m_\ell + m_s = {\rm const.}$, while the lightest ensemble, J250, has a pion mass of $200$ MeV and physical strange quark mass. The J250 ensemble was generated by a subset of the authors, and this work marks its first use.

The lattice spacing for C103 is $a\approx 0.0865$ fm, while for D251 and J250 it is approximately $a\approx 0.0642$ fm. The precise values vary slightly depending on the method used for scale setting~\cite{Bruno:2016plf,Strassberger:2021tsu,RQCD:2022xux}. For the purposes of this work, it suffices to use the quoted numbers. Following Ref.~\cite{Green:2021qol}, we take the lattice spacing for $\beta = 3.70$ from Ref.~\cite{Bruno:2016plf} and scale it to the values of $\beta$ employed in this work using the gradient-flow scale $t_0$ at the SU(3)-symmetric point.
Single-hadron masses of the pion, kaon and nucleon are shown in \Cref{tab:masses_constants}. As can be seen, $M_\pi L > 4$ for all ensembles.

\begin{table*}
  \centering
  \tabcolsep=0.15cm
  \begin{tabular}{c c c c c c c c c c c c}
    \toprule
    & $(L/a)^3 \times (T/a)\phantom{^3}$ & $M_\pi \, [\mathrm{MeV}]$ & $M_K \, [\mathrm{MeV}]$ & $M_N \, [\mathrm{MeV}]$ & $N_\mathrm{cfg}$ & $N_\mathrm{tsrc}$ & $\beta$ & $\kappa_{ud}$ & $\kappa_s$ & $a \, [\mathrm{fm}]$ & $N_\mathrm{ev}$ \\
    \midrule
    C103 & $48^3 \times 96$ & 709.0(8.2) & 709.0(8.2) & 1603(19) & 1490 & 8 & 3.40 & \multicolumn{2}{c}{0.136497611186012} & 0.0865(10) & 384 \\
    D251 & $64^3 \times 128$& 283.2(3.1) & 462.2(5.0) & 1039(12) & 209 & 128 & 3.55 & 0.137140 & 0.13672086 & 0.0642(7) & 160 \\
    J250 & $64^3 \times 192$& 199.4(2.4) & 501.0(5.5) & 969(11) & 265 & 96 & 3.55 & 0.1372067 & 0.136546844 &  0.0642(7) & 160 \\
    \bottomrule
  \end{tabular}
  \caption{Details on the ensembles used in this work: lattice geometry, hadron masses, number of configurations $N_{\rm cfg}$, number of source times $N_\mathrm{tsrc}$, gauge coupling parameter $\beta$, light and strange hopping parameters $\kappa_{ud}$ and $\kappa_s$, lattice spacing $a$, and number of eigenvectors used in distillation $N_{\rm ev}$. All three ensembles have (anti)periodic temporal boundary conditions.
  }
  \label{tab:ensems}
\end{table*}

\begin{table*}
   \centering
   \begin{tabular}{c c c c c c c c}
     \toprule
     & $a M_\pi$ & $a M_K$ & $aM_N$ &  $M_\pi L$ & $M_N L$ \\
     \midrule
     C103 & 0.310810(95) & 0.310810(95) & 0.70263(63) & 14.9189(46) & 33.726(30) \\
     D251 & 0.09214(16)  & 0.15039(10) & 0.3381(11) & 5.897(10) & 21.636(73)
  \\
     J250 & 0.06487(30) & 0.163001(83) & 0.31550(93) & 4.152(19) & 20.192(59)
  \\
     \bottomrule
   \end{tabular}
   \caption{Masses for the ensembles used in this work. $M_\pi$ and $M_K$ are determined from direct fit to the single-pion and kaon correlators, while the nucleon mass $M_N$ is obtained from a conspiracy fit to the lowest lying level in the $T_{1g}$ irrep in the rest frame.}
   \label{tab:masses_constants}
\end{table*}

\subsection{Distillation setup}

Correlation functions are
constructed using the distillation framework~\cite{HadronSpectrum:2009krc}.\footnote{The correlation functions on C103 were computed using the stochastic variant~\cite{Morningstar:2011ka} of distillation. Further details can be found in Refs.~\cite{Horz:2020zvv,BaryonScattering:2025ziz}.} Specifically, we construct single-nucleon operators of azimuthal component of isospin $i$ and spin $m$,
\begin{equation} \label{eq:localN}
    N^{i}_{m}\left(\boldsymbol x, t\right)  = \sum_{\substack{\alpha, \beta, \gamma\\ A, B, C\\ a, b, c}} c^{ABC, i}_{\alpha,\beta,\gamma, m} \,\varepsilon_{abc} \, \tilde{q}_{a \alpha}^A(x) \tilde{q}_{b \beta}^B(x) \tilde{q}_{c \gamma}^C(x),
\end{equation}
where $x=(\boldsymbol x, t)$, $A,B,C$ denote flavor indices, $a,b,c$ color indices and $\alpha,\beta,\gamma$ spin indices.
The coefficients
$c^{ABC,i}_{\alpha\beta\gamma,m}$ project the operator onto the desired spin and isospin quantum numbers.
In the distillation approach, we employ smeared quark fields,
\begin{equation}
\tilde{q}_{a \alpha}^A(\boldsymbol{x},t) = \sum_{b, \boldsymbol{y}} \sum_{n=1}^{N_{\rm ev}} v_a^{(n,t)}(\boldsymbol{x}) v_b^{(n,t)\ast}(\boldsymbol{y}) q_{b\alpha}^A(\boldsymbol{y},t)
\end{equation}
which are obtained by projecting the local quark fields to the space spanned by the lowest $N_{\text {ev}}$ eigenvectors $v_a$ of the covariant three-dimensional Laplacian operator, i.e.\ the Laplacian-Heaviside (LapH) subspace.
Momentum projected nucleon operators, $N^{i}_{m}\left(\boldsymbol p, t\right)$, can be obtained by applying a Fourier transformation to \Cref{eq:localN}.

In order to determine the two-nucleon spectrum, we use two-nucleon interpolators of the form:
\begin{equation}
   \left[N N\right]^{I s m_s}_{\boldsymbol{p}_1, \boldsymbol{p}_2}(t) =  \sum_{m_j, i_j } c^{I sm_s}_{i_1 i_2 m_1 m_2}N^{i_1}_{m_1}\left(\boldsymbol p_1, t\right) N^{i_2}_{m_2}\left(\boldsymbol p_2, t\right),
\end{equation}
where $I$ and $s, m_s$ denote the total isospin and spin quantum numbers.
The two-nucleon interpolator is then projected to definite irreducible representation (irrep) of the finite-volume group as:
\begin{multline}
    \left[NN\right]^{I,s,\Gamma,\lambda,n_\Gamma}
      _{\boldsymbol{p}_1,\boldsymbol{p}_2}(t)
    = \\
    \sum_{\substack{R\in{\rm LG}(\boldsymbol P) \\
                    m_s}}
    c^{m_s}_{\Gamma,\lambda,n_\Gamma}(R)
    \left[NN\right]^{Is\,m_s}
      _{R\boldsymbol{p}_1,R\boldsymbol{p}_2}(t).
\end{multline}
where $\Gamma$ and $\lambda$ label the irrep and its row, $ n_\Gamma$ the multiplicity of each irrep for a particular momentum combination, and ${\rm LG}(\boldsymbol P)$ is the corresponding little group composed of rotations $R$, and it is defined by the total momentum $\boldsymbol P = \boldsymbol p_1 + \boldsymbol p_2$. 

\begin{table*}
  \centering
  \begin{tabular}[t]{ccccc}
    $I$ & $\Gamma(d^2)$ & $(n_1^2,n_2^2)$ & $N_0$ & $N_1$ \\
    \toprule
    0 & $T_{1g}(0)$ & (0,0) &   & 1 \\
      &            & (1,1) &   & 2 \\
      &            & (2,2) &   & 3 \\
      &            & (3,3) &   & 2 \\
    \cmidrule{2-5}
      & $A_2(1)$   & (0,1) &   & 1 \\
      &            & (1,2) &   & 2 \\
      &            & (1,4) &   & 1 \\
      &            & (2,3) &   & 2 \\
    \cmidrule{2-5}
      & $E(1)$     & (0,1) &   & 1 \\
      &            & (1,2) & 1 & 3 \\
      &            & (1,4) &   & 1 \\
      &            & (2,3) & 1 & 3 \\
    \cmidrule{2-5}
      & $A_2(2)$   & (0,2) &   & 1 \\
      &            & (1,1) &   & 1 \\
      &            & (1,3) &   & 2 \\
      &            & (2,2) &   & 1 \\
    \bottomrule
  \end{tabular}\hfill
  \begin{tabular}[t]{ccccc}
    $I$ & $\Gamma(d^2)$ & $(n_1^2,n_2^2)$ & $N_0$ & $N_1$ \\
    \toprule
    0 & $B_1(2)$   & (0,2) &   & 1 \\
      &            & (1,1) &   & 1 \\
      &            & (1,3) & 1 & 1 \\
      &            & (2,2) & 1 & 2 \\
    \cmidrule{2-5}
      & $B_2(2)$   & (0,2) &   & 1 \\
      &            & (1,1) & 1 & 1 \\
      &            & (1,3) &   & 2 \\
      &            & (2,2) & 1 & 2 \\
    \cmidrule{2-5}
      & $A_2(3)$   & (0,3) &   & 2 \\
      &            & (1,2) &   & 1 \\
    \cmidrule{2-5}
      & $E(3)$     & (0,3) &   & 1 \\
      &            & (1,2) & 1 & 3 \\
    \bottomrule
  \end{tabular}\hfill
  \begin{tabular}[t]{ccccc}
    $I$ & $\Gamma(d^2)$ & $(n_1^2,n_2^2)$ & $N_0$ & $N_1$ \\
    \toprule
    0 & $A_2(4)$   & (1,1) &   & 1 \\
      &            & (0,4) &   & 1 \\
      &            & (2,2) &   & 1 \\
      &            & (1,5) &   & 2 \\
      &            & (3,3) &   & 1 \\
    \cmidrule{2-5}
      & $E(4)$     & (1,1) &   & 1 \\
      &            & (0,4) &   & 1 \\
      &            & (2,2) & 1 & 2 \\
      &            & (1,5) & 1 & 3 \\
      &            & (3,3) & 1 & 2 \\
    \midrule
    1 & $A_{1g}(0)$ & (0,0) & 1 &   \\
      &            & (1,1) & 1 &   \\
      &            & (2,2) & 1 &   \\
      &            & (3,3) & 1 &   \\
    \bottomrule
  \end{tabular}\hfill
  \begin{tabular}[t]{ccccc}
    $I$ & $\Gamma(d^2)$ & $(n_1^2,n_2^2)$ & $N_0$ & $N_1$ \\
    \toprule
    1 & $A_1(1)$   & (0,1) & 1 &   \\
      &            & (1,2) & 1 & 1 \\
      &            & (1,4) & 1 &   \\
      &            & (2,3) & 1 & 1 \\
    \cmidrule{2-5}
      & $A_2(2)$   & (0,2) & 1 &   \\
      &            & (1,1) & 1 & 1 \\
      &            & (1,3) & 1 & 1 \\
      &            & (2,2) & 1 & 2 \\
    \cmidrule{2-5}
      & $A_2(3)$   & (0,3) & 1 &   \\
      &            & (1,2) & 1 & 1 \\
    \cmidrule{2-5}
      & $A_1(4)$   & (1,1) & 1 &   \\
      &            & (0,4) & 1 &   \\
      &            & (2,2) & 1 & 1 \\
      &            & (1,5) & 1 & 1 \\
      &            & (3,3) & 1 & 1 \\
    \bottomrule
  \end{tabular}
  \caption{Number of spin-zero interpolating operators $N_0$ and
    spin-one operators $N_1$ for each isospin $I$, irrep $\Gamma$ with
    total momentum $\boldsymbol P = (2\pi/L) \boldsymbol d$, and
    momentum shell $(\boldsymbol p_1,\boldsymbol p_2)$ with
    $p_i^2=(\tfrac{2\pi}{L})^2n_i^2$. Blank entries correspond to zero
    operators.}
  \label{tab:operators}
\end{table*}

There is a one-to-one mapping between noninteracting two-nucleon
states and our interpolating operators.
In each frame and irrep, we
choose a maximum value of $p_1^2+p_2^2$ and construct all two-nucleon
operators up to this maximum. The number of such operators for each
pair of $(p_1^2,p_2^2)$ and two-nucleon spin is listed in
\Cref{tab:operators}; in many cases, the corresponding noninteracting
energy levels are multiply degenerate. In this work, we restrict our attention to irreps sensitive to $S$ waves.

We compute quark propagators projected onto the \(4N_{\rm ev}\)-dimensional LapH subspace, i.e., the \textit{perambulators}. 
The contractions are computed as described in Ref.~\cite{Green:2021qol}.
First, we partially contract one source baryon and one sink baryon to form ``blocks,'' with a cost proportional to \(N_{\rm ev}^4\). Some blocks are reused multiple times; we therefore store them in memory until their last use, avoiding unnecessary recomputation. Contracting two blocks to form a baryon-baryon correlation function is inexpensive, scaling as \(N_{\rm ev}^2\). To maximize statistics, we average over all equivalent total momenta \(\boldsymbol{P}\) and irrep rows.

\subsection{Spectrum fitting strategy}
\label{sec:conspiracy}

We extract finite-volume stationary state energies via the standard Hermitian correlator matrix method~\cite{Michael:1982gb, Michael:1985ne, Fox:1981xz, Ishikawa:1982tb}. For an operator basis of size $N_{\rm op}$ belonging to a given lattice irreducible representation (irrep), the elements of the $N_{\rm op}\times N_{\rm op}$ correlation matrix have spectral representations given by
\begin{equation}\label{eq:correlator_matrix}
\begin{aligned}
    C_{ij}(t) &= \bra{\Omega} O_i(t) O_j^{\dagger}(0)\ket{\Omega}\\
    &= \sum_{n=0}^\infty Z_i^{(n)} Z_j^{\dagger\,(n)}e^{-E_n t},
\end{aligned}
\end{equation}
where $\ket{\Omega}$ is the vacuum state, $Z_i^{(n)} = \bra{\Omega}O_i \ket{n}$ are the overlap factors, and the sum runs over eigenstates with nonzero overlaps, ordered by increasing energy.

We diagonalize the correlator matrix by solving the generalized eigenvalue problem (GEVP)
\begin{equation}
    \label{eq:GEVP}
    C(t_d)\, V = C(t_0)\, V\, \Lambda_d,
\end{equation}
where $t_d\geq t_0$, the columns of $V$ contain the eigenvectors, and $\Lambda_d$ is the diagonal matrix of the corresponding eigenvalues. The matrix $V$ is normalized such that $V^{\dagger} C(t_0) V = I$. The matrices $C(t_d)$ and $C(t_0)$ are determined using the mean values of the matrix elements over all configurations.

If $t_0\geq t_d/2$, then one can define an estimator for $E_k$ with residual excited-state contamination $O(e^{-(E_{N_\text{op}}-E_k)t_d})$~\cite{Blossier:2009kd}.
However, we instead determine finite-volume energy levels from the single-pivot matrix
\begin{equation}
    D(t) = V^{\dagger} C(t) V.
\end{equation}
The two times $t_0$ and $t_d$ are chosen so that $D(t)$ is diagonal within statistical errors. In that case, the states $n$ with $k\neq n<N_\text{op}$ have been approximately eliminated from the diagonal correlators:
\begin{equation}
    D_{kk}(t) \approx |\tilde{Z}_k^{(k)}|^2 e^{-E_k\,t} + \sum_{n=N_{\rm op}}^{\infty} |\tilde{Z}_k^{(n)}|^2 e^{-E_n\,t} \, .
\end{equation}
The overlap factor for the $i^{\rm th}$ operator onto the $k^{\rm th}$ state is then obtained from the state's amplitude in the diagonal correlator via
\begin{equation}
\label{eq:overlap_factors}
Z_{i}^{(k)} \approx C_{ij}(t_0) V_{jk} \tilde{Z}_k^{(k)}.
\end{equation}

\begin{figure*}
    \centering
    \includegraphics[width=0.9\linewidth]{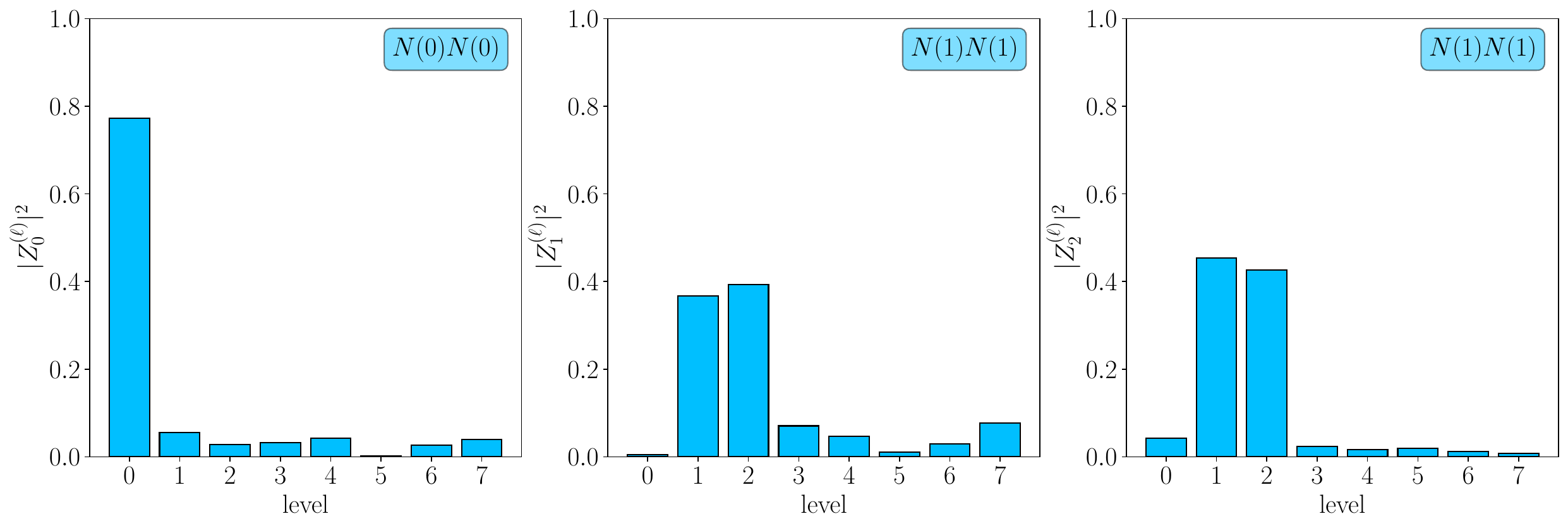}
    \caption{\label{fig:overlaps}Overlap factors for the three lowest momentum operators in the isosinglet $T_{1g}(0)$ irrep in the rest frame on the D251 ensemble. $N(n_1^2)N(n_2^2)$ refers to nucleon interpolators with momentum $p_i^2 = (\frac{2\pi}{L})^2 n_i^2$. See \Cref{tab:operators} for further details on the operators.}
\end{figure*}

To extract the two-nucleon ($NN$) energy levels, we use the recently introduced \textit{conspiracy model}~\cite{BaryonScattering:2025ziz}, in which an $NN$ correlator [$C_{NN}(t)$, given by a diagonal element of $D(t)$] is simultaneously fit alongside two single-nucleon reference correlators. These reference states are selected by identifying the two-nucleon operators that have the largest overlap factors [\Cref{eq:overlap_factors}] with the given $NN$ state.
Overlap factors provide information for identifying energy states and allow us to classify them according to the quantum numbers of the operators with which they have the largest overlap.
In \Cref{fig:overlaps} we show the overlap factors of the three lowest-momentum $NN$ operators for the $T_{1g}(0)$ irrep on the eight deuteron levels in the rest frame from ensemble D251. 
As can be seen, one or two levels dominate for each of the operators, as expected by the degeneracies given in \Cref{tab:operators}.

To describe the residual excited-state contamination that is not removed by the GEVP, i.e.\ starting from $n=N_\text{op}$, the conspiracy fits use a multi-exponential model, which parametrizes the single- and two-nucleon correlators as
\begin{subequations}
    \begin{align}
        \label{eq:conspiracy_N}
        C_{N_i}(t) &= \left(Z_0^{\left(N_i\right)}\right)^2 e^{-E_0^{\left(N_i\right)} t}\left(1 + \sum_{n > 0}|r^{\Ni}_n|^2 e^{-\Delta_{n0}^{\left(N_i\right)}\,t}\right), \displaybreak[0] \\
        \label{eq:conspiracy_NN}
        C_{NN}&(t) = \left(Z_{00}^{\left(NN\right)}\right)^2 e^{-\left(E_0^{\left(N_1\right)} + E_0^{\left(N_2\right)}\right)\,t}\left( e^{-\delta_{00} t} + {} \vphantom{\sum_{\left(n,m\right)\neq(0,0)} |r_{nm}^{\left(NN\right)}|^2 e^{-\left(\Delta_{n0}^{(N_1)} + \Delta_{m0}^{(N_2)} + \delta_{nm}\right)\,t}} \right.\nonumber\\
        & \left.\hspace{1pt}\sum_{\left(n,m\right)\neq(0,0)} |r_{nm}^{\left(NN\right)}|^2 e^{-\left(\Delta_{n0}^{(N_1)} + \Delta_{m0}^{(N_2)} + \delta_{nm}\right)\,t}\right),
    \end{align}
\end{subequations}
where $Z_n^{X}$ is the amplitude for level $n$ and correlator $X$, with ratio amplitudes defined as $r^{X}_n = Z_n^{X}/Z_0^{X}$, similarly for the two-nucleon amplitude $Z_{nm}^{(NN)}$ and ratio $r_{nm}^{(NN)}$. The model includes up to $N_{\rm st}$ single-nucleon states.  The energy gap between the $n$\textsuperscript{th} excited state and the ground state for reference nucleon $N_i$ is 
\begin{equation}
    \Delta_{n0}^{\Ni} = E_n^{\Ni} - E_0^{\Ni}.
\end{equation}
The $NN$ energy is given by 
\begin{equation}
    E^{\left(NN\right)}_{nm} = E^{N_1}_n + E^{N_2}_m + \delta_{nm},
\end{equation}
where $\delta_{nm}$ is the finite-volume lab-frame interaction energy. \Cref{eq:conspiracy_NN} contains $N_{\rm st}^2$ terms when the single nucleons have different momenta, and $N_{\rm st}\left(N_{\rm st} + 1\right)/2$ otherwise.
Typically, $N_{\rm st} = 2$ or 3.
Note that the excited-state energies $E_n^{(N_i)}$ and $E^{(NN)}_{nm}$, $n, m > 0$, are model energies that capture the contributions of the full set of residual excited states and should thus not be interpreted as accurate estimates of physical states.

As such, the conspiracy model allows us to directly parametrize the interaction energies and profit from the cancellation of excited states and correlated noise between the single- and two-nucleon correlators.
In contrast to fits to ratios of correlation functions, which also exploit this correlation, our approach directly fits the original correlation functions and thus preserves the positivity of the overlap factors.
For a detailed comparison of this model with other approaches, such as fitting ratios of correlation functions, see Ref.~\cite{BaryonScattering:2025ziz}.

To stabilize the fits and allow for a sufficient number of exponentials to control excited states, we use Bayesian priors as follows:
\begin{subequations}\label{eq:priors}
    \begin{align}
        E_0 &\sim m_{\rm eff}\left(t_{\rm ref}\right)\times \mathcal{N}\left(1,0.1\right)\\
        Z_0, Z_{00} &\sim Z_{\rm eff}\left(t_{\rm ref}\right)\times\mathcal{N}\left(1,\,1\right)\\
        a\Delta_{n,n-1} &\sim \exp\left[\mathcal{N} \left(\log\left( 2aM_\pi\right), 0.5 \right)\right]\\
        r_n, r_{nm} &\sim \mathcal{N}\left(1, 0.25\right)\\
        \delta_{00} &\sim m_{\rm eff}^{R}\left(t_{\rm ref}\right)\times \mathcal{N}\left(1,\, 1\right),
    \end{align}
while for $\delta_{nm}$ we use two different choices:
\begin{align}
    \delta_{nm}^{(\mathrm{Set}\,1)}
    &\sim \delta_{00}
    + m_{\mathrm{eff}}^R(t_{\mathrm{ref}})
      \times\mathcal{N}(0,1),\\
    \delta_{nm}^{(\mathrm{Set}\,2)}
    &\sim m_{\mathrm{eff}}^R(t_{\mathrm{ref}})
      \times\mathcal{N}(0,1),
\end{align}
where sets 1 and 2 denote the two analysis strategies described
below. Here, $\mathcal{N}(\mu,\sigma)$ denotes the Gaussian normal centered at $\mu$ with width $\sigma$, and the effective masses and amplitudes are
    \begin{align}
        m_{\rm eff}(t) &= \log \left[C(t)/C(t+1)\right]\\
        Z_{\rm eff}(t) &= \sqrt{e^{m_{\rm eff}(t)\,t}\,C(t)}.
    \end{align}
  \end{subequations}
Here, $m_{\rm eff}^R(t)$ is the effective mass of the ratio between the $NN$ correlator and the reference single-nucleon correlators, $C^R(t) = C_{NN}(t)/\left[C_{N_1}(t)\, C_{N_2}(t)\right]$, and $t_{\rm ref}/a$ has been chosen in the range $[17,26]$ where the effective mass of the ratio of correlators exhibits the onset of a plateau. The pion masses for each ensemble are listed in \Cref{tab:masses_constants}. To propagate prior fluctuations into the statistical uncertainty, for each bootstrap resampling we randomly sample new prior means from the posterior distribution with covariance matrix rescaled by a constant factor. Further details on this procedure, as well as an analysis of the stability of the resulting bootstrap distributions, are given in \Cref{app:stability}.
\subsection{Extracted spectrum}
\label{sec:spectrum}
To ensure a robust extraction of the finite-volume spectrum, we have analyzed
the stability of the fitted energy levels against variations in the correlator analysis choices, particularly the minimum
fitting time, $\tmin$, and the number of states, $\nstates$.
To this end, we defined two independent strategies to extract the spectrum.
The spectra resulting from these analyses are denoted ``Set\,1'' and ``Set\,2'' in the following. Bootstrap results for the energy levels analyzed in this work can be found as ancillary material to the arXiv submission; see \cref{app:stability}.

\subsubsection{Set 1}
The $t_{\rm min}$ and $N_{\rm st}$ choices for ``Set 1'' are determined by probing 
the stability of the fit results across several different choices of ($t_{\rm min}$, $N_{\rm st}$).
For the final D251 spectrum,
we chose the conspiracy model with $\tminnstates = (6a,\,3)$ after evaluating its performance across all irreps.
On this ensemble, different choices of $\tminnstates$ induce variations in the
energy shifts $\delta_{00}$ which are well within a standard deviation of one another.

\begin{figure}
    \centering
    \includegraphics[width=\columnwidth]{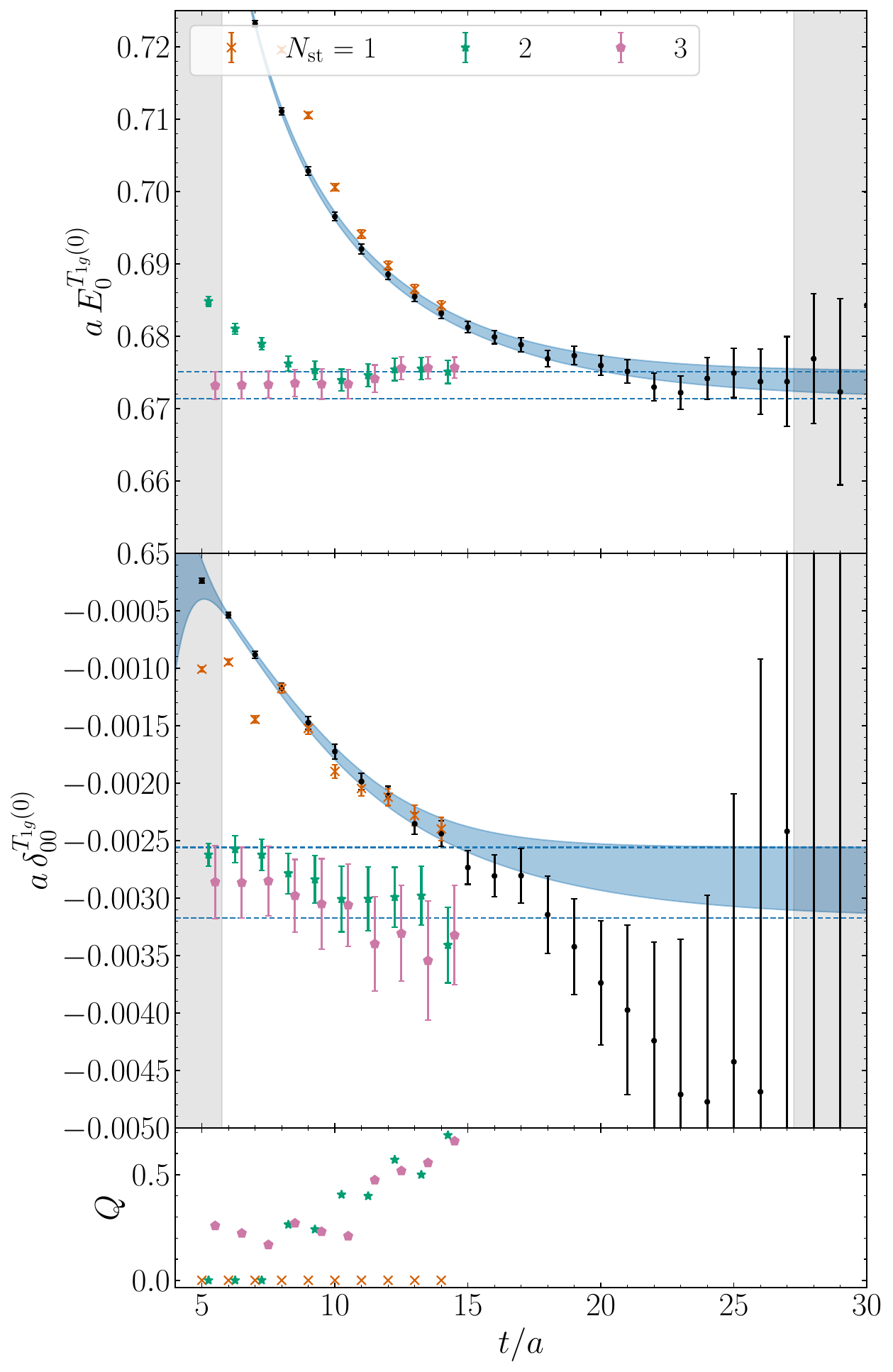}
    \caption{\label{fig:D251_T1g_stability}Determination of the D251 $T_{1g}(0)$ ground-state energy versus Euclidean time, in lattice units: total energy (top), interaction energy $\delta_{00}$ (middle), and fit quality $Q$, also known as the $p$-value in a frequentist framework (bottom). The colored points in all three panels show results from fits performed with varying $t_\text{min}$ and varying $N_{st}$ (see legend). In the upper two panels, the black points show the effective (interaction) energy of the diagonal correlator versus $t$ and the blue band illustrates the posterior fit for our final chosen parameters $\tminnstates = (6a,\,3)$. The chosen fit range is bounded by the gray regions.}
\end{figure}

To illustrate the robustness of the extraction, in \Cref{fig:D251_T1g_stability},
the effective energy and interacting energy shift are shown for 
the ground state in the $T_{1g}(0)$ irrep on the D251 ensemble.
This energy level extraction exhibits stability across a wide range of $\tminnstates$ combinations,
and the ground state energy levels and interaction energies remain statistically consistent.

\begin{figure}
    \centering
    \includegraphics[width=\columnwidth]{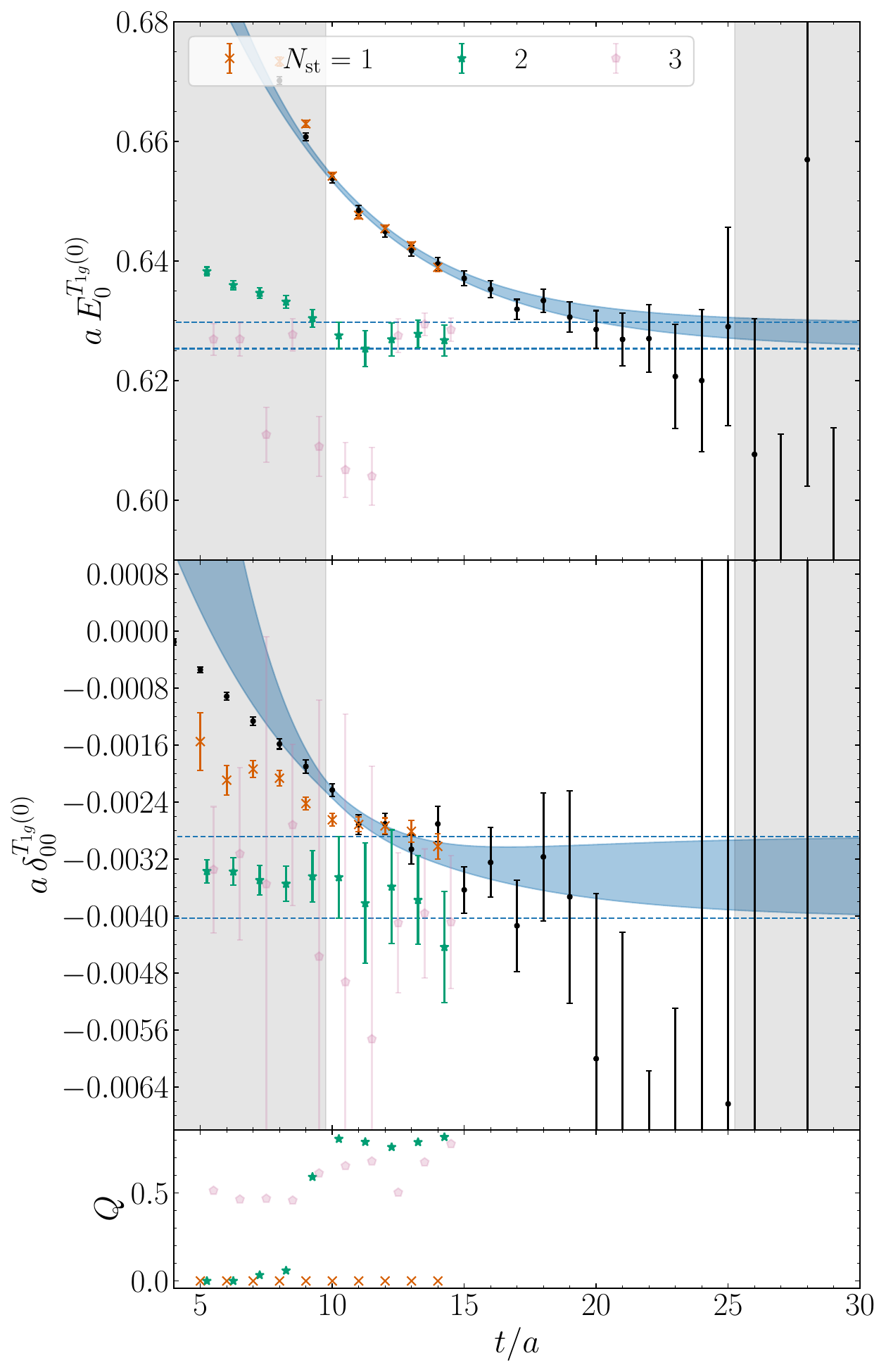}
    \caption{\label{fig:J250_T1g_stability}Stability for the J250 $T_{1g}(0)$ finite-volume ground state; all details are analogous to \Cref{fig:D251_T1g_stability}. The transparent data correspond to $\nstates=3$, demonstrating the instability of our fits when attempting to isolate this number of states. Therefore, only two-state fits at sufficiently late times are considered. The final chosen fit is $\tminnstates=(10a,\,2)$, which yields the blue band.}
\end{figure}

\begin{figure*}[h!]
    \centering
    \includegraphics[width=\linewidth]{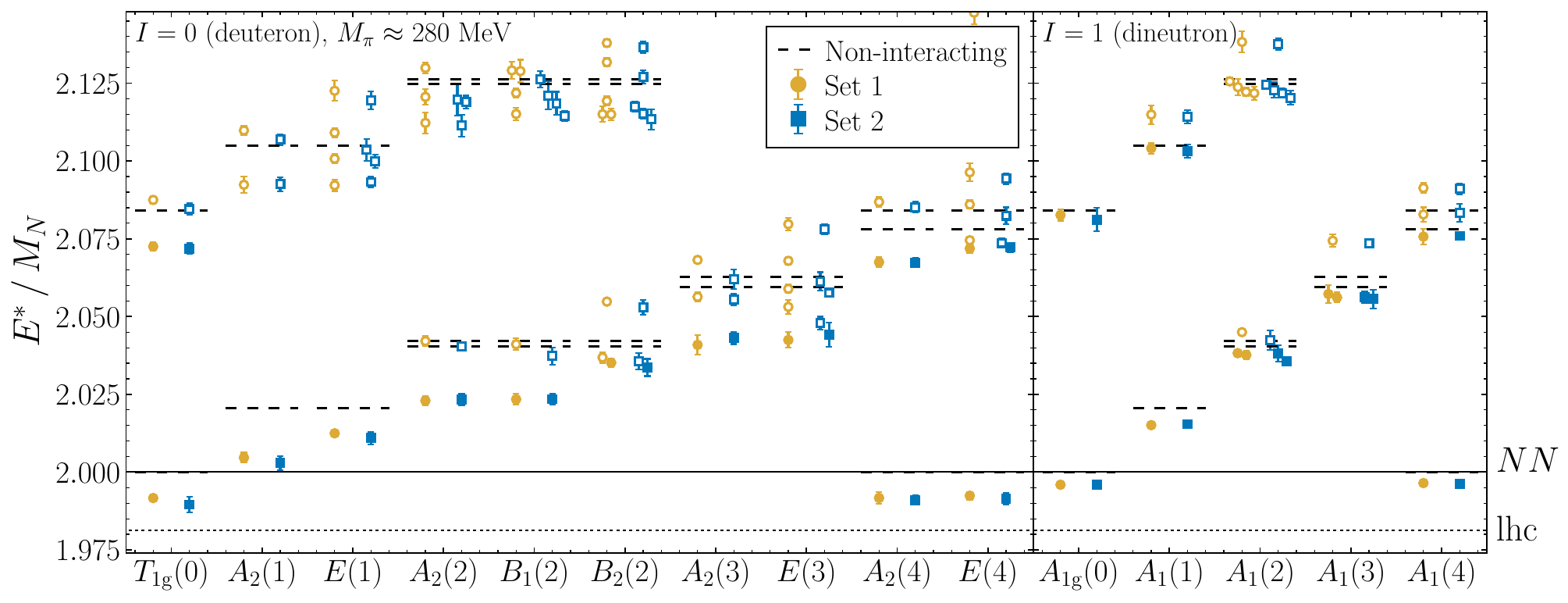} \\  \vspace{0.5cm}
    \includegraphics[width=\linewidth]{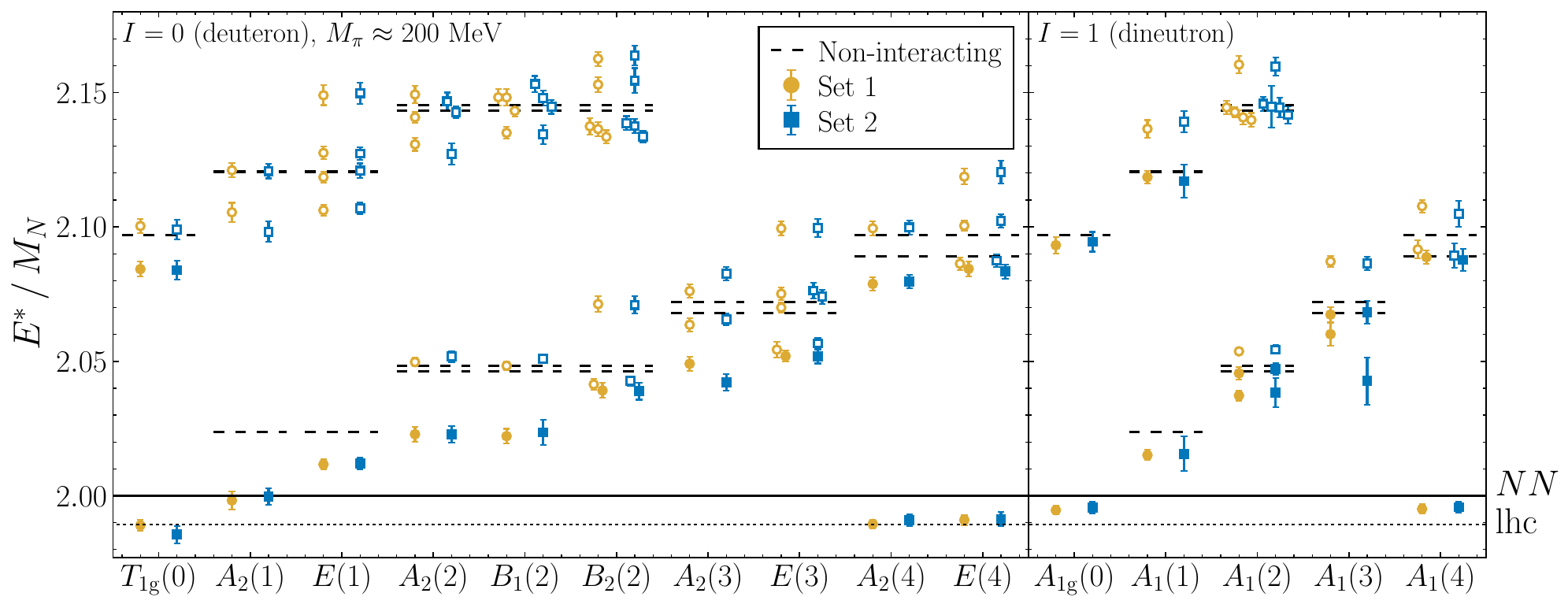}
    \caption{\label{fig:fullspectrum} Full spectra on the D251 ensemble (top) and J250 ensemble (bottom). Two versions of the spectrum analysis are shown, labeled as Set 1 (orange circles) and Set 2 (blue squares). Open markers denote energy levels that are not included in the analysis, while filled markers indicate levels used in the determination of the scattering amplitudes.
    }
\end{figure*}

Also on J250, the extracted energies are stable
under the variation of $\tmin$; however, for a few levels, in particular in the $I = 0$ channel, the fits
with $\tmin = 11a$ exhibit larger errors, so that stability beyond $\tmin = 10a$ cannot be judged in these cases; see \Cref{fig:J250_stability} for a more detailed analysis.
Similarly, our stability analysis shows that we are not able to resolve three states on this ensemble. 
Consequently, we chose a conspiracy model
with $\tminnstates=(10a,\,2)$ for our final spectrum extraction, which
yields results consistent with those obtained from other $\tminnstates$ choices.
The stability analysis for the ground-state energy level in the $T_{1g}(0)$ irrep is shown in \Cref{fig:J250_T1g_stability}. 
We have included $\nstates=3$ fits to explicitly demonstrate how the noise present in our data prevents us from reliably extracting more than two states. In the top panel, notice that the three-state fits fluctuate sporadically. We observe that these three-state fits are highly sensitive to priors and can isolate erroneous states. Notice also the dramatically larger error bars for the three-state $\delta_{00}$ extraction. Conversely, the two-state fits are insensitive to priors and remain robust throughout the entire parameter range.
The spectrum, for both $I=0$ and $I=1$, obtained through the stability analysis described above is shown as ``Set 1'' in \Cref{fig:fullspectrum}.  
More details for both D251 and J250 are given in \Cref{app:stability}.

\subsubsection{Set 2}
The second methodology aims to assess the systematic error deriving from variations in $\tmin$ and $\nstates$ and to include it in the total error budget, by performing model and data-subset averaging, based on the  Akaike Information Criterion
(AIC)~\cite{Akaike1,Jay:2020jkz,Frison:2023lwb,Neil:2022joj,Neil:2023pgt}. We performed several fits to the single-nucleon
correlators with different $N_{\rm st}$, by varying $t_{\rm min}$ at fixed $t_{\rm max}$.
The \textit{optimal} description, i.e.\ a $(t_{\rm min},N_{\rm st})$ pair, is identified as the fit with the smallest AIC value.
Once the optimal description for single-nucleon correlators is identified, the conspiracy model is applied as detailed
earlier in this section, by performing $N_{\rm fit}$ fits that differ only in the choice of $t_{\rm min}$ for the two-nucleon correlators,
at fixed $t_{\rm max}$ and numbers of exponentials, i.e.\ $N_{\rm st}^2$ or $N_{\rm st}(N_{\rm st} + 1) /2$ in the case
of distinguishable and indistinguishable scatterers, respectively. The central value for the fit parameters is determined by taking
a weighted average over the $N_{\rm fit}$ fits, with the weight of the $i^{\rm th}$ fit given by
\begin{equation}
\label{eq:AIC_weight}
    w_i = \frac{e^{-\frac{1}{2} {\rm AIC}_i}}{\sum_{i=1}^{N_{\rm fit}} e^{-\frac{1}{2} {\rm AIC}_i}} ,
    \qquad
    \mathrm{AIC}_i = \chi_i^2 - 2\,\mathrm{dof}_i,
\end{equation}
where ${\rm AIC}_i$ is the AIC value obtained from the $i^{\rm th}$ fit. For each bootstrap resampling, one of the $N_{\rm fit}$
models is sampled with probability $w_i$ and is used to fit the given resampling. This procedure provides the estimate of the fit
parameters on each bootstrap resampling and the error is then obtained by applying standard bootstrap methods.
This procedure aims at accounting for the systematic uncertainty arising from the choice of $t_{\rm min}$.
The spectrum, for both $I=0$ and $I=1$, obtained through the AIC-based analysis is labeled as ``Set 2'' in \Cref{fig:fullspectrum}.  
Further details on the spectrum extraction analyses are given in \Cref{app:stability}.

\section{Scattering amplitudes from finite-volume energies}
\label{sec:FV}

The lattice QCD results for the $NN$ finite-volume spectrum can now be used to constrain the infinite-volume $NN$ scattering amplitude. In this section, we will first outline some general properties of the $NN$ system and then describe the different methods we have used to constrain the scattering amplitude.

\subsection{\texorpdfstring{$NN$}{NN} scattering amplitudes}
\label{sec:NNscatteringamplitudes}

Before explaining how the $NN$ scattering amplitudes are described, we summarize the set of quantum numbers that characterize the two-nucleon system, and how they allow the analysis of lattice spectra to be separated into multiple sectors. The first quantum number that characterizes the $NN$ system is total isospin, as we are working with $m_u=m_d$. Total isospin can take two values: $I=0$ (known as the deuteron channel) and $I=1$ (dineutron channel). On the other hand, we have the quantum numbers related to angular momentum. These include orbital angular momentum, $\ell$, and its third component, $m_\ell$, as well as spin angular momentum, characterized by  $s$ and $m_s$. As each nucleon has spin $1/2$, the two-nucleon system can appear with spin $s=0$ or $s=1$. Therefore, the angular momentum quantum numbers of the $NN$ system are fully determined by $\ell, m_\ell, s, m_s$, or alternatively in terms of the total angular momentum: $J ,m_J ,\ell, s$. 

The complete orbital, spin and isospin wave function of the $NN$ system must be antisymmetric, which imposes restrictions on the allowed combinations of quantum numbers. Moreover, the total parity of the system is conserved and given by $P=(-1)^\ell$, which also restricts the allowed mixings. As a consequence of these selection rules, the total spin of the two-nucleon system is also conserved. However, different values of $\ell$ are allowed to mix, as long as the total $J$ and parity are the same. \Cref{tab:NNquantumnumbers} summarizes some allowed amplitudes with $J \leq 2$ for each isospin channel. The focus of this work is on the $(I,J^P) = (0, 1^+)$ and  $(I,J^P) = (1, 0^+)$ cases. Note that it is also customary to refer to partial waves using the spectroscopic notation, i.e. $^{2s+1}\ell_J$.  In this case, the relevant waves are $^3S_1$ and $^3D_1$ for $I=0$, and $^1S_0$ for $I=1$.

\begin{table}[h!]
\centering
\begin{tabular}{c c c c c c c}
\hline\hline
\(J^P\)
& \(0^{+}\)
& \(0^{-}\)
& \(1^{+}\)
& \(1^{-}\)
& \(2^{+}\)
& \(2^{-}\) \\
\hline
\(I=0\)
& --
& --
& \(\{{}^{3}S_1,{}^{3}D_1\}\)
& \({}^{1}P_1\)
& \({}^{3}D_2\)
& -- \\
\(I=1\)
& \({}^{1}S_0\)
& \({}^{3}P_0\)
& --
& \({}^{3}P_1\)
& \({}^{1}D_2\)
& \({}^{3}P_2\) \\
\hline\hline
\end{tabular}
\caption{Allowed \(NN\) partial waves with \(J\leq2\) and \(\ell\leq2\),
expressed in spectroscopic notation. The deuteron channel $I=0$, $J^P=1^+$ features mixing between different values of $\ell$. }
\label{tab:NNquantumnumbers}
\end{table}

Elastic two-particle scattering amplitudes can be described by the K-matrix formalism, which implements S-matrix unitarity. The on-shell scattering amplitude, $\cM$, is defined from the S matrix as
\begin{equation}
    S^{J^P}_{\ell,\ell'}(E_{\rm cm})=\delta_{\ell, \ell'}+2 i \rho \, \cM^{J^P}_{\ell,\ell'}(E_{\rm cm})\,,
\end{equation}
where  $\rho=p/(8\pi E_{\rm cm})$ with $p$ being the relative momentum in the cm frame, $E_{\rm cm}^2 = 4(M_N^2 + p^2)$. From here, the K matrix, $\cK$, is defined as follows,
\begin{equation}
    [\cM^{-1}(E_{\rm cm})]^{J^P}_{\ell,\ell'} =     [\cK^{-1}(E_{\rm cm})]^{J^P}_{\ell,\ell'} - i \rho \delta_{\ell, \ell'} \,.
    \label{eq:M_Kmat_rel}
\end{equation}
 Unitarity implies that $[\mathcal K^{-1}(E_{\rm cm})]^{J^P}_{\ell,\ell'}$ is a real-valued, meromorphic function in the elastic region and just below the threshold.

The dynamics of the $NN$ interactions is encoded in the energy dependence of the K matrix. Choosing a suitable parametrization for the latter is thus an important topic. Based on the threshold behavior, one expects
\begin{equation}\label{eq:thresholdsuppresionpartialwaves}
     [\cK^{-1}]^{J^P}_{\ell, \ell'} = p^{-(\ell +\ell')} \left(c^{J^P}_{\ell, \ell'} + O(p^2) \right),
\end{equation}
where $c^{J^P}_{\ell, \ell'}$ labels a constant, and the next order correction is of order $p^2$. 

In the single-channel limit, the K matrix is directly given by the phase shift $\delta$. 
For instance, in the $S$-wave-only case
\begin{equation}
    [\mathcal K^{-1}]^{s^+}_{0, 0}=  \frac{p \cot \delta^I_0}{8\pi E_{\rm cm}}.
    \label{eq:K_delta_rel}
\end{equation}
The $S$-wave phase shift is frequently parametrized using an effective range expansion (ERE) as:
\begin{equation}
    \label{eq:ERE}
    p \cot \delta^I_0 = -\frac{1}{a_0} + \frac{1}{2} r_0 p^2 + P_0 p^4,
\end{equation}
where $a_0$, $r_0$ and $P_0$ are known as the scattering length, effective range and shape parameter, respectively. More complicated parametrizations are, however, also possible.
One particular form, which allows for a divergence in $p\cot\delta_0$, i.e.\ a zero in the K matrix, is a three-parameter rational parametrization,
\begin{equation}
    \label{eq:rational3}
    p^{-1}\tan\delta_0^I = \frac{C_0 + C_1 p^2}{1 + C_2 p^2} .
\end{equation}
We use this model for some of the isospin-1 fits, and call it ``rational3''.

In the $I=0$, $J^P=1^+$ case, there is physical mixing between $\ell=0$ and $\ell=2$. (Note that the sign of our off-diagonal amplitude with $|\ell'-\ell|=2$ is the opposite of \rrcite{Briceno:2013lba, Briceno:2013bda, Briceno:2014oea, Green:2022rjj} in order to be consistent with $NN$ phenomenology~\cite{Blatt:zz1952a, Blatt:1952zz, Stapp:1956mz, Rijken:1991at}; see \rcite{Baeza_inprep}.) One option for incorporating this mixing into the parametrization of the K matrix is the well-known Blatt-Biedenharn decomposition \cite{Blatt:zz1952a}:
\begin{equation}
   \label{eq:BBparam}   \left[\mathcal{K}^{-1}\right]^{J^P=1^+}= R^T(\epsilon_1) \left(\begin{array}{cc}
 \frac{p \cot \delta_{1\alpha}}{ 8 \pi E_{\rm cm}} & 0 \\
0 & \frac{p \cot \delta_{1\beta}}{ 8 \pi E_{\rm cm}}
\end{array}\right)R(\epsilon_1),
\end{equation}
where $R(\epsilon_1)$ is a rotation matrix defined as:
\begin{equation}
    R(\epsilon_1) = \left(\begin{array}{cc}
\cos (\epsilon_1) & \sin (\epsilon_1) \\
-\sin (\epsilon_1) & \cos (\epsilon_1)
\end{array}\right).
\end{equation}
Near threshold and for small mixing angle, the ``$\alpha$ wave'' and ``$\beta$ wave'' phase shifts are dominated by $\ell=0$ and $\ell=2$, respectively. In particular, the expected threshold behavior is
\begin{equation}
     \delta_{1\alpha} = O(p),\quad  \delta_{1\beta} = O(p^5),\quad \epsilon_1 = O(p^2).
\end{equation}
This decomposition is especially useful, as the deuteron arises as a subthreshold pole in the first Riemann sheet of the $\alpha$ wave.

An alternative to the Blatt-Biedenharn decomposition is the so-called Stapp decomposition~\cite{Stapp:1956mz}. For this, the scattering matrix for $J^P=1^+$ quantum numbers is decomposed as
\begin{equation}
\label{eq:stappparametrization}
\begin{aligned}
S^{J^P=1^+}
&=
\begin{pmatrix}
\text{e}^{i\bar\delta_{{}^3S_1}} & 0 \\
0 & \text{e}^{i\bar\delta_{{}^3D_1}}
\end{pmatrix}
\begin{pmatrix}
\cos 2\bar\epsilon_1 & i\sin 2\bar\epsilon_1 \\
i\sin 2\bar\epsilon_1 & \cos 2\bar\epsilon_1
\end{pmatrix}
\\[-2pt]
&\quad\times
\begin{pmatrix}
\text{e}^{i\bar\delta_{{}^3S_1}} & 0 \\
0 & \text{e}^{i\bar\delta_{{}^3D_1}}
\end{pmatrix}\,,
\end{aligned}
\end{equation}
where $\bar\delta_{{}^3S_1}$, $\bar\delta_{{}^3D_1}$ and $\bar{\epsilon}_1$ are the so-called bar-nuclear phase shifts and mixing angle, respectively. Compared to the Blatt-Biedenharn decomposition, this has the advantage that the mixing angle is suppressed close to threshold for higher total angular momentum. For $J^P=1^+$ considered here, $\bar\epsilon_1 = O(p^3)$. 
However, the condition for poles in $\mathcal M$ is more complicated in this decomposition than in the Blatt-Biedenharn one, where pole positions can be readily determined from the eigenphases.
It is worth noting that the two decompositions can be related via simple analytic relations~\cite{Stapp:1956mz},
\begin{equation}\label{eq:BB_Stapp}
    \begin{array}{rcl}
     \delta_{1\alpha}+\delta_{1\beta} & = & \bar\delta_{{}^3S_1} + \bar\delta_{{}^3D_1}\,,  \\[5pt]
      \sin(\bar\delta_{{}^3S_1}-\bar\delta_{{}^3D_1}) & = & \tan2\bar\epsilon_1\,/\,\tan2\epsilon_1\,, \\[5pt]
      \sin(\delta_{1\alpha}-\delta_{1\beta}) & = & \sin2\bar\epsilon_1\,/\,\sin2\epsilon_1\,.
    \end{array}
\end{equation}

The $NN$ scattering amplitudes may contain either virtual- or bound-state poles. In the $I=1$ channel, these are obtained as solutions of
\begin{equation}
    p\cot\delta_0^I-ip=0,
\end{equation}
with the analogous condition for the $\alpha$ wave in the $I=0$ channel (using the Blatt-Biedenharn decomposition). Solutions can be written as $p_{\rm pole}=i\kappa$, where $\kappa$ is the signed binding momentum. A negative value of $\kappa$ corresponds to a virtual state, while a positive value corresponds to a bound state.

A relevant feature of $NN$ scattering is the presence of left-hand cuts in the scattering amplitudes, which are induced by one-pion exchange in the $t$ channel. The pole at $t=M_\pi^2$ turns into a cut along the negative energy axis after partial wave projection, beginning at:
\begin{equation}
    E_{\rm lhc} = 2\sqrt{ M_N^2 - \frac{M_\pi^2}{4}}.
\end{equation}
Below that energy, the K matrix becomes complex. While the locations of the deuteron and dineutron poles in this work turn out to be above the left-hand cut (lhc), which is also true at the physical point, this can have implications in the analysis of finite-volume lattice data, as discussed below in more detail. Moreover, the non-analyticity in the K matrix limits the range of convergence of simple polynomial parametrizations, such as in the case of the effective-range expansion.

The presence of a left-hand cut imposes an important constraint on the analytic structure of the scattering amplitude. Because the amplitude has a singularity at the lhc, it must approach the bound-state condition from above at the branch point. Consequently, either the virtual-state or the bound-state condition must be crossed between the lhc and threshold, i.e.\ it is guaranteed that a pole exists.

\subsection{Standard two-particle formalism}
\label{sec:qc2}

The finite-volume two-nucleon spectrum determined from lattice QCD carries information about two-nucleon interactions. In particular, the finite-volume energy shifts with respect to the noninteracting energies can be related to the two-nucleon scattering amplitude in the elastic region. These shifts, which are power-law-dependent on the lattice volume, serve as input to the amplitude analysis.\footnote{
The derivation of finite-volume formalisms typically neglects corrections that are exponentially suppressed with the lattice volume, as well as discretization effects. See, however, \rcite{Hansen:2024cai} for a generalization of the Lüscher formalism that takes into account discretization effects.}

The standard two-particle formalism~\cite{Luscher:1986pf,Luscher:1990ux,Kim:2005gf,Briceno:2014oea} comes in the form of a quantization condition (QC) that relates the two-particle K matrix to a calculable geometric function, $F$, which does not depend on the details of the interaction. Specifically, this QC reads
\begin{equation}
\prod_{I, s} \det_{J, m_J, \ell} \left[ F\left( E, \boldsymbol{P}, L \right) + \cK(E_{\rm cm}) ^{-1} \right] = 0\,,
\label{eq:QC2}
\end{equation}
The QC depends on the total energy of the system in the lattice frame, $E$, the total momentum of the two-hadron system, $\boldsymbol{P}= (2 \pi / L) \boldsymbol{d}$ with  $\boldsymbol{d} \in \mathbb{Z}^3$, and the size of the box, $L$. The center-of-mass (cm) energy is then given as $E_{\rm cm} = \sqrt{E^2 - \boldsymbol{P}^2}$. The finite-volume energies correspond to those values of $E$ for which \Cref{eq:QC2} is satisfied. 

Note that the matrices in \Cref{eq:QC2} all have angular momentum indices, and the determinant runs, in general, over all quantum numbers characterizing the $NN$ system. While the nucleon-nucleon K matrix is diagonal in total angular momentum, total angular momentum conservation does not apply to $F$, since it is a finite-volume object. Thus, a priori, all values of $J^P$ need to be included when evaluating the QC. However, $F$ is also diagonal in spin, as is the K matrix for identical $NN$ particles, so the QC can be factorized into four independent spin-isospin sectors, each with definite parity. In order to render the QC finite-dimensional, a truncation of $\ell < \ell_{\rm max}$ must be applied, which is usually motivated by the suppression of higher partial waves close to threshold---see \Cref{eq:thresholdsuppresionpartialwaves}. Note however that the proximity of a left-hand cut may render this assumption valid for only a very narrow kinematic range.

Information on the energy dependence of the two-nucleon K matrix can then be extracted from the finite set of two-nucleon energies determined from lattice QCD, using the QC.
A common approach for this purpose is to choose some explicit parametrization of the K matrix, $\mathcal{K}^{-1}(E_\mathrm{cm}; \boldsymbol{\theta})$, where $\boldsymbol{\theta}$ denotes free parameters of the model---see the previous section for specific examples of models. These parameters are then adjusted via a fit to the lattice data.
Concretely, one proceeds by minimizing a $\chi^2$ function relating the lattice-QCD energy levels, $\boldsymbol E_\mathrm{lat}$, and the predicted energy levels $\boldsymbol E_\mathrm{QC}(\boldsymbol \theta)$ obtained from solving \Cref{eq:QC2} for specific parameter values $\boldsymbol \theta$,
\begin{align}
    \label{eq:chi2Fit}
    \chi^2(\boldsymbol \theta) &= (\boldsymbol E_\mathrm{lat} - \boldsymbol E_\mathrm{QC}(\boldsymbol \theta))^T \Sigma_\mathrm{d}^{-1} (\boldsymbol E_\mathrm{lat} - \boldsymbol E_\mathrm{QC}(\boldsymbol \theta)) , \\[10pt]
    \label{eq:chi2opt}
    \boldsymbol \theta_\mathrm{opt} &= \argmin_{\boldsymbol \theta} \chi^2(\boldsymbol \theta) ,
\end{align}
where $\Sigma_\mathrm{d}$ denotes the full covariance matrix of the lattice data.
The results of this procedure will inherently depend on the chosen model for the K matrix.

An issue of the standard finite-volume formalism presented above is the presence of left-hand cuts, discussed in the previous section. On the cut, the complex-valued K matrix  invalidates one of the assumptions in the derivation of the QC, and even for energies that lie above the cut, neglected exponentially suppressed finite-volume corrections can become large. In the finite volume, a strong attractive interaction may translate into finite-volume energies that lie on top of the left-hand cut, even if physical poles are above the cut~\cite{Green:2021qol}. In addition, as already discussed above, the presence of a cut limits the convergence of the ERE, and therefore the region for which \Cref{eq:thresholdsuppresionpartialwaves} is valid. This means that, even for energies very close to the cut, higher partial waves may be needed for a correct description of the lattice data. While these issues are absent for heavy pion masses, they all become increasingly relevant as the quark mass is reduced and the lhc approaches the two-nucleon threshold. Approaches to overcome this problem will be discussed below. 

\subsection{Relativistic left-hand cut formalism}
\label{sec:lhcRelativistic}

As already mentioned, the standard L\"uscher formalism presented in \Cref{sec:qc2}
is expected to fail for energies in the vicinity or on top of a left-hand cut. In recent years, several alternatives have been proposed to incorporate the effects of the left-hand cut in the finite-volume formalism~\cite{Raposo:2023oru, Raposo:2025dkb,Dawid:2024oey,Hansen:2024ffk,Dawid:2024dgy,Bubna:2024izx, Yu:2025gzg}.

The proposal from \rrcite{Raposo:2023oru,Raposo:2025dkb}, which we call the \textit{relativistic left-hand cut formalism} here, is based on a re-derivation of the standard L\"uscher QC through an all-order expansion of the scattering amplitude in finite volume that explicitly takes one-particle exchanges into account. A short-range K-matrix-like object, $\hat\cK$, can be constrained from finite-volume data using the re-derived QC. This quantity is then used to obtain the scattering amplitude and phase shift through relations involving integral equations.

\subsubsection{Quantization condition}

Following the kinematic notation of \Cref{sec:qc2}, the modified QC, which we call LHCQC for short, takes the form
\begin{align}
\prod_{I, s} \det_{J, m_J, \ell} \left[ \hat{\cK}^I(E_{\rm cm})^{-1} + F(E, \boldsymbol P, L) + \mathcal C^I_L(E, \boldsymbol P)\right] = 0 \,,
\label{eq:rel_lhcQC}
\end{align}
where $F$ is the same known function appearing in the standard L\"uscher QC of \Cref{eq:QC2}, while $\mathcal C_L^I$ is an additional function encoding the one-pion exchange (OPE) effects in finite volume. As before, the QC factorizes in spin and isospin. 
In the limit of no OPEs, $\mathcal C_L^I$ vanishes and $\hat{\cK}^I$ reproduces the standard two-particle K-matrix $\mathcal K^I$, such that \Cref{eq:rel_lhcQC} reduces to the usual L\"uscher QC.

The new function $\mathcal C_L^I$ is constructed from an explicit form of the OPE, denoted $\mathcal E^I$ below. In the following, we treat the nucleons as distinguishable and thus keep only the $t$-channel OPE.
\footnote{This is equivalent within the formalism to treating them as identical and keeping both $t$- and $u$-channel OPE contributions.}
Starting from the general relativistic $t$-channel pion exchange, we bring it to the form
\begin{multline}
\mathcal E ^ I (\boldsymbol p', \boldsymbol p)
= - \, c_I \, g_{\pi NN}^2 ~\times \\ \frac{\boldsymbol\sigma_1 \cdot (\boldsymbol p'^{\star} - \boldsymbol p^{\star}) ~ \boldsymbol\sigma_2 \cdot (\boldsymbol p'^{\star} - \boldsymbol p^{\star})}
{(\omega_N(\boldsymbol p') - \omega_N(\boldsymbol p))^2 - (\boldsymbol p' -\boldsymbol p)^2 - M_\pi^2 + i\epsilon}  \,,
\label{eq:rel_OPE}
\end{multline}
where $\boldsymbol p$ ($\boldsymbol p^\prime$) is the incoming (outgoing) spatial momentum of one of the nucleons and $\omega_N(\boldsymbol p)= \sqrt{\boldsymbol p^2 + M_N^2}$. Boosts of the spatial momenta to the cm frame are denoted by $\star$ above. Instead of the more standard boost with velocity $\boldsymbol \beta = \boldsymbol P/E$, here we use the $E$-independent Wu boost \cite{Wu:2014vma},
\begin{align}
\boldsymbol \beta_{\rm Wu} \pprime = \frac{\boldsymbol P}{\omega_N(\boldsymbol p \pprime ) + \omega_N(\boldsymbol P - \boldsymbol p \pprime )} \,.
\label{eq:Wu_boost}
\end{align}
Consequently, $\mathcal E^I$ carries implicit dependence on $\boldsymbol P$ but not on $E$.\footnote{
In obtaining this form of the OPE from the general relativistic expression, we make a few replacements based on the two-particle state on-shell condition, ~$E = \omega_N(\boldsymbol p) + \omega_N(\boldsymbol P - \boldsymbol p)$, without loss of generality of the underlying formalism: we use it in defining the modified Wu cm frame boost prescription and to simplify kinematic prefactors of the relativistic spinors.
}

The coupling $g_{\pi NN}$ in \Cref{eq:rel_OPE} is the pion-nucleon coupling, directly related to the residue of the pion pole in the $NN$ amplitude, while $c_I$ is a numerical factor arising from the projection to total isospin. This takes the value $-3$ for the $I=0$ channel and $+1$ for the $I=1$ channel. The Pauli matrices $\boldsymbol\sigma_1, \boldsymbol\sigma_2$ carry spin indices for each nucleon, and standard Clebsch-Gordan machinery can be used to project to total spin quantum numbers $s , m_s $.

In building $\mathcal C_L^I$, we treat $\mathcal E^I$ as a matrix in the finite-volume momenta $\boldsymbol p,  \boldsymbol p' \in (2\pi/L) \,  \mathbb Z^3$ and spin indices, with elements $\mathcal E^I_{\boldsymbol p' s' m_s';\, \boldsymbol p s m_s}$.
We also define the two-particle on-shell pole $\Delta_{2,L}$ in the same matrix space,
\begin{multline}
(\Delta_{2,L})_{\boldsymbol p' s' m_s';\, \boldsymbol p s m_s} 
= \\
\frac{1}{L^3} \, \delta_{\boldsymbol p' \boldsymbol p} \, \delta_{s's} \, \delta_{m_s' m_s} \frac{H (\boldsymbol p_{\rm cm}^2) }{4 \omega_{N}(\boldsymbol p) \big[ q^2 - \boldsymbol p_{\rm cm}^2 \big] } \,,
\label{eq:Delta2L_def}
\end{multline}
with $q = \sqrt{E_{\rm cm}^2/4 - M_N^2}$ being the cm-frame scattering momentum and $\boldsymbol p_{\rm cm}$ being the cm-frame-boosted momentum $\boldsymbol p$. We use ``cm'' here to distinguish the standard boost from the Wu boost used above. The function $H(\boldsymbol p^2_{\rm cm})$ is a UV regulator for two-nucleon loops, as used in the L\"uscher $F$ function. We choose an exponential function \cite{Kim:2005gf}
\begin{equation}
H(\boldsymbol p_{\rm cm}^2) = \exp \left[\alpha (q^2 - \boldsymbol p_{\rm cm}^2) \right] \,,
\label{eq:KSSreg}
\end{equation}
with the cutoff parameter $\alpha$ controlling the suppression, associated with an energy scale $\Lambda = 1/\sqrt{\alpha}$.

We then obtain $\mathcal C_L^I$ through
\begin{align}
\mathcal C_L^I = - \mathcal Y^T \Delta_{2,L} \left[ 1 + \mathcal E^I \Delta_{2,L} \right]^{-1} \mathcal E^I \, \Delta_{2,L} \, \mathcal Y \,,
\label{eq:CL_def}
\end{align}
with the dependence of each matrix on $E,\boldsymbol P$ kept implicit. We employ matrices of spherical harmonics, 
\begin{align}
\cY_{\boldsymbol p' s' m_s'; \, \ell m_\ell s m_s} 
&= \delta_{s's} \delta_{m_s' m_s}
\left( \frac{|\boldsymbol p'_{\rm cm}|}{q} \right)^{\!\ell}
\sqrt{4\pi} \, Y_{\ell m_\ell} (\hat{\boldsymbol p}'_{\rm cm}) \,,
\label{eq:Yc_def1} \\
\cY^T_{\ell' m_\ell's' m_s'; \, \boldsymbol p s m_s} 
&= \delta_{s's} \delta_{m_s' m_s} 
\left( \frac{|\boldsymbol p_{\rm cm}|}{q} \right)^{\!\ell'}
\sqrt{4\pi} \, Y_{\ell' m_\ell'}^* (\hat{\boldsymbol p}_{\rm cm}) \,,
\label{eq:Yc_def2}
\end{align}
where $\hat{\boldsymbol p}_{\rm cm} = \boldsymbol p_{\rm cm} /|\boldsymbol p_{\rm cm}|$ and $\hat{\boldsymbol p}'_{\rm cm} = \boldsymbol p'_{\rm cm}/|\boldsymbol p'_{\rm cm}|$. These operate a change of basis between the finite-volume momentum (plane-wave) basis and the orbital angular momentum basis. A further change of basis is needed to bring $\mathcal C_L^I$ to the total angular momentum $J m_J \ell s$ basis used in \Cref{eq:rel_lhcQC}. Projection to finite-volume irreps follows in an identical manner to the standard L\"uscher method.

Given a parametrization for $\hat\cK$, the LHCQC of \Cref{eq:rel_lhcQC} can be used to fit the spectrum. This can be used to constrain the $\hat\cK$ parameters and also the OPE coupling $g_{\pi NN}$, as is done in this work. The next subsection describes how these are used to reconstruct the amplitude.

\subsubsection{Scattering amplitude}

As per Ref.~\cite{Raposo:2025dkb}, the on-shell amplitude in the $(I,J^P)$ channel can be written as
\begin{multline}
\cM^{J^P} =  \cM_{\cE}^{J^P} + 
\big( 1 + \cL ^{J^P} \big) \\
\times ~ \Big[ \big(\hat{\cM}^{J^P} \big)^{-1}  + \cC ^{J^P} \Big]^{-1} \big( 1 + \cR ^{J^P} \big) \,.
\label{eq:Mfull}
\end{multline}
Each object here is a matrix in the space of allowed $(\ell,s)$ pairs for the given $(I,J^P)$ channel. The cases considered in this work are diagonal in the spin, thus we omit the trivial spin index in the expressions below.

The fundamental building blocks of \Cref{eq:Mfull} are the two auxiliary amplitudes $\hat{\cM}^{J^P}$ and $\cM^{J^P}_{\cE}$. The former is related algebraically to $\hat{\cK}^{J^P}$ through
\begin{equation}
\big( \hat{\cM}^{J^P} \big)^{-1}_{\ell', \ell} = \big( \hat{\cK}^{J^P} \big)^{-1}_{\ell', \ell } - i\rho \, \delta_{\ell'\ell}  \,,
\label{eq:M0_def}
\end{equation}
and can loosely be understood as an ``amplitude without OPEs''. The latter is defined to obey (a set of) integral equations:
\begin{multline}
\big( \cM_{\cE}^{J^P} \big)_{\ell' , \ell } (p',p) = \cE^{J^P}_{\ell' , \ell } (p',p) \\
- \sum_{\ell''} \int_0^\infty \! \frac{k^2 \,dk}{2\pi^2} ~ \cE^{J^P}_{\ell' , \ell'' } (p',k) \, \Delta_2 (k) \big( \cM_{\cE}^{J^P} \big)_{\ell'', \ell } (k,p) \,,
\label{eq:ME_int_eq}
\end{multline}
where the arguments $p,p',k$ are magnitudes of spatial momenta in the cm frame and $\Delta_2(k)$ is the two-particle propagator pole,
\begin{align}
\Delta_2(k) = \frac{H(k^2) }{4 \omega_N(k) \big[ q^2 - k^2 + i\epsilon \big] } \,,
\label{eq:Delta2}
\end{align}
i.e.~an infinite-volume analog of $\Delta_{2,L}$. The partial-wave projections of the OPE kernels $\cE^I$, denoted above by $\cE^{J^P}_{\ell',\ell}$, are obtained as described in Ref.~\cite{Jackura:2023qtp}.

The remaining objects in \Cref{eq:Mfull} are defined from integrals of $\mathcal M_{\mathcal E}^{J^P}$,
\begin{align}
\mathcal L^{J^P}_{\ell',\ell } 
&= - \int_0^\infty \!\frac{k^2\, dk}{2\pi^2} 
\, \big( \cM_{\cE}^{J^P} \big)_{\ell',\ell} (q, k) \, \Delta_2 (k) \bigg( \frac{k}{q} \bigg)^{\!\ell} \,, 
\label{eq:Ldef}\\
\mathcal R^{J^P}_{\ell',\ell} &= -\int_0^\infty \!\frac{k^2\, dk}{2\pi^2}
\bigg( \frac{k}{q} \bigg)^{\!\ell'} \! \Delta_2(k) \, \big( \cM_{\cE}^{J^P} \big) _{\ell',\ell } (k, q) \,,
\label{eq:Rdef} \\
\nonumber
\mathcal C^{J^P}_{\ell',\ell } &=
- \int_0^\infty \!\frac{k'^2\, dk'}{2\pi^2} \int_0^\infty \!\frac{k^2\, dk}{2\pi^2} \,
\bigg( \frac{k'}{q} \bigg)^{\!\ell'} \! \Delta_2(k') \\
& \qquad\qquad\quad \times ~ \big( \cM_{\cE}^{J^P} \big)_{\ell',\ell} (k', k) \, \Delta_2(k) \bigg( \frac{k}{q} \bigg)^{\!\ell}
\,.
\label{eq:Cdef}
\end{align}
Note that $\cM_{\cE}^{J^P}$ has one of its arguments off-shell in the definition of $\cL^{J^P}$ and $\cR^{J^P}$, and both arguments off-shell in the definition of $\cC^{J^P}$.

To obtain the full amplitude $\cM^{J^P}$ given a $\hat{\cK}$ parametrization and corresponding parameters, we solve the integral equations for $\cM_{\cE}^{J^P}$ and hence obtain $\cL^{J^P}$, $\cR^{J^P}$ and $\cC^{J^P}$ from the integrals above. \Cref{eq:Mfull,eq:M0_def} then provide the algebraic relations to construct $\cM^{J^P}$. One can then use the relations to the standard K-matrix and phase shift, \Cref{eq:M_Kmat_rel,eq:K_delta_rel}, to extract these quantities.

\subsection{Lippmann-Schwinger equation and the plane-wave quantization condition}
\label{sec:qc2PlaneWave}

Another version of a QC that makes it possible to incorporate one-particle exchange effects is the so-called \textit{plane-wave quantization condition}, first proposed in \rcite{Meng:2021uhz}. This is based on the formulation of the Lippmann-Schwinger equation (LSE) in a finite volume. Before presenting the form of this QC, however, we first introduce the LSE and describe the particular form of the two-nucleon interaction potential that we will be using in this work. Details about the numerical implementation of the formalism and how the infinite-volume LSE is solved will be presented elsewhere~\cite{Baeza_inprep}. 

\subsubsection{Lippmann-Schwinger equation in infinite volume}\label{sec:qc2PlaneWaveLSE}

In infinite volume, the LSE relates the off-shell scattering amplitude to a scheme-dependent interaction potential, $V$,
\begin{samepage}
\begin{multline}\label{eq:LippmannSchwingerInfiniteVolume}
     \cM(\boldsymbol{p}',\boldsymbol{p};E)= -V(\boldsymbol{p}',\boldsymbol{p})\\
     \displaystyle +\int\frac{\text{d}^3\boldsymbol{k}}{(2\pi)^3}V(\boldsymbol{p}',\boldsymbol{k})G(k^2,E)\cM(\boldsymbol{k},\boldsymbol{p};E)\,,
\end{multline}
\end{samepage}
where the off-shell amplitude depends separately on the three-momenta of the initial and final nucleons in the cm frame, $\boldsymbol{p}$ and $\boldsymbol{p}'$, and on the total cm energy of the system, $E$, and reduces to the physical scattering amplitude when both initial and final states are on shell, $E^2/4=\boldsymbol{p}^2+M_N^2=\boldsymbol{p}^{\prime 2}+M_N^2$. Note that the off-shell amplitude is not uniquely defined; in this case its particular form depends on the scheme of the interaction potential, which can be modified via field redefinitions~\cite{Birse:2012ph, Epelbaum:2025aan}. Here, we have also introduced the nucleon propagator, $G(k^2,E)$. In this work, we will use a relativistic propagator based on the Blankenbecler-Sugar equation~\cite{Blankenbecler:1965gx, Meng:2023bmz},
\begin{equation}\label{eq:nucleonpropagator}
    G(k^2,E)=\frac{1}{\omega_N(k)}\frac{1}{E^2-4\omega_N^2(k)+i\varepsilon}\,.
\end{equation}

The interaction potential encodes the form of the interactions between the two nucleons  and is usually computed from an effective field theory~\cite{Weinberg:1990rz, Weinberg:1991um}. In general, the potential can be split into two different parts,
\begin{equation}\label{eq:potentiallongshort}
    V=V_\text{long}+V_\text{short}\,,
\end{equation}
where the first term, $V_\text{long}$, contains long-range interactions which are mediated via the exchange of one or more mesons between the two nucleons, while the last one, $V_\text{short}$, contains short-range point-like interactions. To ensure the LSE is convergent and the scattering amplitude finite, the potential needs to be regularized. This is achieved by multiplying the naive EFT potential by some cutoff function that decays exponentially in the UV. In particular, we use the semi-local momentum-space regularization from \rcite{Reinert:2017usi} for the long-range part of the potential, while the short-range interaction is regularized with a function depending on the squared incoming and outgoing momenta.

To illustrate this, we summarize the form of the regularized potential at leading order (LO) in the chiral expansion, and comment on the next-to-leading order (NLO). The long-range part of the potential is dominated by one-pion exchange processes, and takes the form
\begin{multline}\label{eq:potentiallongLO}
    V_\text{long}^\text{LO}=-c_I g_{\pi NN}^2\biggl\{\frac{[(\boldsymbol{p}'-\boldsymbol{p})\cdot\boldsymbol{\sigma}_1][(\boldsymbol{p}'-\boldsymbol{p})\cdot\boldsymbol{\sigma}_2]}{(\boldsymbol{p}'-\boldsymbol{p})^2+M_\pi^2}\\+C(M_\pi,\Lambda)\boldsymbol{\sigma}_1\cdot\boldsymbol{\sigma}_2\biggr\}\times\text{exp}\left[-\frac{(\boldsymbol{p}'-\boldsymbol{p})^2+M_\pi^2}{\Lambda^2}\right]\,,
\end{multline}
where $\sigma_i$ is a vector of Pauli matrices acting in spin space between the $i^{\rm th}$ particle of the initial and final states, $c_I$ is as in \Cref{sec:lhcRelativistic}, and $\Lambda$ is the cutoff parameter. Also, we have introduced a known coefficient that ensures the spin-spin part of the potential in position space vanishes at the origin,
\begin{multline}\label{eq:potentiallongLOcutoffcontactterm}
    C(M_\pi,\Lambda)=\\-\frac{\Lambda(\Lambda^2-2M_\pi^2)+2\sqrt{\pi}M_\pi^3\text{e}^{M_\pi^2/\Lambda^2}\text{erfc}(M_\pi/\Lambda)}{3\Lambda^3}\,.
\end{multline}
Up to this added contact term, the regularization, and relativistic corrections, the long-range potential [\Cref{eq:potentiallongLO}] is the same as the OPE $\mathcal{E}$ given in \Cref{eq:rel_OPE} in the center-of-mass frame.
At NLO, the long-range potential contains the two-pion-exchange contributions. For simplicity and since the left-hand cut associated with the exchange of two pions lies well below the finite-volume energies analyzed in this work, we only consider the LO OPE potential for this study. Even if two-pion exchange has an effect, in the range of energies we study, it can be effectively reabsorbed in the contact terms of the short-range potential, which we now discuss.

The short-range potential at LO takes the form
\begin{equation}\label{eq:potentialshortLO}
    V_\text{short}^\text{LO}(\boldsymbol{p}',\boldsymbol{p})=\left[C_\text{S}\mathbbm{1}+C_\text{T}\sigma_1\cdot\sigma_2\right]H_\text{cut}({p}',{p})\,.
\end{equation}
where $\mathbbm{1}$ is the identity matrix in spin space, and we have defined the regulator,
\begin{equation}
   H_\text{cut}({p}',{p}) = \text{exp}\left[-\frac{p'^2+p^2}{\Lambda^2}\right]
\end{equation}
For simplicity, we keep the cutoff $\Lambda$ to be the same as in the long-range part of the potential. At NLO, the number of operators increases to a total of seven---see \rcite{Epelbaum:2004fk} for further details.

To solve the LSE in infinite volume, it is convenient to work in angular momentum space. As already discussed in \Cref{sec:NNscatteringamplitudes}, the two-nucleon scattering amplitude becomes block diagonal in total angular momentum, and so does the potential. Already at LO, the long-range potential from \Cref{eq:potentiallongLO} contributes to all partial waves.

A summary of the regularized OPE potential projected to all partial waves can be found in \rrcite{Rijken:1991at, Epelbaum:2004fk, Baeza_inprep}. The contact term potential, on the other hand, only contributes to some partial waves at each order. In particular, the LO potential in \Cref{eq:potentialshortLO} only contributes with a constant to the $S$-wave for both spin channels. Using the notation $V^{sJ}_{\ell\ell'}=\langle {}^{2s+1}\ell_J|V|{}^{2s+1}\ell'_J\rangle$, we write
\begin{equation}\label{eq:potentialshortLOpartialwave}
   \begin{array}{rl}
    (V_\text{short}^\text{LO})^{00}_{00} & =  \tilde{C}_{{}^1 S_0}H_\text{cut}({p}',{p}) \\[5pt]
    & =  \displaystyle \frac{1}{2\pi^2}\left({C}_\text{S}-3{C}_\text{T}\right)H_\text{cut}({p}',{p})\,,\\[12pt]
    (V_\text{short}^\text{LO})^{11}_{00} & =  \tilde{C}_{{}^3 S_1}H_\text{cut}({p}',{p}) \\[5pt]
    & =  \displaystyle \frac{1}{2\pi^2}\left({C}_\text{S}+{C}_\text{T}\right)H_\text{cut}({p}',{p})\,,
    \end{array}
\end{equation}
where the dependence on $p$ and $p'$ on the left-hand side is left implicit for readability. Note the appearance of $1/(2\pi^2)$ factors, coming from the normalization chosen for the partial-wave projection.

At NLO, the most general short-range potential contains contributions to $S$  and $P$ waves, as well as to mixing between $S$ and $D$ waves with $J=1$. In this work, we are interested in those that contribute to the deuteron and dineutron channels,
\begin{equation}\label{eq:potentialshortNLOpartialwave}
    \begin{array}{rcl}
         (V_\text{short}^\text{NLO})^{00}_{00} & = & {C}_{{}^1 S_0}(p'^2+p^2)H_\text{cut}({p}',{p})\,,  \\[5pt]
         (V_\text{short}^\text{NLO})^{11}_{00} & = & {C}_{{}^3 S_1}(p'^2+p^2)H_\text{cut}({p}',{p})\,, \\[5pt]
         (V_\text{short}^\text{NLO})^{11}_{20} & = & {C}_{{}^3S_1-{}^3 D_1}p'^2H_\text{cut}({p}',{p})\,,\\[5pt]
         (V_\text{short}^\text{NLO})^{11}_{02} & = & {C}_{{}^3S_1-{}^3 D_1}p^2H_\text{cut}({p}',{p})\,.
    \end{array}
\end{equation}
These can also be expressed in the plane-wave basis as a linear combination of dimension-two operators~\cite{Reinert:2017usi}. 

There are a few points that merit attention.
First, we note that the low-energy coefficients that appear in the long- and short-range potentials play a different role. As already mentioned, the coefficient in the OPE part is a physical observable, as it is proportional to the residue of the scattering amplitude at the pion pole. Thus, it is expected to be independent of changes to the cutoff. This is not the case for the contact terms from the short-range potential, which instead act as counterterms of the regularization. Second, we note that, while the cutoff acts as a UV regulator of the theory, it is in general not possible to take the limit $\Lambda\rightarrow\infty$. Instead, one only expects the model to correctly describe our lattice results for a limited range of cutoff values~\cite{Epelbaum:2004fk}. Finally,  we note that alternative formulations of the LSE are also possible. Of particular interest is the non-relativistic limit that is often used to describe experimental results~\cite{Reinert:2017usi,Epelbaum:2004fk, Epelbaum:2008ga,
Machleidt:2011zz, Hammer:2019poc}. 

\subsubsection{Plane-wave quantization condition}\label{sec:qc2PlaneWaveQC}

The plane-wave quantization condition~\cite{Meng:2021uhz} is based on the formulation of the LSE in finite volume in a basis of plane waves. It allows the long-distance effects of particle-exchange potentials to be accounted for, while also preventing problems with partial-wave truncation by explicitly working in the basis of plane waves. Following \rrcite{Meng:2021uhz,Meng:2023bmz}, the finite-volume energies in the cm are the solutions to the following quantization condition,
\begin{equation}
    \label{eq:PWQC}
    \prod_{I,s}\det_{\boldsymbol{p},m_s}(\mathbb{G}^{-1}(E)-\tilde{\mathbb{V}})=0\,,
\end{equation}
where the determinant runs over all states in the plane wave basis, characterized by their relative three-momentum and the spin of the two nucleons, although it factorizes in isospin and total spin. Here, we have introduced two different matrices. First, we have
\begin{equation}
    \mathbb{G}^{-1}(E)=E^2\mathbbm{1}-\mathbb{H}\,,
\end{equation}
with $\mathbbm{1}$ denoting here the identity in the plane wave basis, and $\mathbb{H}$ being a diagonal matrix that contains the free energies of the system in the cm,
\begin{equation}
    \mathbb{H}_{nn'}=E_{\rm cm}^2(\boldsymbol{p}_n)\delta_{nn'}\,.
\end{equation}
On the other hand, $\tilde{\mathbb{V}}$ is a modified version of the potential evaluated in the plane-wave basis,
\begin{equation}
    \tilde{\mathbb{V}}=\frac{1}{L^3}\sqrt{\mathbb{N}} \mathbb{V} \sqrt{\mathbb{N}} \,,
\end{equation}
where $\mathbb{V}$ is the potential, evaluated for the states of the plane-wave basis, and $\mathbb{N}$ is a diagonal matrix related to the Jacobian of the transformation from the lattice frame, where the plane wave basis is defined, to the cm frame, where the potential is evaluated. To ensure the free energies predicted by the formalism agree with the free energies of the QCD simulations, these transformations are always performed using relativistic boosts. Using the Wu boost~\cite{Wu:2014vma} for off-shell kinematics, the $\mathbb{N}$ factor reads
\begin{equation}
    \mathbb{N}=\frac{\omega(\boldsymbol{p_{n,1}})+\omega(\boldsymbol{p_{n,2}})}{\omega(\boldsymbol{p_{n,1}})\omega(\boldsymbol{p_{n,2}})}\delta_{n'n}\,,
\end{equation}
where $\boldsymbol{p}_{n,i}$ are the momenta in the lattice frame of particle $i$ in the $n^{\rm th}$ state of the plane-wave basis.

In the case of the $NN$ system, the non-relativistic potential presented in \Cref{eq:potentiallongshort,eq:potentiallongLO,eq:potentiallongLOcutoffcontactterm,eq:potentialshortLO,eq:potentialshortLOpartialwave,eq:potentialshortNLOpartialwave} does not depend on the total energy of the system. In this case, \Cref{eq:PWQC} can be simplified to an eigenvalue problem,
\begin{equation}
    (\mathbb{H}+\tilde{\mathbb{V}})v_n=E_n^2 v_n\,.
\end{equation}

As in the case of the standard QC, the plane-wave QC factorizes into the different isospin and spin sectors, but mixes different angular momenta, as the rotational symmetry is broken into the cubic group by the finite volume. However, all the matrices that enter the formalism can be decomposed into the different irreducible representations $\Gamma_i$ of the cubic group, as is the case for the standard QC, and so each of these can be analyzed separately.

\section{Strategy for the scattering analysis}
\label{sec:methodology}

Having determined the lattice QCD energies in \Cref{sec:setup}, the next step is to fit the spectrum via the quantization conditions presented in the previous section. To do so, the spectrum is fitted by minimizing the following \(\chi^2\) function:
\begin{equation}
\label{eq:chi2}
\begin{aligned}
\chi^2 ={} \sum_{ij} &
\left[\Delta E_{\mathrm{lab}}^{\mathrm{data}}
      -\Delta E_{\mathrm{lab}}^{\mathrm{QC}}(\boldsymbol \theta)\right]_i
\left[\Sigma_d^{-1}\right]_{ij}
\\
& \ \times
\left[\Delta E_{\mathrm{lab}}^{\mathrm{data}}
      -\Delta E_{\mathrm{lab}}^{\mathrm{QC}}(\boldsymbol \theta)\right]_j ,
\end{aligned}
\end{equation}
where \(\Delta E_{\mathrm{lab}}\) is the lab-frame energy shift relative to the corresponding noninteracting energy\footnote{Note that $\Delta E_\mathrm{lab}^\mathrm{data}$ corresponds to the fit parameter $\delta_{00}$ in the conspiracy model; in the latter we use lattice units and in \Cref{eq:chi2} nucleon-mass units.}. The superscript ``data'' denotes the lattice-QCD result, whereas ``QC'' denotes the prediction obtained from the quantization condition for a given set of parameters $\boldsymbol \theta$. Finally, $\Sigma_d$ is the covariance matrix of the lattice-QCD energy shifts.
We convert all these quantities to nucleon-mass units before plugging them into \Cref{eq:chi2}.

Note that \Cref{eq:chi2} implicitly applies a correction for discretization effects due to the dispersion relation. We find that these effects are small, as shown in \Cref{app:disp}.
Since the derivation of quantization conditions assumes the continuum dispersion relation, this is the appropriate choice for the finite-volume analysis, as shown in Ref.~\cite{Hansen:2024cai}. The nucleon mass $M_N$ entering the dispersion relation and in all subsequent analysis is determined from a conspiracy fit to the $T_{1g}$ lowest level in the rest frame.

Each ensemble is fitted separately, but we consider both two-nucleon isospin channels simultaneously. This allows us to maximally utilize the information, as one-pion exchange affects both channels with a shared coupling.  

Due to the treatment of the left-hand cut, the fits can become rather costly, especially so in the case of the LHCQC, and an evaluation of the fit on each resampling is not feasible. As such, in order to determine the covariance of the parameters $\Sigma_{\boldsymbol{\theta}}$, we use the derivative method, 
\begin{equation}
    (\Sigma_{\boldsymbol{\theta}}^{-1})_{n m}=\frac{\partial \Delta E_i^{\mathrm{lab}, \mathrm{QC}}}{\partial \theta_n}\left(\Sigma_d^{-1}\right)_{i j} \frac{\partial \Delta E_j^{\mathrm{lab}, \mathrm{QC}}}{\partial \theta_m},
\end{equation}
where the derivatives of the energy shifts with respect to the parameters are taken analytically for the PWQC and numerically with finite differences for the other cases. For a representative PWQC fit, we have verified that this procedure yields uncertainties on physical observables that differ by less than \(20\%\) from those obtained using bootstrap resampling.

We employ several parametrizations of the short-range couplings for both QCs used. In the LHCQC approach, this corresponds to the parametrization of the energy-dependent modified K matrix. In the deuteron sector, we apply the Blatt-Biedenharn decomposition given in \Cref{eq:BBparam} to $\hat\cK$. The modified \(\alpha\) wave is described by either a two- or three-parameter effective-range expansion, denoted ERE2 and ERE3, respectively, as defined in \Cref{eq:ERE}. The modified \(\beta\) wave is always fixed to the noninteracting case, i.e.\ $\hat\delta_{1\beta}=0$. In some fits, we fix the mixing angle to zero, while in others we include $S-D$ mixing through a single additional parameter $\hat\theta$, such that $\hat \epsilon_1 = \hat\theta p^2$. In the dineutron channel, we consider both an ERE2 and a three-parameter rational parametrization as defined in \Cref{eq:rational3}, which we call rational3. The most complicated fits contain three short-range parameters in each isospin channel, together with one OPE coefficient, for a total of seven parameters. 
We denote the pair of $I=0$ and $I=1$ models used for $\hat{\mathcal K}$ like isoscalar/isotriplet, e.g.\ ERE3/rational3.

On the other hand, in the PWQC, the deuteron \(^{3}S_1\) and dineutron \(^{1}S_0\) partial waves are described either by a constant term at leading order (LO) or by a constant plus a momentum-dependent term at next-to-leading order (NLO). We also explore the inclusion of an $S-D$ mixing term in the deuteron channel, and the corresponding fit model is called NLOwSD in the following. The corresponding forms of the potential are given in \Cref{eq:potentialshortLOpartialwave,eq:potentialshortNLOpartialwave}. The most complicated PWQC fit thus contains five contact terms and, including the OPE coupling, six parameters in total. 

An important point is that both left-hand-cut formalisms involve an intrinsic choice of cutoff. This choice defines a scheme for the short-range quantities—either the modified K matrix in the LHCQC or the potential couplings in the PWQC—while the one-pion-exchange coupling should remain scheme independent, as it is related to the residue of the scattering amplitude at the pion pole. In each formalism, the cutoff is characterized by an energy scale \(\Lambda\).

In our analysis, we do not treat the cutoff scale as a fit parameter. Instead, we keep it fixed within each fit. In principle, physical quantities, such as energy levels, scattering amplitudes, or the one-pion-exchange coupling, should be cutoff independent. In other words, any variation of the cutoff must be reabsorbed into the short-range quantities. In practice, however, residual cutoff dependence may arise when the short-range quantities are described using parametrizations with only a few parameters. To account for this effect, we repeat the analysis for several choices of the cutoff scale and use the resulting variation to estimate the associated systematic uncertainty.

\begin{figure*}
     \centering
     \includegraphics[width=\textwidth]{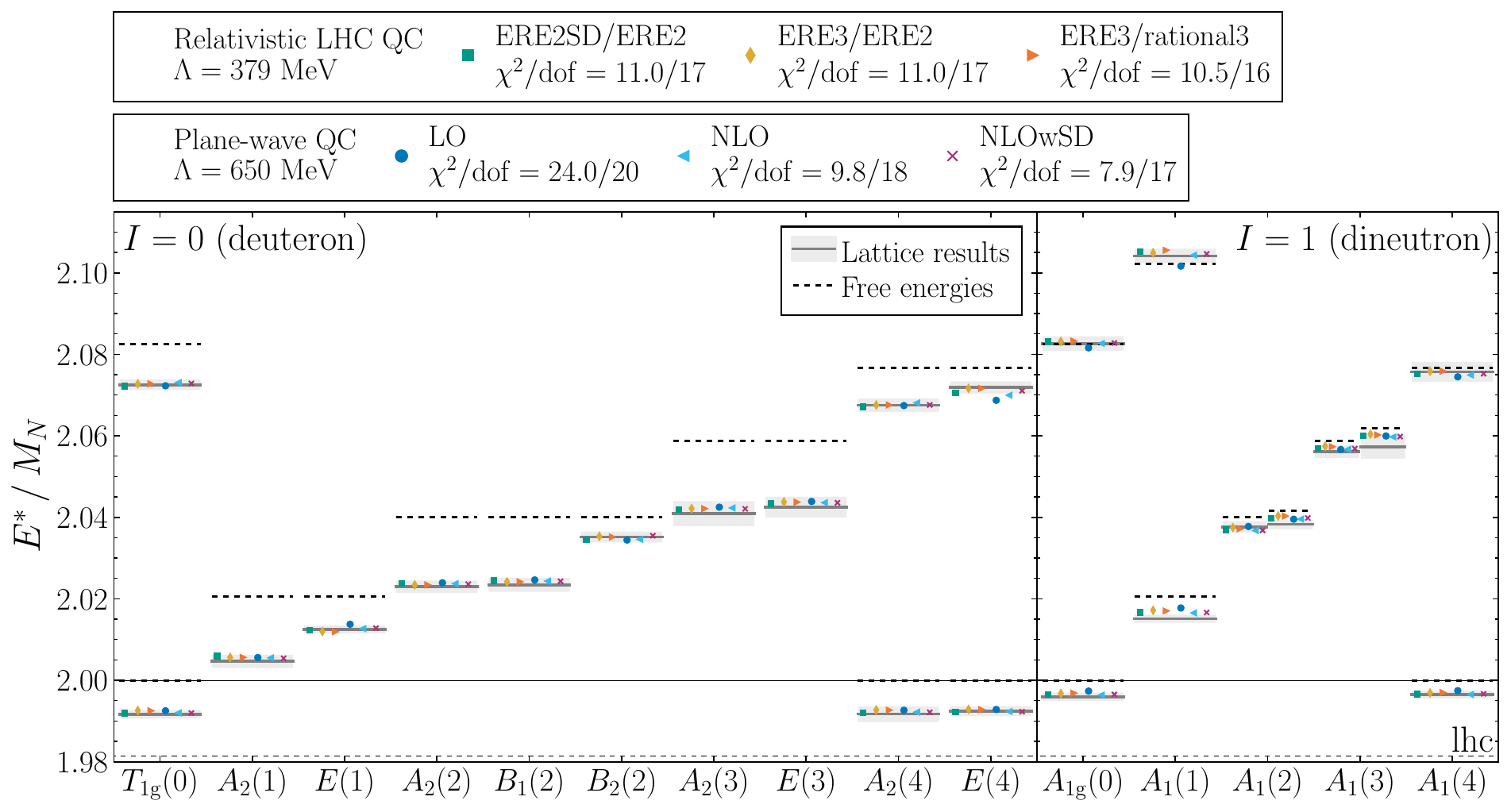}\vspace{0.8cm}
     \includegraphics[width=\textwidth]{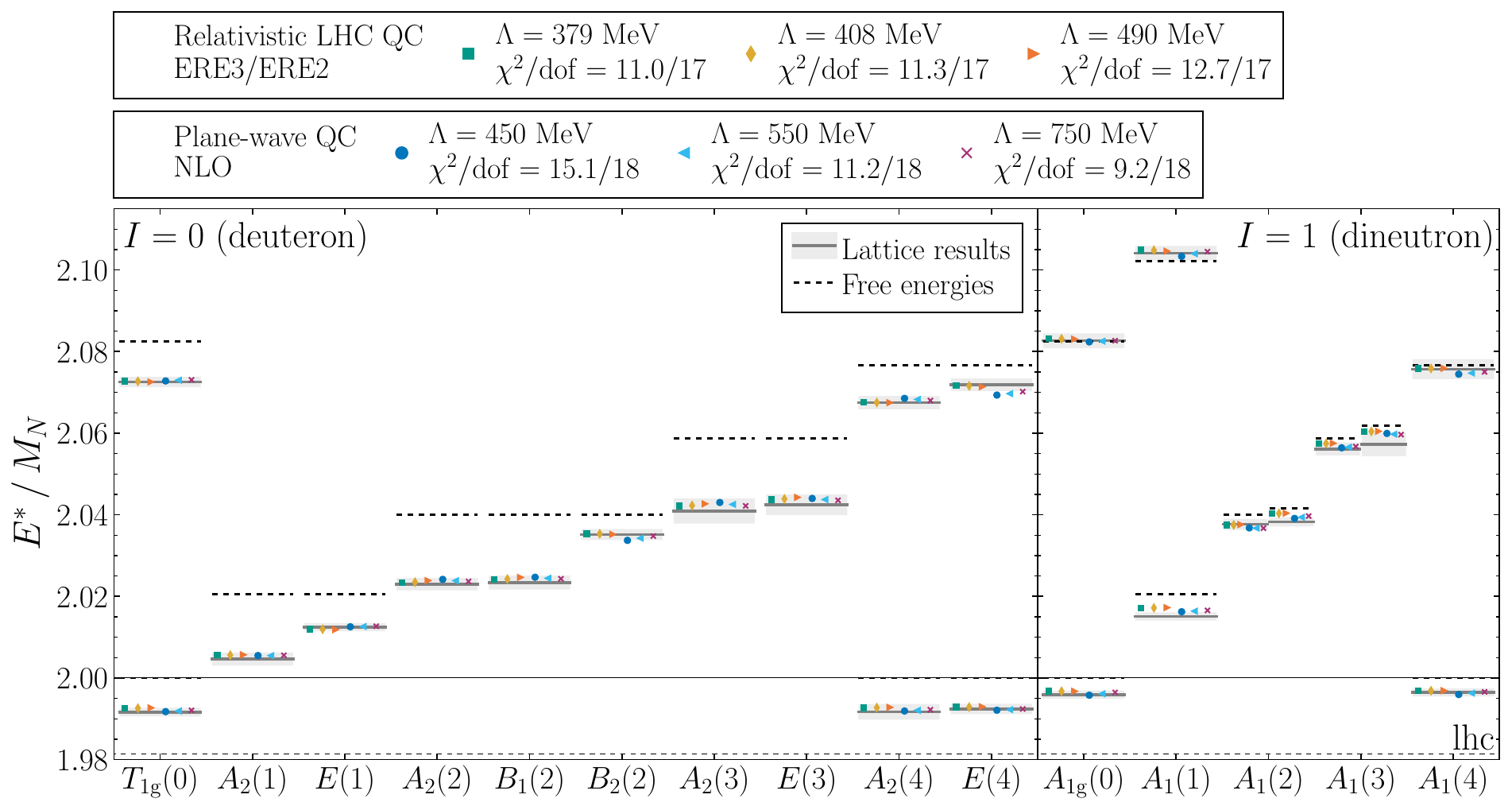}\vspace{0.3cm}
     
     \caption{\label{fig:D251energyfit} Best-fit predictions to the finite-volume energies of D251 (Set 1) obtained with different fit models with a fixed cutoff (top) and with varying cutoff and a fixed model (bottom), as indicated in the figure legends. The parameters used in each fit can be inferred from \Cref{tab:RFTlhcTableD251Part1,tab:RFTlhcTableD251Part2,tab:PWQCtableD251}. Note that the predictions from the PWQC fit performed with the NLOwSD model correspond to a cutoff not shown in \Cref{tab:PWQCtableD251} and not used in the final analysis presented later; the fit parameters are similar to those for $\Lambda=600$ MeV.}

 \end{figure*}

The fit results for both ensembles and using both QCs and sets of energies as described in \Cref{sec:setup} can be found in \Cref{app:tables}, where we present the results for all parametrizations and values of the cutoff considered. All combinations of fit models for the two isospin channels considered in the LHCQC approach, together with the corresponding results, can be found in \Cref{tab:RFTlhcTableD251Part1,tab:RFTlhcTableD251Part2,tab:RFTlhcTableJ250Part1,tab:RFTlhcTableJ250Part2}, while the results from fits using the PWQC are in \Cref{tab:PWQCtableD251,tab:PWQCtableJ250}. As an example, we present in \Cref{fig:D251energyfit} the best-fit predictions for the finite-volume energies of the D251 ensemble using the energy levels from Set 1. The top panel presents a comparison of the predictions from different models at fixed values of the cutoff for each formalism, while the bottom panel fixes the model and presents the dependence on the cutoff. We observe moderate fluctuations in the predicted energies between the different fit models, most of which are within $1\sigma$ of the lattice results. Similar figures for the other set of energies and the J250 ensemble can be found in \Cref{app:tables}. 

One of the parameters fitted by both methods is the pion-nucleon coupling, $g_{\pi NN}$,
which has an experimental value $g_{\pi NN}^\text{phys}\approx13.1$~\cite{Baru:2010xn,Reinert:2017usi}. 
While we observe that our results fall in the correct ballpark, with $g_{\pi NN}$ ranging from about 10 to 15, we also find significant model dependence,  and opt not to quote a final result for this quantity. 

In addition to the analysis of the two new ensembles, we also reanalyzed the lattice energies for the C103 ensemble determined in \rcite{BaryonScattering:2025ziz}. Due to the increased numerical cost of both the PWQC and the LHCQC with increased hadron masses and increased cutoff, we opt to redo the analysis only using the PWQC. At this heavy pion mass, there is no need to use a formalism that includes the effects of the left-hand cut, and the standard QC presented in \Cref{sec:qc2} should be sufficient. Still, the use of the PWQC makes it possible to incorporate the effect of higher partial waves and also fit both isospin channels simultaneously. Therefore, we use the result from these fits. As for the other ensembles, variations of the fit model and cutoff are used to assign a systematic uncertainty to the results. A comparison between the PWQC and the standard QC is given in \Cref{app:standardFV}, specifically \Cref{fig:kcotI0I1C103Luscher}.

Scattering amplitudes are obtained from the fit results by solving integral equations. These relate short-range and OPE couplings to scattering amplitudes. The integral equations in each formalism have slightly different forms [see \Cref{eq:Mfull,eq:M0_def,eq:ME_int_eq,eq:LippmannSchwingerInfiniteVolume}], but basic details on how to solve them numerically are shared. Specifically, momentum integrals are discretized to sums, and the integral equation is turned into a matrix equation. The error of the discretization is reduced via Gauss-Legendre quadratures, and is kept at least an order of magnitude smaller than the statistical error. Analyticity allows the integration contour to be deformed, which improves the convergence properties of the integral, and allows potential singularities of the kernels to be avoided. This methodology for solving the integral equations used in the LHCQC approach is described in detail in Refs.~\cite{Jackura:2020bsk, Dawid:2023jrj, PitangaLachini:2026lyd}. We solve the Lippmann-Schwinger equation following \rrcite{Gloeckle1983, Baeza_inprep} by using a principal-value integral equation for $\mathcal{K}$, which is converted to an ordinary integral equation via an on-shell subtraction.
Statistical errors from the fit parameters are also propagated to the scattering amplitude by using error propagation via derivatives.

We ultimately quote physical observables: the phase shifts and mixing angle, complemented by the scattering lengths, effective ranges, and binding momenta. These are determined independently for each determination of the energy levels, quantization condition used, as well as fit model and cutoff value. Results for the scattering lengths, effective ranges, shape parameters and binding momentum for each of these combinations can be found in \Cref{tab:LHCQCtableD251scattering_isospin0,tab:LHCQCtableD251scattering_isospin1,tab:LHCQCtableJ250scattering_isospin0,tab:LHCQCtableJ250scattering_isospin1,tab:PWQCtableD251scattering_isospin0,tab:PWQCtableD251scattering_isospin1,tab:PWQCtableJ250scattering_isospin0,tab:PWQCtableJ250scattering_isospin1} in \Cref{app:tablesscattering}.

The final results are obtained from averaging all the different combinations, and we present them in the next section together with both statistical and systematic uncertainties. The averaging procedure, including the quantification of systematic uncertainties, is as follows. For each fixed dataset-formalism combination, the different fits are averaged at the level of physical observables using model averaging based on the AIC \cite{Akaike1,Jay:2020jkz,Neil:2022joj,Neil:2023pgt}, with weights $w_i$ as defined in \Cref{eq:AIC_weight}. Following \rrcite{Jay:2020jkz,Neil:2022joj}, observables are averaged as
\begin{equation}
    \langle \mathcal O \rangle = \sum_i w_i \mathcal O_i.
    \label{eq:avg}
\end{equation}
The statistical uncertainty is computed as
\begin{equation}
    \sigma^2_{\mathrm{stat}} = \sum_i w_i \sigma^2_i,
    \label{eq:statavg}
\end{equation}
i.e., as the weighted average of the individual statistical variances, while the systematic uncertainty is
\begin{equation}
    \sigma^2_{\mathrm{syst}}
    = \sum_i w_i \mathcal O_i^2
    - \left(\sum_i w_i \mathcal O_i\right)^2,
    \label{eq:systavg}
\end{equation}
which accounts for the variation among models. This procedure yields four determinations—two datasets times two formalisms—of each observable, each with a statistical uncertainty and a systematic uncertainty associated with the cutoff choice and parametrization.

The final step consists of averaging these four determinations into a single result. We adopt the conservative approach of assigning a weight of \(1/4\) to each determination, corresponding to no preference among the different analyses and finite-volume formalisms. This final average is also computed using \Cref{eq:avg,eq:statavg,eq:systavg}. The final results are quoted with the statistical and systematic uncertainties given separately. They are added in quadrature to obtain the total error.

\section{Results for scattering amplitudes}
\label{sec:results}

We now present observables obtained from the analysis of the lowest-lying finite-volume energies of the $NN$ system, in both the deuteron and dineutron channels on both ensembles. The analysis includes both datasets as well as both formalisms that deal with the left-hand cut, as introduced in \Cref{sec:FV}. Here we focus on the results, while the methodology has been presented in \Cref{sec:methodology}.
The complete set of individual fits is collected in \Cref{app:tables}; in this section, we provide only an overview of the main results and sources of uncertainties.

\subsection{Lowest partial waves, threshold parameters, and pole positions}

The primary output of our analysis is the energy dependence of the \(NN\) scattering phase shifts, with particular emphasis on the lowest partial waves, in which virtual- or bound-state poles occur. These are the \(S\)-wave in the dineutron system and the \(\alpha\)-wave in the deuteron system, following the Blatt-Biedenharn decomposition. In addition to the full energy-dependent phase shifts, we examine the associated threshold parameters, including the scattering length, effective range, and shape parameters. Finally, we determine the corresponding pole positions. Results for both ensembles are shown; supplemented by C103 at a heavier pion mass, they are used to discuss the pion mass dependence.

\begin{figure*}[p!]
    \centering
    \begin{subfigure}{\textwidth}
        \centering
        \includegraphics[width=\textwidth]{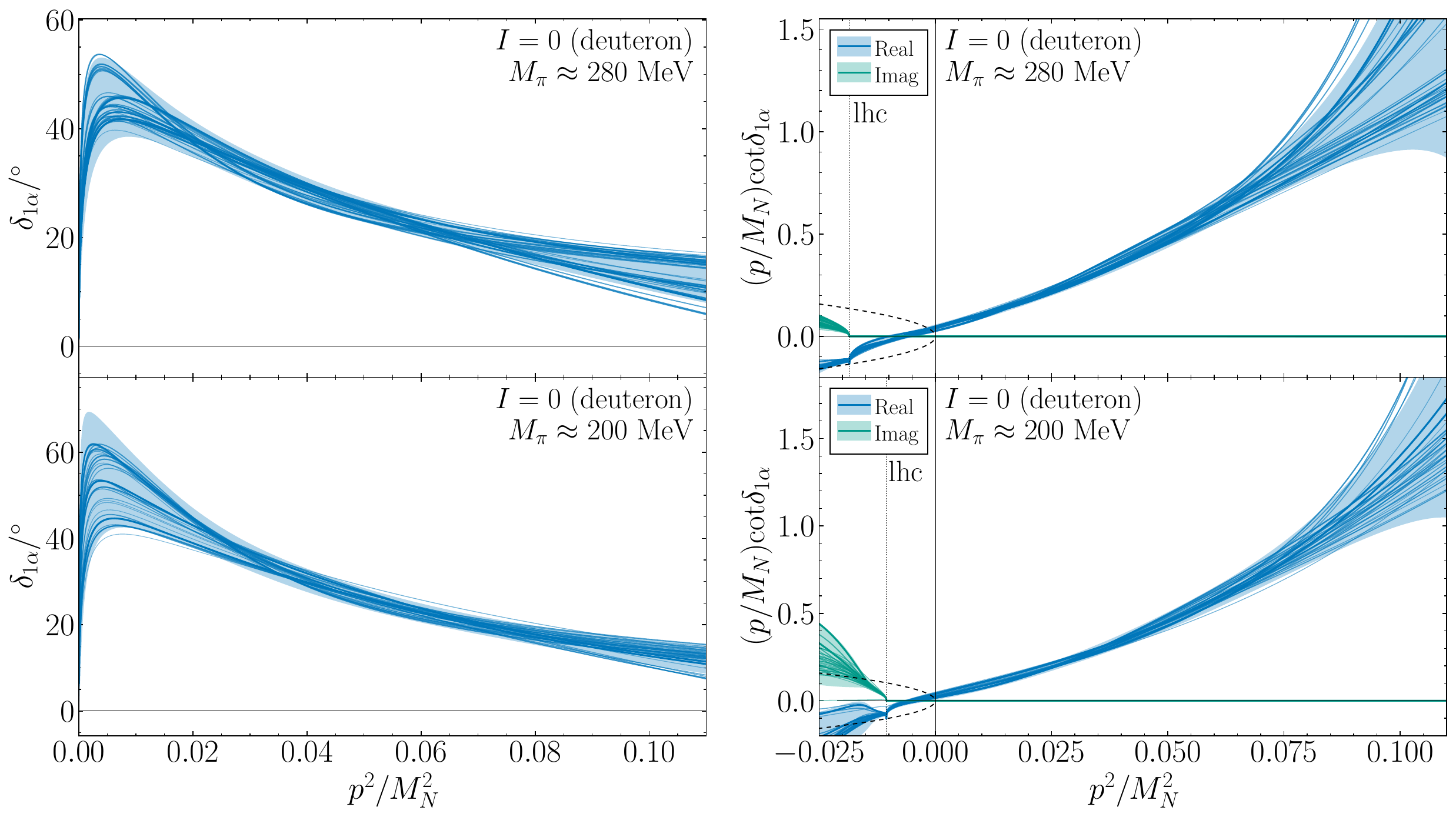}
        \caption{Isospin-0 (deuteron) channel, in the Blatt-Biedenharn decomposition.}
    \end{subfigure}\vspace{0.4cm}
    
    \begin{subfigure}{\textwidth}
        \centering
        \includegraphics[width=\textwidth]{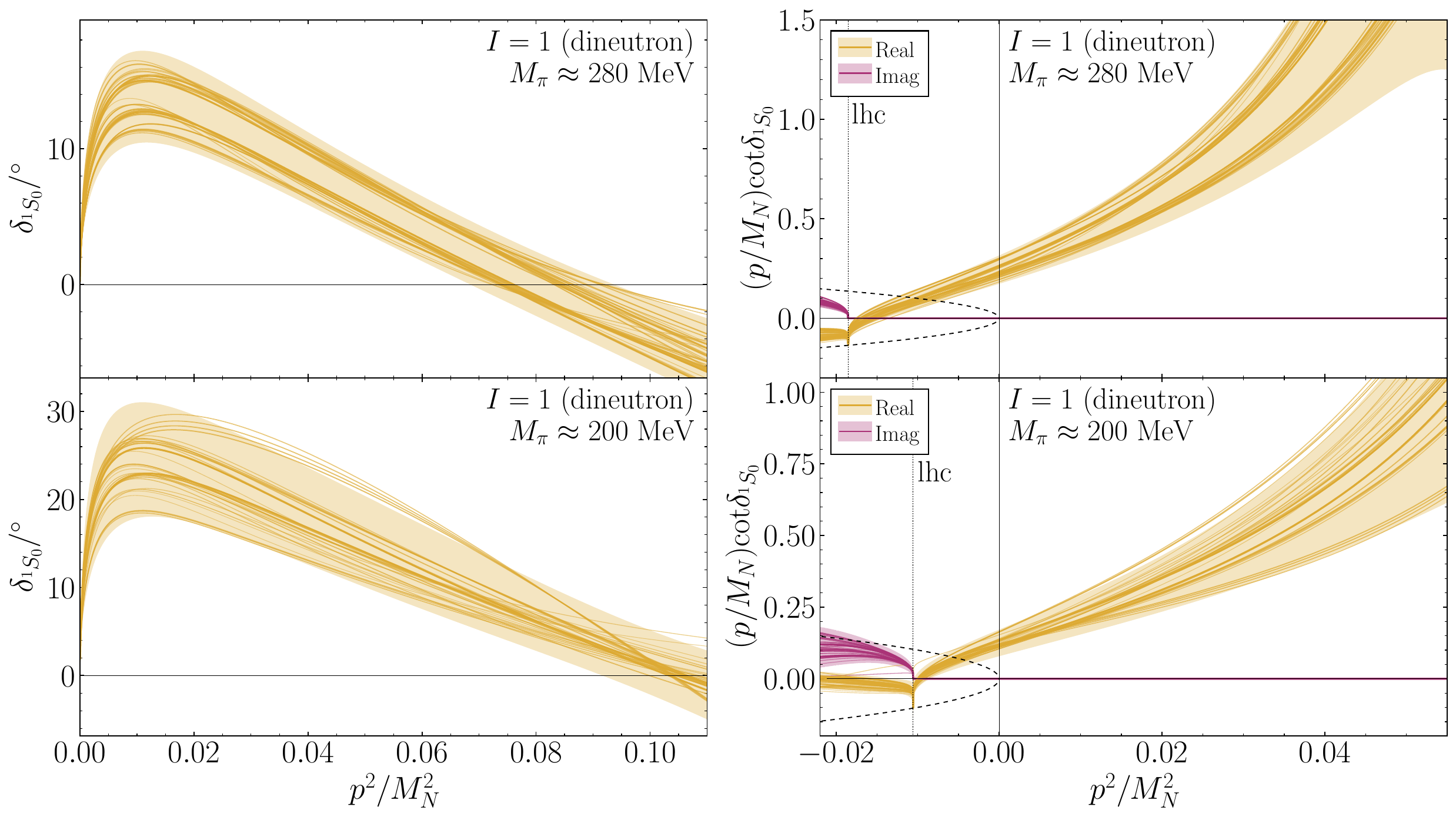}
        \caption{Isospin-1 (dineutron) channel.}
    \end{subfigure}
    \caption{\label{fig:phaseshiftI0I1withfits} Averaged results for the scattering phase shift of the lowest partial wave in the D251 ($M_\pi\approx280$ MeV) and J250 ($M_\pi\approx200$ MeV) ensemble for the $I=0$ and $I=1$ channels, together with results from all the models that enter the AIC average. Only models that contribute with a weight $w>10^{-3}$ are represented, with thiner/lighter lines corresponding to lower weights.
    }
\end{figure*}

The phase shifts as functions of the center-of-mass scattering momentum are shown in \Cref{fig:phaseshiftI0I1withfits} for both D251 and J250. These are presented both as the phase shift, \(\delta\), and in the \(p\cot\delta\) representation. Each panel includes an uncertainty band that incorporates statistical uncertainties and systematic variations arising from the energy-determination procedure, the left-hand cut formalism, and the parametrization and cutoff used. The individual fits entering the weighted average are displayed as solid curves with their opacity indicating their relative weight. Only those fits that contribute to the final result with a weight larger than $10^{-3}$, including the 1/4 factor, are shown. In all cases, we note that statistical uncertainty dominates the model-dependent systematic variation.

The phase shifts exhibit similar behavior in both isospin channels and on both ensembles. In particular, they rise rapidly immediately above threshold, indicating the presence of a nearby virtual state. This interpretation becomes more transparent in the \(p\cot\delta\) representation. According to the effective-range expansion, the intercept at \(p^2=0\) is given by \(-1/a_0\), where \(a_0\) is the scattering length. In all cases, this intercept is positive, corresponding to a negative scattering length in our convention and further supporting the virtual-state interpretation. The right panels of \Cref{fig:phaseshiftI0I1withfits} also display the bound- and virtual-state pole conditions; an intersection with the \(p\cot\delta\) curve signals a pole in the scattering amplitude. For both isospin channels and on both ensembles, the central curve intersects the positive, virtual-state branch. Thus, all of our results favor a virtual-state pole in the corresponding \(NN\) scattering amplitude.

\begin{figure*}[t!]
    \centering
    \includegraphics[width=\textwidth]{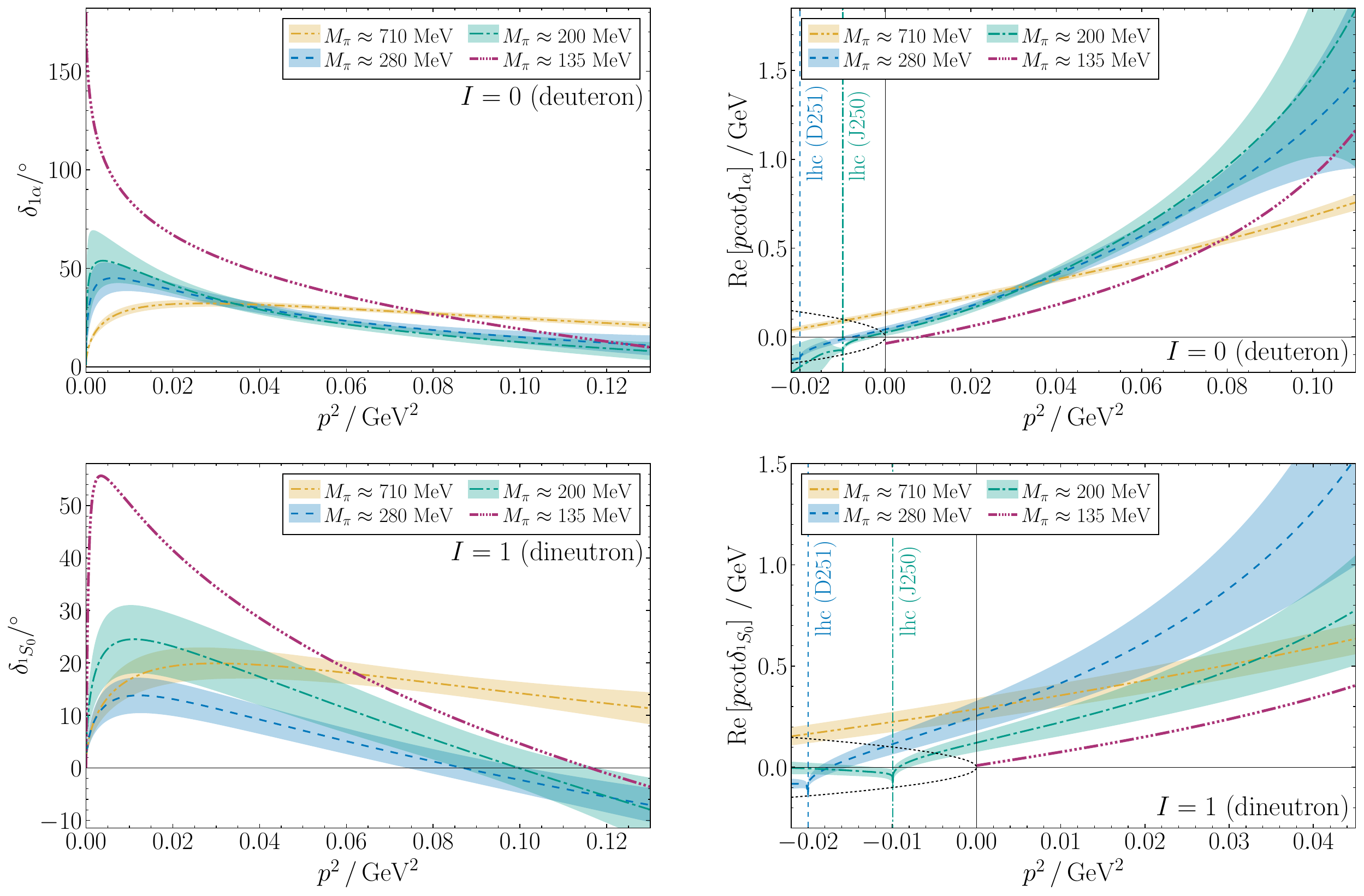}
    \caption{\label{fig:kcotI0I1withphysical} Results, in physical units, for the phase shift (left) and real part of the corresponding $p \cot \delta$ (right) of the lowest partial wave in the deuteron ($\alpha$ wave, top) and dineutron ($S$-wave, bottom) channels, for all ensembles considered in this work, together with experimental results at physical pion mass. }
\end{figure*}

We now compare the shapes of the phase shifts at different pion masses. This comparison is shown in the left column of \Cref{fig:kcotI0I1withphysical}, which also includes physical-point results obtained from the partial-wave analysis of experimental data \cite{nnonline,Stoks:1993tb} and lattice results at a heavier pion mass, $M_\pi\approx 710$ MeV. In the deuteron channel, the \(\delta_{1\alpha}\) phase shift differs qualitatively between the physical point and the heavier-than-physical ensembles. This difference arises primarily because the physical deuteron is a bound state rather than a virtual one, and its phase shift equals \(180^\circ\) at threshold. By contrast, the dineutron-channel phase shifts are qualitatively more similar across the range of pion masses considered. The near-threshold enhancement becomes more pronounced as the pion mass is lowered, suggesting that the virtual-state pole moves toward threshold.

The pion-mass dependence is more transparent in the \(p\cot\delta\) representation shown in the right panels of \Cref{fig:kcotI0I1withphysical}. As the pion mass is lowered, the threshold intercept of \(p\cot\delta\) decreases in both isospin channels. Comparing our lattice results with the known physical-point values shows that the evolution is generally in the expected direction.
In the deuteron channel, the approach of the threshold intercept of \(p\cot\delta\) toward zero signals proximity to a virtual-to-bound-state transition.

This picture can be made more quantitative by examining the pion-mass dependence of the threshold parameters and pole positions. The results are summarized in \Cref{tab:finalScatteringParameters}, where both statistical and systematic uncertainties are indicated. In the deuteron channel, the inverse scattering length approaches zero as the pion mass is lowered. On the J250 ensemble, it lies within \(1.5\sigma\) of zero, corresponding to the unitary limit at which the bound-state pole crosses threshold. A similar trend is observed in the pole momentum. Consequently, while a bound deuteron at \(M_\pi \approx 200\)~MeV is not favored by the central values, it cannot be excluded. On the D251 ensemble, at \(M_\pi \approx 280\)~MeV, the bound-state scenario is disfavored by more than \(2.5\sigma\). In contrast, the dineutron remains unbound with a significance exceeding \(5\sigma\) in all cases.

\begin{figure*}[t!]
    \centering
    \includegraphics[width=\textwidth]{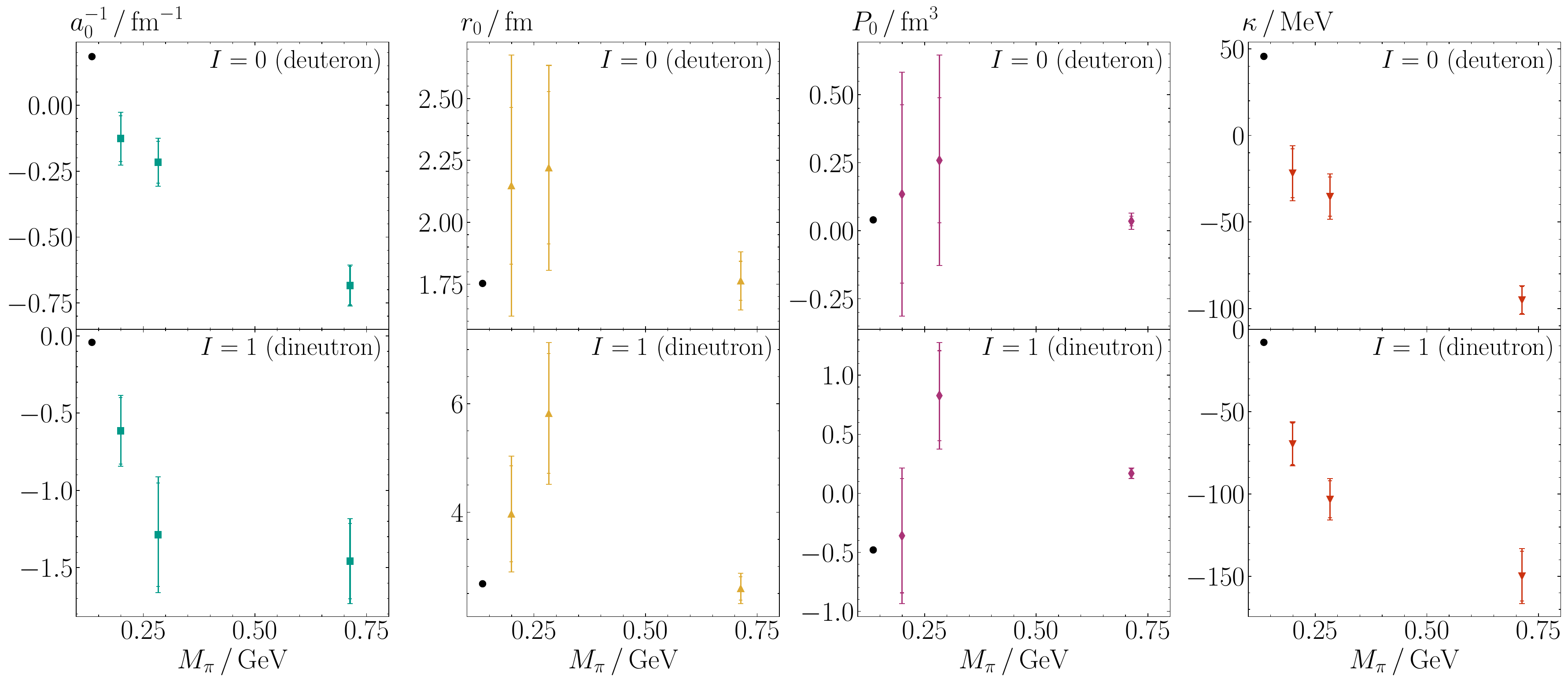}
    \caption{\label{fig:pionmassdependence} Results for the inverse scattering length, effective range, shape parameter and binding momentum for the deuteron (top) and dineutron (bottom) channels. Physical results are represented as black dots.
    }
\end{figure*}

The pion-mass dependence of the threshold parameters is summarized in \Cref{fig:pionmassdependence}. Both the inverse scattering length and the pole position evolve toward their physical-point values as the pion mass is lowered. Their relatively steep dependence suggests a virtual-to-bound-state transition in the deuteron channel at a pion mass of around \(M_\pi \approx 200\,\text{MeV}\). 
By contrast, the effective range and the shape parameter appear to be already consistent with their physical-point values, although the uncertainties are comparatively larger than for the scattering length.

\begin{table*}[h]
  \centering
  \tabcolsep=0.15cm
      \begin{tabular}{c c c c c c c c}
    \toprule
     Isospin & Ensemble & $M_\pi$ / MeV&  $a_0^{-1}$ / fm${}^{-1}$ & $a_0$ / fm &  $r_0$ / fm &  $P_0$ / fm${}^3$ & $\kappa$ / MeV \\
    \midrule
     \multirow{4}{*}{0} 
& C103 & 710  & $-0.685(73)(27)$  & $-1.46(16)(6)$  & $1.763(79)(86)$  & $0.035(18)(23)$  & $-95.1(7.7)(3.2)$ \\  
& D251 & 280  & $-0.216(79)(44)$  & $-4.9(2.7)(1.3)$  & $2.22(31)(28)$  & $0.26(23)(31)$  & $-35(11)(6)$ \\  
& J250 & 200  & $-0.126(87)(50)$  & $-9(11)(3)$  & $2.15(32)(42)$  & $0.13(33)(31)$  & $-22(14)(7)$ \\ 
& \rcite{NavarroPerez:2014ovp} 
& 135 & 0.185 & 5.420 & 1.753 & 0.040 & 45.67  \\
\midrule \multirow{4}{*}{1} 
& C103 & 710  & $-1.46(24)(13)$  & $-0.69(12)(6)$  & $2.59(22)(17)$  & $0.169(40)(20)$  & $-150(15)(7)$ \\  
& D251 & 280  & $-1.29(33)(17)$  & $-0.79(21)(10)$  & $5.8(1.1)(0.7)$  & $0.83(38)(24)$  & $-103(11)(5)$ \\  
& J250 & 200  & $-0.61(21)(8)$  & $-1.65(58)(19)$  & $3.97(89)(60)$  & $-0.36(48)(31)$  & $-70(13)(4)$ \\ 
& \rcite{NavarroPerez:2014ovp} & 135 & $-0.042$ & $-23.735$ & 2.673 & $-0.50$ & $-7.892$  \\  
\bottomrule
    \end{tabular}
    
  \caption{ Summary table with the scattering parameters determined from the lattice QCD calculations in this work. 
  The last row in each isospin channel indicates the physical point results taken from Table II and Table III in~\rcite{NavarroPerez:2014ovp}.
  While the treatment of isospin corrections for $NN$ scattering is a nontrivial subject, see \rcite{Miller:2006tv}, for the sake of qualitative comparison we simply use the $np$ scattering results for the physical point.}
  \label{tab:finalScatteringParameters}
\end{table*}

\subsection{Partial-wave mixing in the deuteron channel}

The finite-volume analysis performed in this work constrains not only the dominant $\alpha$ wave in the deuteron channel, but also the $\beta$ wave and the mixing angle. This is possible even for those fits where no explicit higher-partial-wave contact terms are included in the EFT potential or modified K matrix, as one-pion exchange generates nontrivial dynamics in the higher partial waves.

\begin{figure*}[t!]
    \centering
    \includegraphics[width=\textwidth]{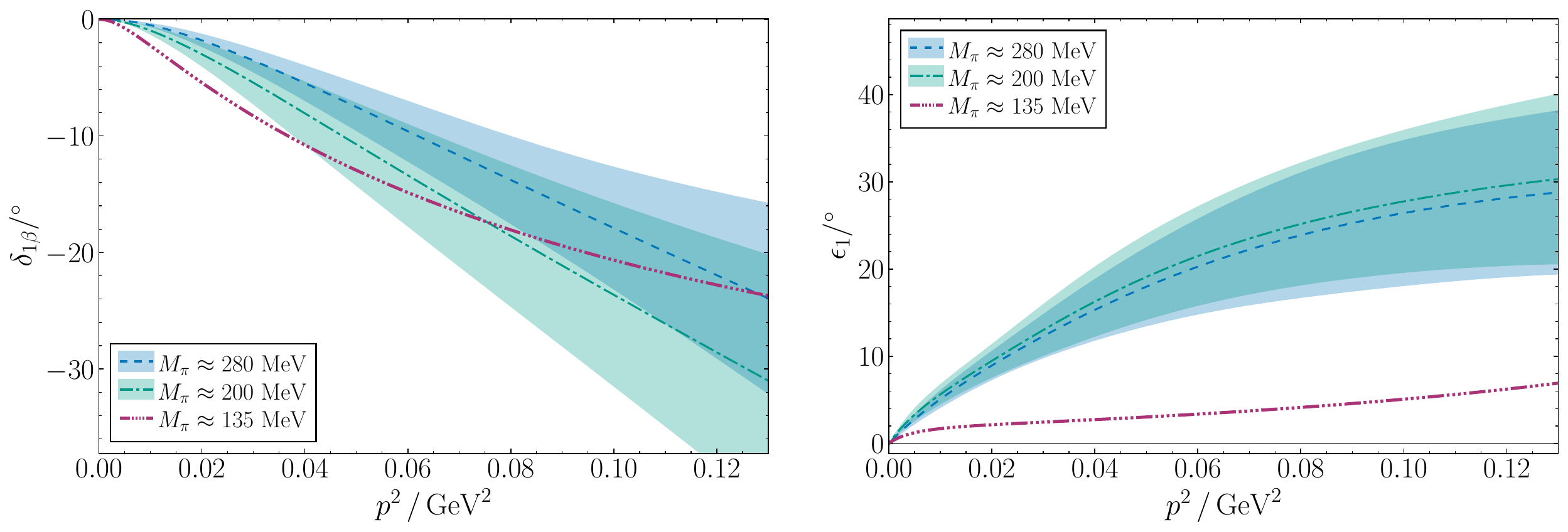}
    \caption{\label{fig:phaseshiftdwaveI0I1withphysical} Results, in physical units, for the $\beta$ wave (left) and the $S-D$ mixing angle (right) in the deuteron channel, for both D251 and J250, together with experimental results at physical pion mass.}
\end{figure*}

For a more complete comparison to phenomenological results, we present in \Cref{fig:phaseshiftdwaveI0I1withphysical} the $\beta$ wave and the mixing angle, which are defined following the Blatt-Biedenharn decomposition described in \Cref{eq:BBparam}. We compare our lattice results with the physical-point values~\cite{nnonline,Stoks:1993tb}. 

The $\beta$-wave results, displayed in the left panel, have large uncertainties. Nevertheless, at low $p^2$ the lattice $\delta_{1\beta}$ trend toward the physical curve as the pion mass is decreased, while at higher $p^2$ we see consistency within the large errors. The situation is somewhat different for the mixing angle, shown in the right panel. The results at heavier-than-physical pion masses yield a larger mixing angle than at the physical point, although they agree in sign. Despite the large uncertainties, this behavior suggests some tension between the lattice and physical-point results. The origin of this tension is unclear: it may reflect a nontrivial pion-mass dependence or indicate that a more flexible parametrization of the higher-partial-wave dynamics is needed. This latter option, however, would require additional data to be properly constrained, and so is left for future work.

\begin{figure*}[t!]
    \centering
    \includegraphics[width=\textwidth]{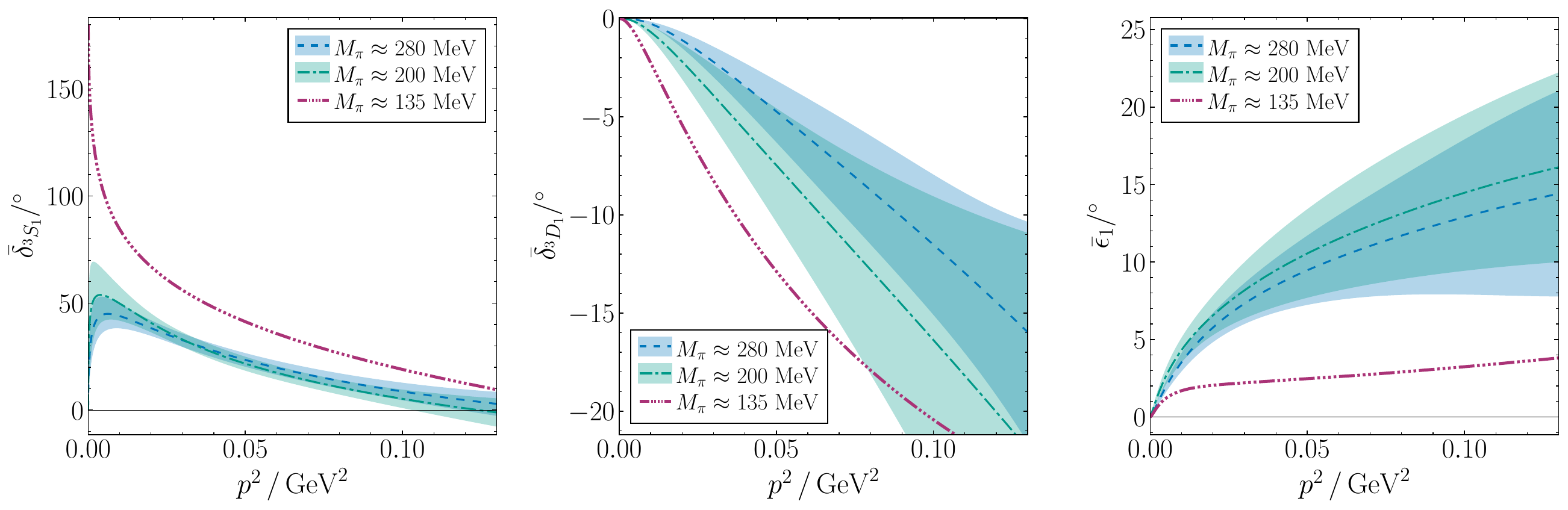}
    \caption{\label{fig:phaseshiftI0I1withphysicalStapp} Results, in physical units, for the bar phase shifts and mixing angle in the deuteron channel, as defined from the Stapp decomposition. Results are presented for both D251 and J250, together with experimental results at the physical pion mass.}
\end{figure*}

It is also informative to present the coupled partial-wave results in an alternative convention, namely the Stapp decomposition defined in \Cref{eq:stappparametrization}. In this version, the phase shifts and mixing angle differ from their Blatt-Biedenharn counterparts, although the two sets of variables are related as described in \Cref{eq:BB_Stapp}. The coupled-channel amplitudes expressed in the Stapp decomposition are shown in \Cref{fig:phaseshiftI0I1withphysicalStapp} for both ensembles. Despite the differences in their numerical values, the results exhibit the same qualitative behavior as in the Blatt-Biedenharn decomposition.

\subsection{Discussion of systematics\label{sec:systematics}}

In this analysis, we have considered two variants of the spectrum determination and two finite-volume formalisms. The results reported in \Cref{tab:finalScatteringParameters} and shown in \Cref{fig:phaseshiftI0I1withfits,fig:phaseshiftdwaveI0I1withphysical,fig:pionmassdependence,fig:kcotI0I1withphysical,fig:phaseshiftI0I1withphysicalStapp} already incorporate the systematic variations associated with these choices. In the following, we examine how the different analysis variants contribute to the combined systematic uncertainty.

\begin{figure*}[t!]
    \centering
    \begin{subfigure}[t]{\textwidth}
        \centering
        \includegraphics[width=\textwidth]{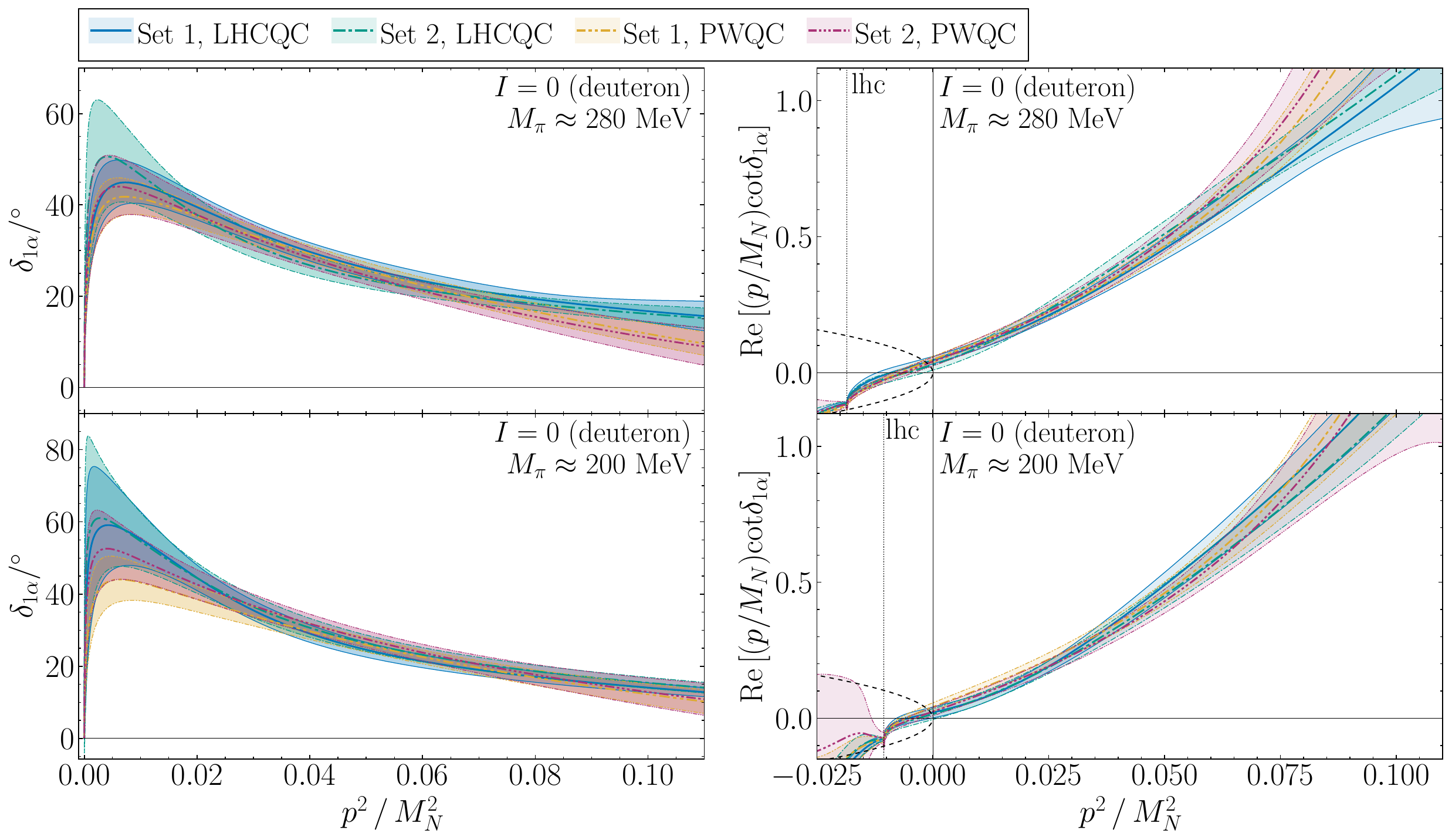}
        \caption{Isospin-0 (deuteron) channel, in the Blatt-Biedenharn decomposition.}
    \end{subfigure}\vspace{0.4cm}
    
    \begin{subfigure}[t]{\textwidth}
        \centering
        \includegraphics[width=\textwidth]{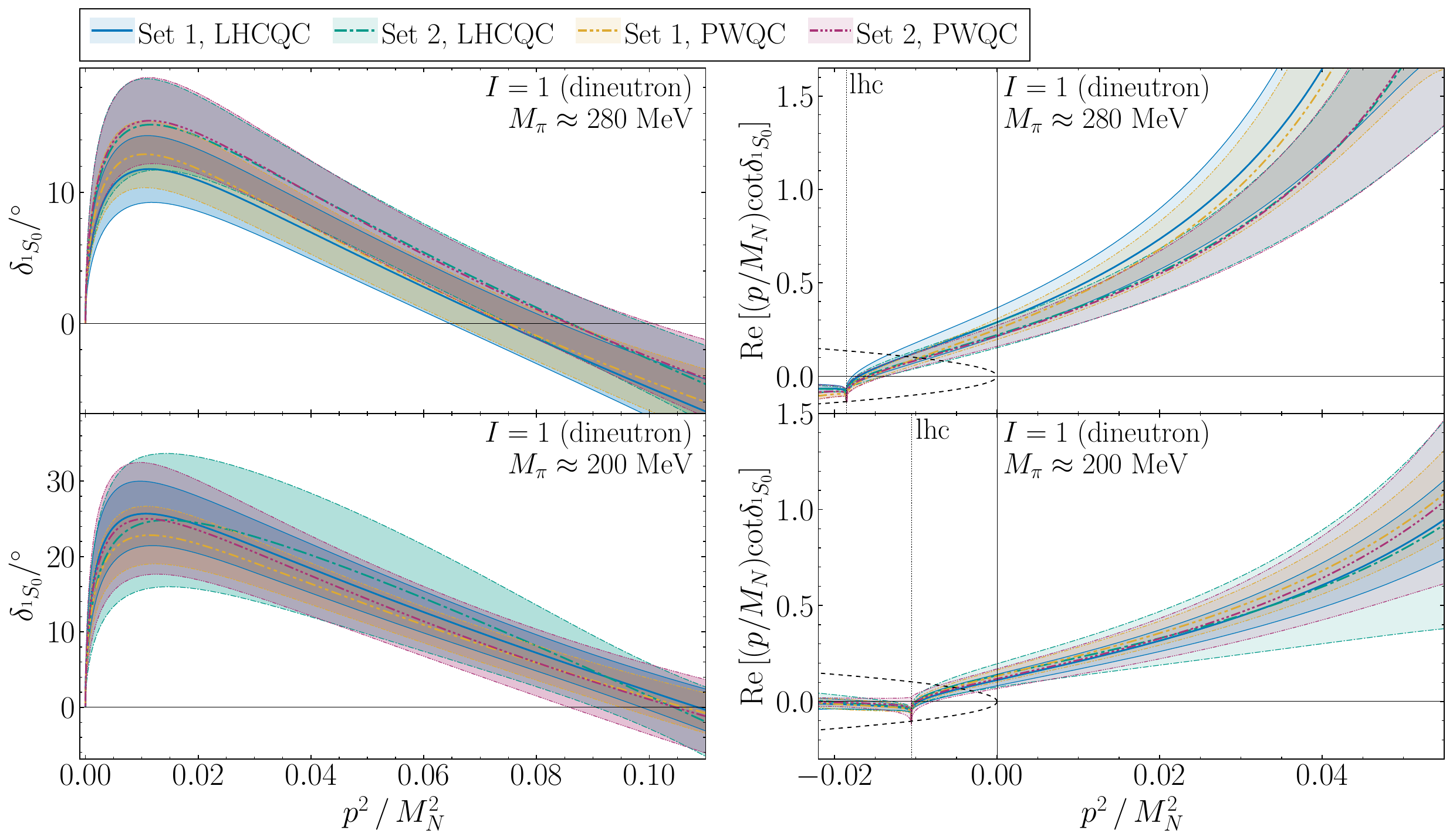}
        \caption{Isospin-1 (dineutron) channel.}
    \end{subfigure}
    \caption{\label{fig:phaseshiftI0I1allmethods} Comparison of the scattering phase shift of the lowest partial wave for all the methods used to analyze the lattice correlators and energy spectrum, as indicated in the legend. Results are presented for both D251 ($M_\pi\approx 280$ MeV) and J250 ($M_\pi\approx 200$ MeV) ensembles, and both isospin channels, as indicated in each panel.}
\end{figure*}

We begin by comparing the \(S\)-wave (dineutron) and \(\alpha\)-wave (deuteron) results. In \Cref{fig:phaseshiftI0I1allmethods}, we show four bands for each of the D251 and J250 ensembles and each isospin channel, corresponding to the combinations of the two spectrum-determination strategies and the two analysis formalisms. Each band represents an AIC-weighted model average over the fits within a given analysis variant. All four yield mutually consistent results, with appreciable differences emerging only at large cm momenta, beyond the region directly constrained by the data. This happens in particular for the deuteron $\alpha$ wave, where $p\cot\delta_{1\alpha}$ tends to have greater curvature at large cm momenta when using the PWQC. For the dineutron, on the other hand, the two formalisms yield very similar results, and the largest effect is between Set~1 and Set~2 on D251, at about the $1\sigma$ level. Consequently, the final average over the four variants, performed with equal weights, leads to only a moderate increase in the total uncertainty. More concretely, in the energy range where the lattice data lie, the inclusion of systematic effects only increases the uncertainty of the phase shifts by 20\% or less. 

\begin{figure*}[t!]
    \centering
    \begin{subfigure}{\textwidth}
        \centering
        \includegraphics[width=\textwidth]{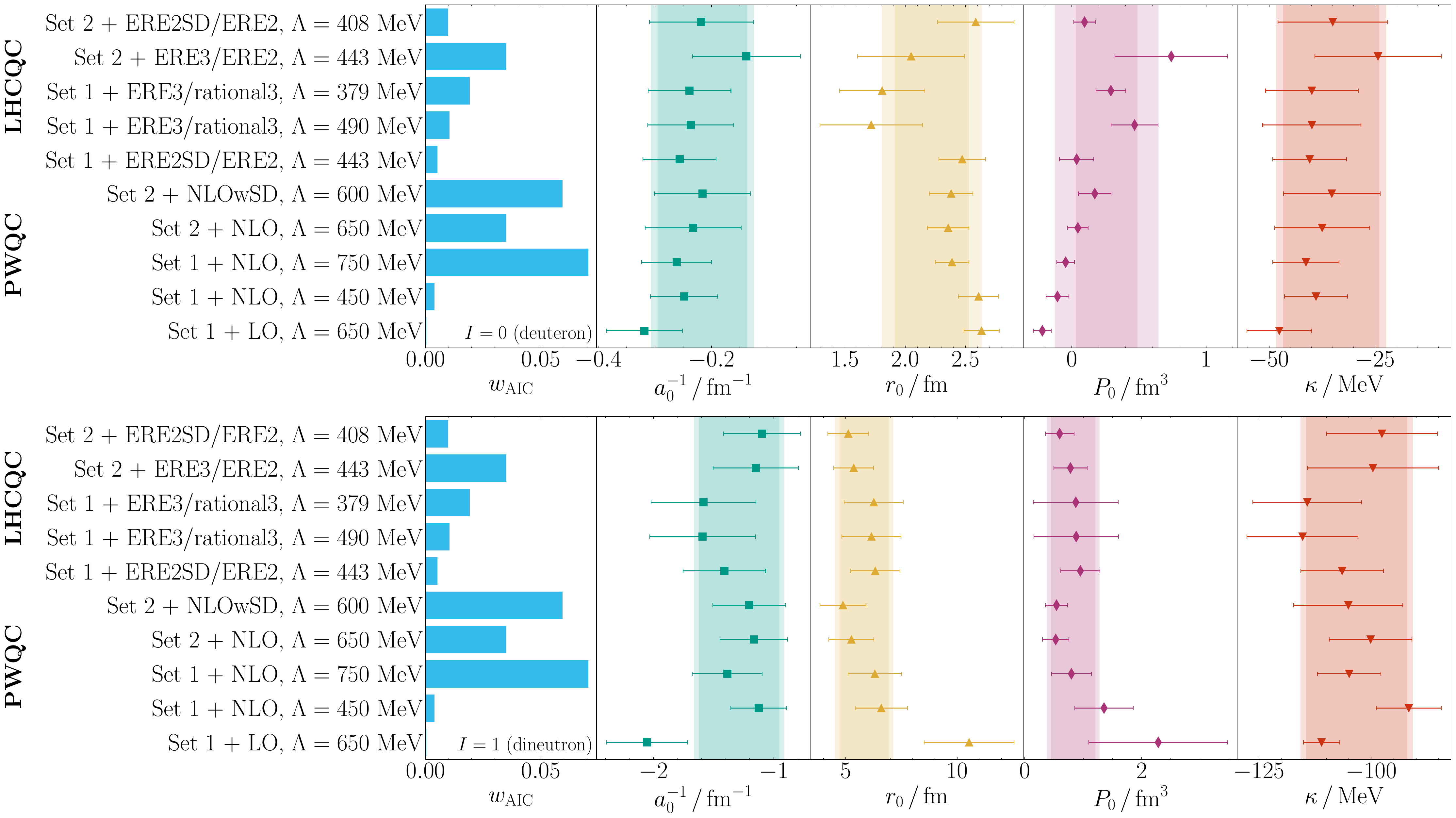}
        \caption{D251 ensemble ($M_\pi\approx280$ MeV).}
    \end{subfigure}\vspace{0.7cm}
    
    \begin{subfigure}{\textwidth}
        \centering
        \includegraphics[width=\textwidth]{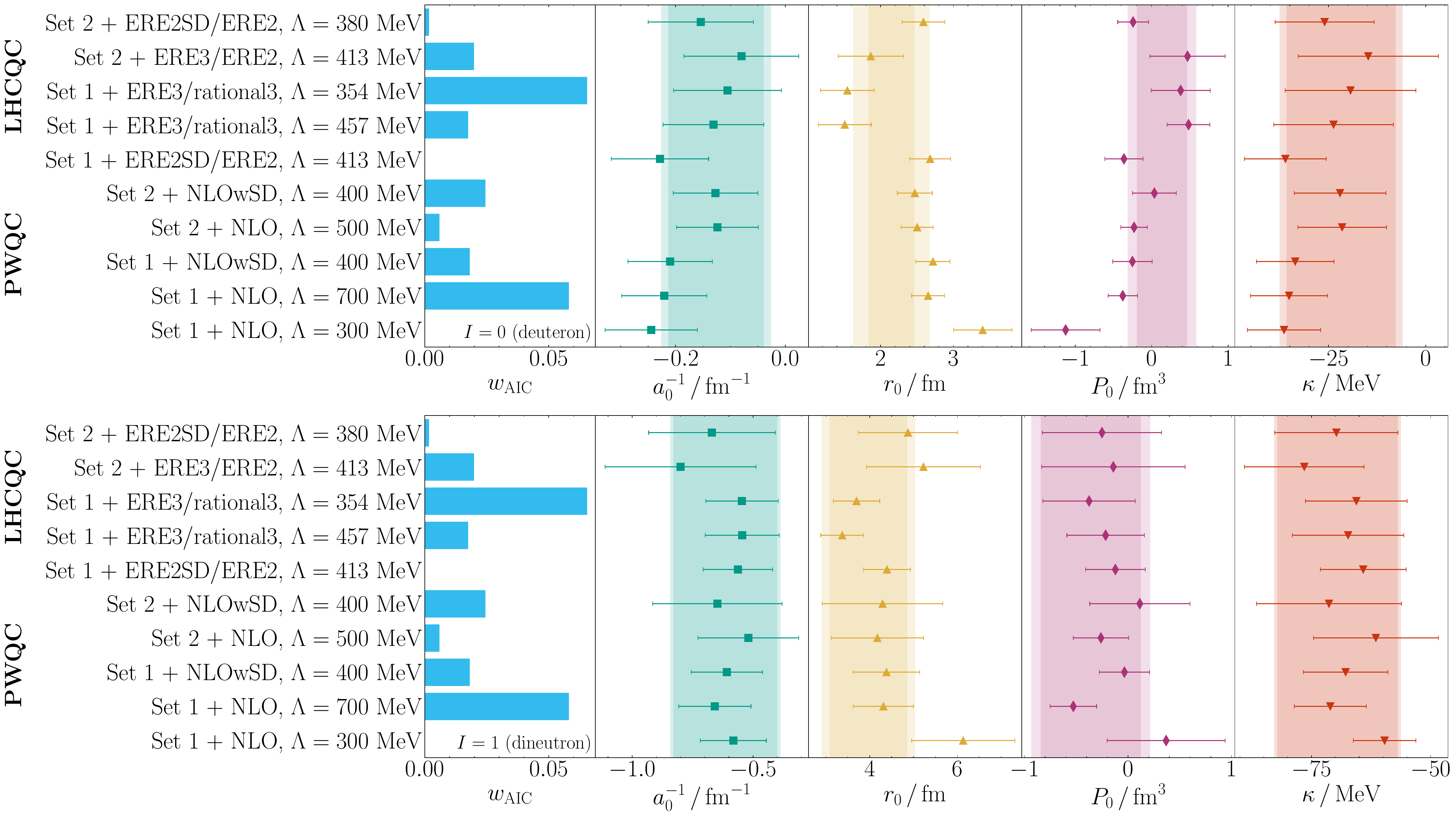}
        \caption{J250 ensemble ($M_\pi\approx200$ MeV).}
    \end{subfigure}
    \caption{\label{fig:scatteringobservablessystematics} Subset of results for scattering observables for the D251 and J250 ensembles from multiple combinations of the energy analysis, QC, fit model and cutoff, together with the averaged result. The dark and light bands represent the statistical and total uncertainties, respectively. The weight associated with each model is also indicated.
     }
\end{figure*}

It is also informative to examine the systematic variation of the scattering parameters reported in \Cref{tab:finalScatteringParameters}. This comparison is shown in \Cref{fig:scatteringobservablessystematics} for both the D251 and the J250 ensembles. The figures display the scattering observables obtained from a broad selection of fits of varying quality, including both spectrum determinations, both analysis formalisms, and different choices of cutoff and model parametrization. The individual results are presented together with their assigned weights, including the 1/4 factor used for combining the four different strategies, and the final model-averaged result. As can be seen, the final uncertainty band adequately captures the spread among the individual fit results, while fits that deviate substantially from the average are assigned relatively small weights.
The errors of most of the scattering parameters are clearly statistics dominated, with the exception of the effective ranges and shape parameters in the isospin-0 channel, where the statistical and systematic uncertainties are of similar size.

\section{Conclusion and outlook}
\label{sec:conclusions}

In this work, we have presented the first study of \(NN\) scattering at close-to-physical quark masses. We determined the low-energy \(NN\) scattering amplitudes in the lowest partial waves of both isospin channels using two CLS ensembles with pion masses of approximately \(200\) and \(280\)~MeV. In particular, the results at \(M_\pi\approx200\)~MeV provide the lattice-QCD determination of the corresponding \(NN\) scattering parameters closest to the physical point to date.

Two-nucleon correlation functions were computed using the distillation framework and a large set of operators. The energy levels were extracted using Bayesian multi-exponential fits to the principal correlators after solving the GEVP. We performed two independent analyses, which allow us to estimate systematics of the energy determination. 

The finite-volume formalism provides the bridge between the finite-volume energies and the scattering amplitudes.  In that regard, a central methodological development of this work is the use of finite-volume formalisms that account explicitly for the nearby left-hand cut induced by pion exchange processes. This treatment is essential because the left-hand cut lies close to the \(NN\) threshold and, in one ensemble, the beginning of the left-hand cut is even higher than some finite-volume energies included in the analysis. We employed two complementary approaches: the plane-wave quantization condition and the relativistic left-hand-cut formalism. Together with the two independent spectrum determinations and variations of the amplitude parametrization and cutoff, these approaches allow the systematic uncertainties associated with the finite-volume analysis to be assessed in a controlled manner. The consistency among the different analysis variants demonstrates the robustness of the resulting amplitudes and scattering parameters.

By analytically continuing the constrained amplitudes, we investigated their pole structure and identified a near-threshold pole in every channel and on both ensembles, as required by analyticity and the presence of a left-hand cut. The central results favor poles on the second Riemann sheet, corresponding to virtual states. As the pion mass is lowered, the pole in the deuteron channel moves toward threshold, consistent with its eventual transition into the physical deuteron bound state on the first Riemann sheet. Although a bound state at \(M_\pi\approx200\)~MeV is not favored by the central values, it cannot be excluded within the present uncertainties. By contrast, the dineutron remains a virtual state throughout the pion-mass range considered.

Previous work~\cite{BaryonScattering:2025ziz} computed the scattering amplitude at $M_\pi \sim 710$ MeV. Combining these results with the new ones in this work, there is a clear evolution of the pole positions and threshold parameters toward their physical values as the pion mass is lowered. The strongest pion-mass dependence is observed in the inverse scattering length and the pole position, whereas the effective range and shape parameters exhibit a milder dependence and are already broadly consistent with their physical-point values. 

The left-hand-cut analysis in this work is sensitive to the pion-nucleon coupling $g_{\pi NN}$. We find that the extracted $g_{\pi NN}$ is qualitatively consistent with the expected phenomenological value, although it exhibits significant model dependence. This may reflect the limited ability of the scattering data to constrain this quantity and suggests that the analysis would benefit from an independent determination of the pion-nucleon coupling to provide an additional constraint on the fits.

The pion-mass range covered here also enables a direct connection with effective-field-theory predictions. In a companion paper~\cite{Baeza-Ballesteros:2026qxo}, we investigate the pion-mass dependence of the scattering parameters in greater detail and compare the lattice QCD results with EFT expectations.

Two important directions remain for future work. Calculations at multiple lattice spacings are required to determine the continuum limit of the scattering amplitudes, while simulations directly at the physical point would remove the need to infer the final stage of the pion-mass evolution. Together, these studies will provide a more precise picture of the emergence of the physical deuteron bound state and the near-threshold dynamics of the \(NN\) system.

\begin{acknowledgments}
We acknowledge helpful discussions with Ra\'ul Brice\~no, Sebastian Dawid, Evgeny Epelbaum, Andrew Jackura, Gerald A. Miller, and Nelson Pitanga Lachini.

JBB thanks the University of Bern and the Albert Einstein Center for Fundamental Physics for their hospitality during his visit.

The work of FRL and MS was supported in part by the Swiss
National Science Foundation (SNSF) through grant No. 200021-236432, and the Platform for
Advanced Scientific Computing (PASC) project ``ALPENGLUE''. 
The work of ABR, JB, DL is supported by the European Research Council (ERC) consolidator grant StrangeScatt-101088506.
The work of AN is supported by the NSF Faculty Early Career Development Program (CAREER) under award PHY-2047185.
The work of TRR was supported by the NSF through cooperative agreement 2020275. CM acknowledges support from the NSF under awards PHY-2514831, OAC-2311430, and OAC-2608348.
This work was supported in part by the US Department of Energy, Office of Science under grant contract numbers DE-AC02-05CH11231 (ML, TRR, AWL), through the
“Nuclear Theory for New Physics” Topical Collaboration award No.\ DE-SC0023663 (ML, TRR, AWL).
The work of HW was supported by Deutsche Forschungsgemeinschaft
(German Research Foundation, DFG) through the Cluster of Excellence “Precision Physics, Fundamental Interactions and Structure of Matter” (PRISMA++ EXC 2118/2), funded within the German Excellence
Strategy (Project No.\ 390831469). DM is funded by the Heisenberg Programme of the Deutsche Forschungsgemeinschaft (DFG, German Research Foundation) project number 454605793.

For the J250 ensemble, the authors gratefully acknowledge the computing time provided by HLRS on ``Hawk'' through the project ``BarScattChi'' and resources of the National Energy Research Scientific Computing Center (NERSC), a Department of Energy User Facility using NERSC award ERCAP0024058. The former was used for production, while the latter contributed to the thermalization. This work presents the first use of this ensemble.

The perambulators were computed using either
QUDA-LapH~\cite{Clark:2025cuz}, which is built on
QUDA~\cite{Clark:2009wm, Babich:2011np, Clark:2016rdz, Clark:2017ekr,
  Clark:2025cuz}, or a combination of QDP++~\cite{Edwards:2004sx} and
QUDA. The mode triplets necessary for contractions involving baryons
were computed using QDP++ and PRIMME~\cite{PRIMME}. The contractions
were performed using NumPy~\cite{numpy} and CuPy~\cite{cupy}. Correlator analysis utilized the \texttt{gvar}~\cite{gvar} and \texttt{lsqfit}~\cite{lsqfit} libraries.

Part of the analysis was performed on UBELIX (\url{https://www.id.unibe.ch/hpc}), the HPC cluster at the University of Bern, and on grants from the Swiss National Supercomputing Centre (CSCS) under project IDs \texttt{c36} and \texttt{lp151} on Alps, as well as on the HPC cluster Elysium
of the Ruhr-Universit\"at Bochum, subsidized by the DFG(INST213/1055-1). Some of the spectrum fits were performed in the PAX cluster at DESY in Zeuthen. The perambulators on D251 used a grant from CSCS under project ID \texttt{lp53} on Alps. The perambulators on J250 and the contractions for
J250 and D251 were computed using resources provided by the Gauss
Centre for Supercomputing e.V.\ (\url{www.gauss-centre.eu}) on
JUWELS~\cite{juwels} at Jülich Supercomputing Centre.
We are grateful to our colleagues within the CLS initiative for sharing the D251 ensemble.

\end{acknowledgments}

\clearpage
\appendix
\crefalias{section}{appendix}
\widetext

\section{Lattice dispersion relation}
\label{app:disp}

In general, single-nucleon energy levels differ between the lattice and the continuum. This can be quantified by modeling the single-nucleon energy as a function of its lab-frame momentum $|\boldsymbol{p}|^2 = \left(\frac{2\pi}{L}\right)^2|\boldsymbol{n}|^2$, using a lattice dispersion relation
\begin{equation}
    \label{eq:dispersion}
    (aE_n)^2 = (aM_N)^2 + \xi \left(\frac{2\pi a}{L}|\boldsymbol{n}|\right)^2,
\end{equation}
where the subscript of $E_n$ denotes the magnitude of $\boldsymbol{n}$ in the momentum, and $\boldsymbol{n}$ is a vector of integers.

\begin{figure}[htbp!]
    \centering
    \includegraphics[width=0.5\columnwidth]{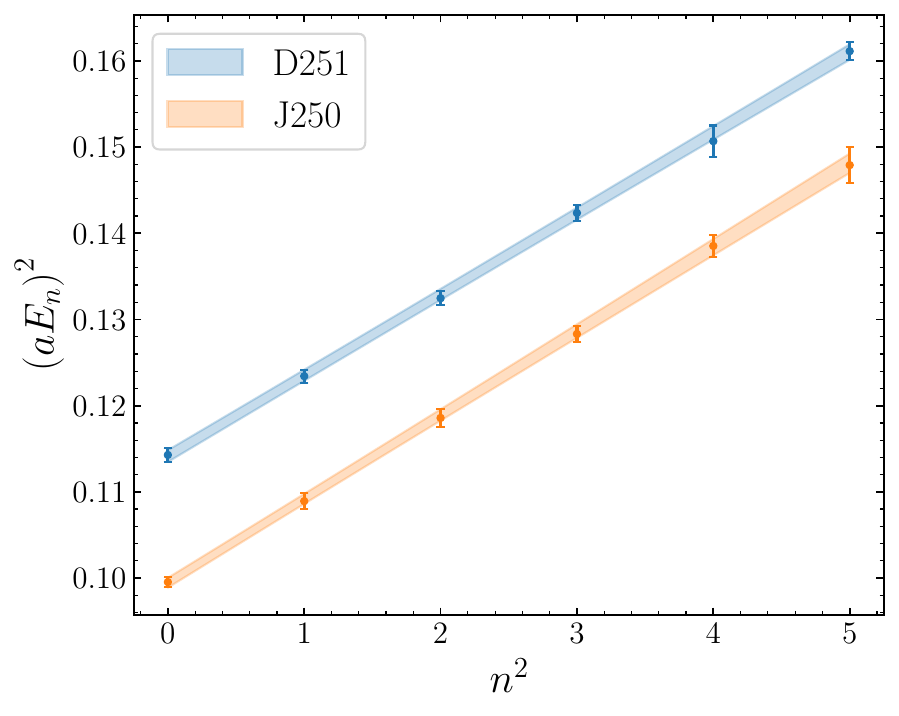}
    \caption{\label{fig:dispersion}Dispersion relation for the D251 (blue) and J250 (orange) single-nucleon spectrum. 
    The data are obtained using the conspiracy model with the reference states listed in Table~\ref{tab:ref_states}. The fit is performed using \Cref{eq:dispersion}.}
\end{figure}

\Cref{fig:dispersion} shows the data for $E_n$ and fits to \Cref{eq:dispersion} for D251 as well as J250.
The data are obtained from the single-nucleon sector of the conspiracy model; see \Cref{eq:conspiracy_N}. The corresponding two-nucleon energy levels fitted together with the single nucleon are shown in \Cref{tab:ref_states}.
We fit the obtained data for $E_n$ to \Cref{eq:dispersion}, using loose priors for $M_N$ and $\xi$.
The bands in \Cref{fig:dispersion} show the corresponding posterior predictions, and the parameter posteriors are
\begin{align}
    &aM_N^{\rm D251} = 0.3379(10),&\xi^{\rm D251} = 0.973(18) ,\\
    &aM_N^{\rm J250} = 0.31536(91),&\xi^{\rm J250} = 1.010(22).
\end{align}
Thus, the observed discretization effects on the single-nucleon dispersion relation are at most at the few-percent level, with the fitted values of $\xi$ being less than $1.5\sigma$ away from unity.

\begin{table*}[h!]
   \centering
   \begin{tabular}{c c c c c c c c c}
     \toprule
     & $N(0)$ & $N(1)$ & $N(2)$ &  $N(3)$ & $N(4)$ & $N(5)$ \\
     \midrule
      & $T_{1g}(0)$ level 0  & $T_{1g}(0)$ level 1 & $A_{2g}(0)$ level 0 & $A_{2g}(0)$ level 1 & $A_1(4)$ level 0 & $A_1(4)$ level 1  \\
     \bottomrule
   \end{tabular}
   \caption{Two-nucleon energy levels that have $N(n^2)$ as a single-nucleon reference state in the conspiracy fits (see Sec.~\ref{sec:conspiracy}). The single-nucleon energies obtained from these combined fits are used for the dispersion relation analysis. The same selection is used for both J250 and D251. Note that the reference states for $N(2)$ and $N(3)$ have not been used in the main analysis.}
   \label{tab:ref_states}
\end{table*}

\section{Details on the spectrum extraction}
\label{app:stability}

This appendix provides additional implementation details on the spectrum extraction presented in \Cref{sec:conspiracy,sec:spectrum} and an in-depth analysis of its stability. Details on the structure of the ancillary data file are also given.

\subsection{Stability of the spectrum extraction}
\begin{figure}[htbp]
    \centering
    \includegraphics[width=\columnwidth]{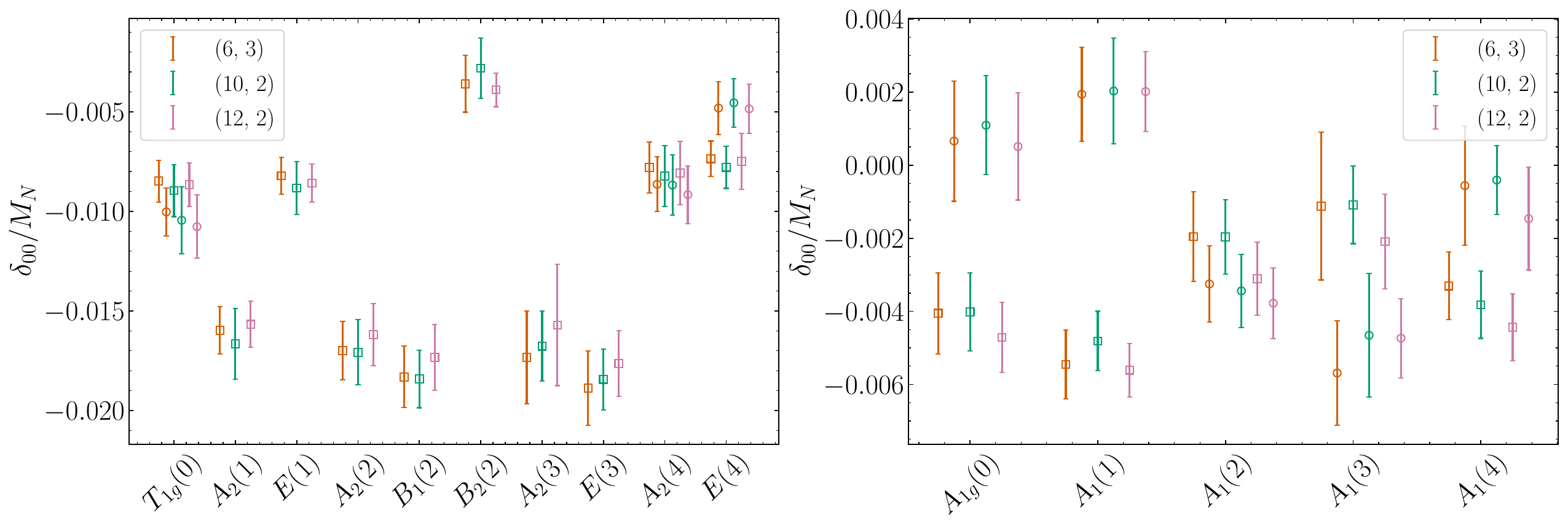}
    \caption{\label{fig:D251_stability}Left: stability of the D251 $I=0$ finite-volume interaction energies. Right: stability of the D251 $I=1$ finite-volume interaction energies. In both panels energy shifts are normalized by the ensemble's nucleon mass. Squares and circles represent the ground state and first excited state for various $\tminnstates$ combinations. Our final analysis uses $\tminnstates = (6a,\,3)$ (orange points).}
\end{figure}

The robustness of the spectrum extraction in ``Set 1'' is assessed 
by analyzing its stability under variations in the $\tminnstates$ choices.
On the D251 ensemble, we performed fits with $\nstates=2$ and $3$ for several values of $\tmin$ in the range $\tmin/a \in [6,12]$ for both the $I=0$ and the $I=1$ channels.
The stability of the interacting energy shift, for a selection of models, is shown in \Cref{fig:D251_stability} for all the irreps relevant for the finite-volume analysis.
Overall, the extracted energy shifts on D251 are stable and fits performed
with different $\tminnstates$ provide results for $\delta_{00}$ with a
similar statistical uncertainty, which are compatible within one standard deviation.

\begin{figure}[htbp]
    \centering
    \includegraphics[width=\columnwidth]{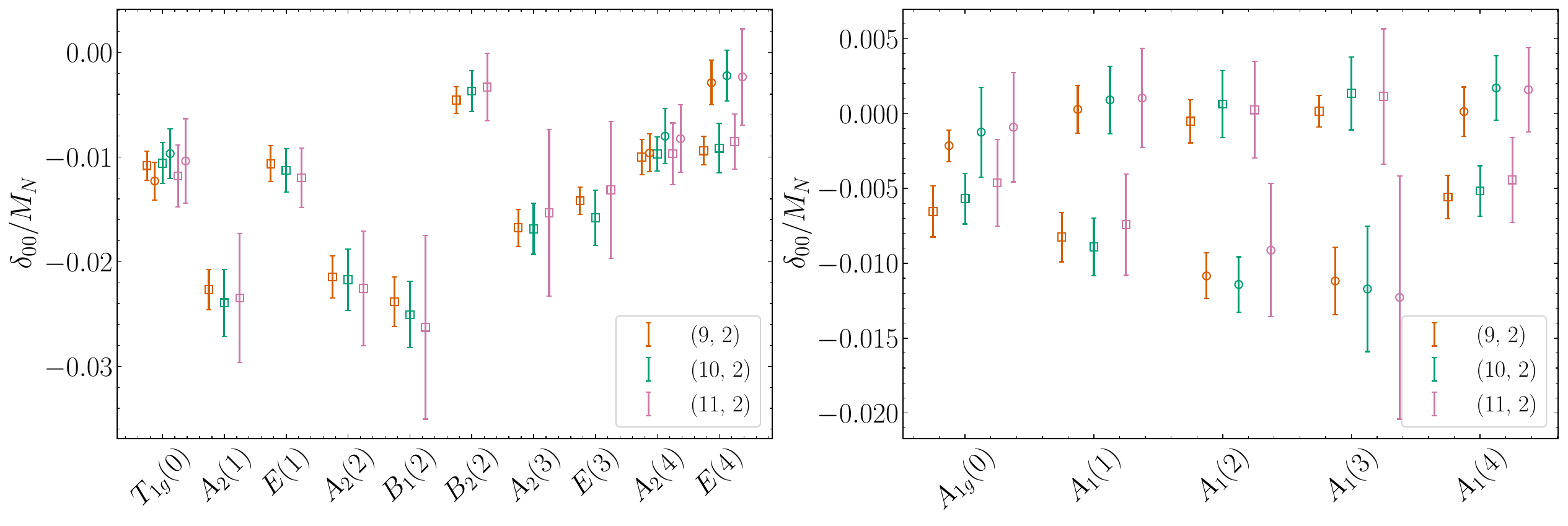}
    \caption{\label{fig:J250_stability}Same as in \Cref{fig:D251_stability} but for the J250 ensemble. Three-state fits are omitted from the stability analysis; see the discussion in the main text. Our final analysis uses $\tminnstates = (10a,\,2)$.}
\end{figure}
The corresponding stability analysis for the energy shifts
relevant for the $I=0$ and $I=1$ channels
on the J250 ensemble is shown in \Cref{fig:J250_stability}.
As discussed in \Cref{sec:spectrum}, the extracted energies on the J250 ensemble are stable under the variation of $\tmin$ except for a few levels, especially
in the $I = 0$ channel, where fits with $\tmin = 11$ exhibit
larger statistical uncertainties.
This behavior is visible in \Cref{fig:J250_T1g_stability} for the $T_{1g}(0)$ irrep and in \Cref{fig:J250_stability} for all the irreps relevant for the finite-volume 
analysis both for the $I=0$ and $I=1$ sectors. Moreover, the more severe 
degradation of the signal at large time separations and the reduced energy gaps on
the J250 ensemble do not allow the clear resolution of three states in our multi-state fits. For this reason, in the stability analysis on the J250 ensemble we only 
compare $\nstates=2$ fits on time separation windows starting at $\tmin/a\in[9,11]$.

\begin{figure}
    \centering
    \includegraphics[width=0.49\textwidth]{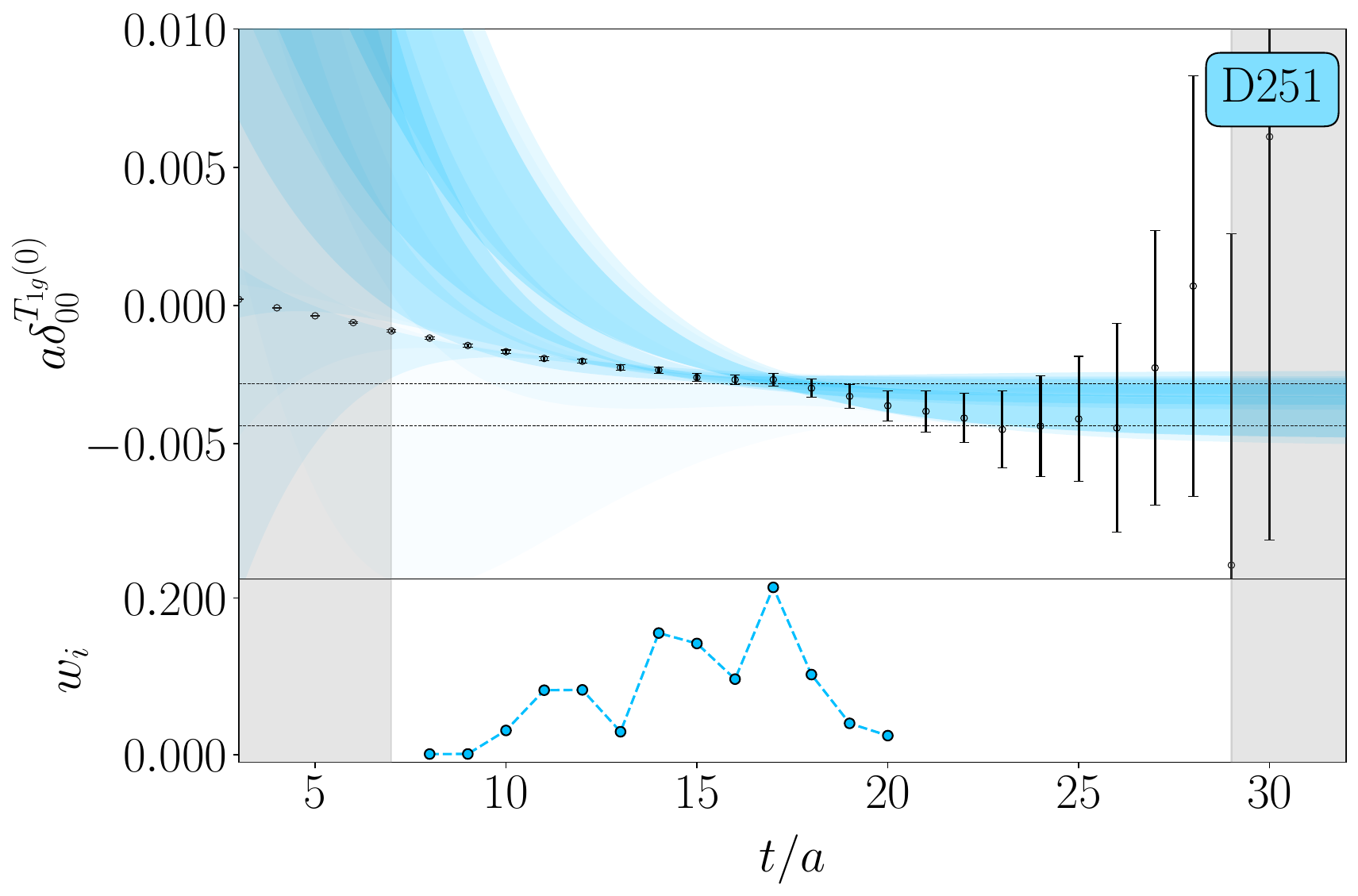}
    \includegraphics[width=0.49\textwidth]{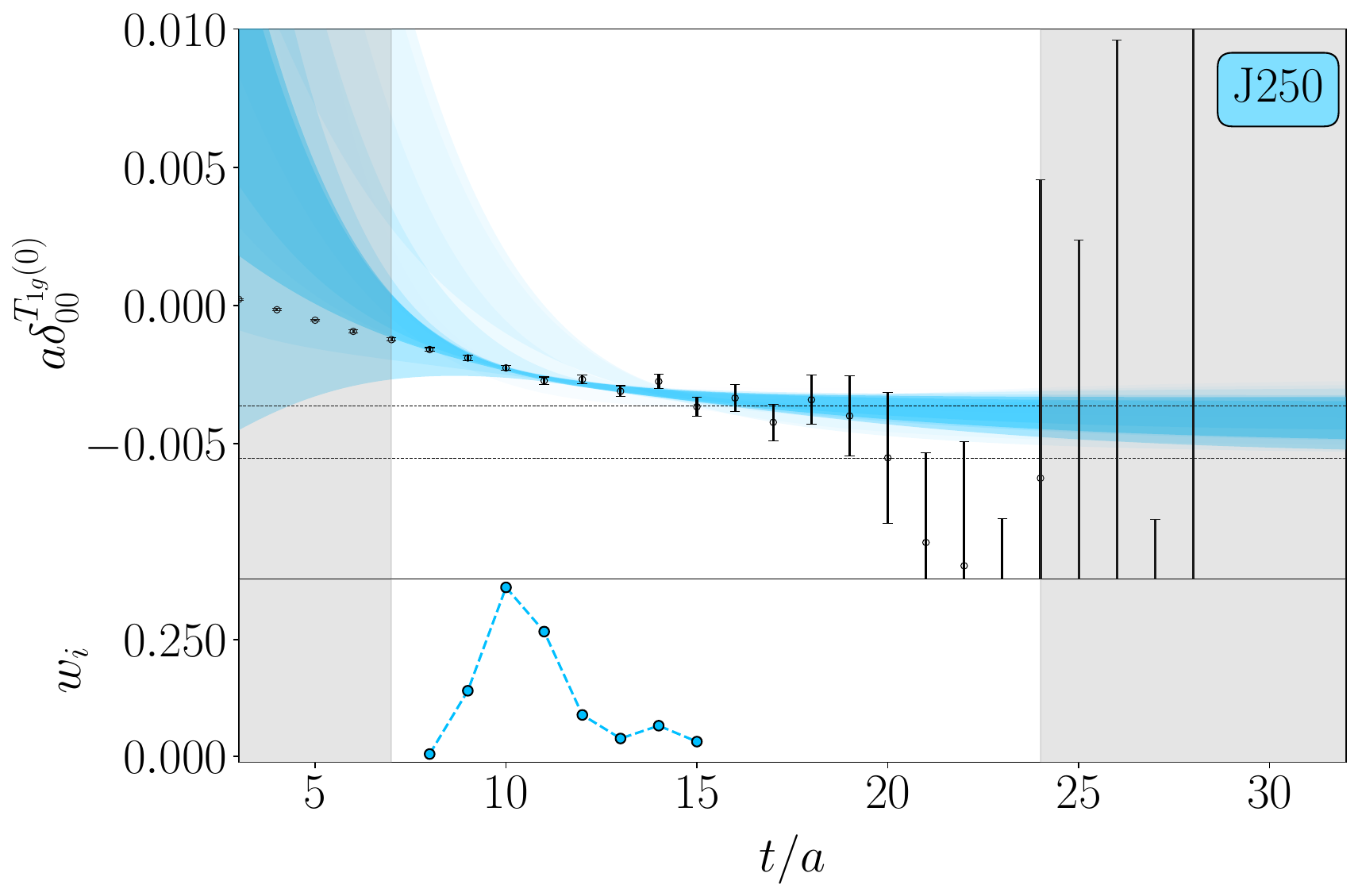}
    \caption{\label{fig:model_avg}Model averaging for the $T_{1g}(0)$ ground-state energy shift on the D251 (left) and J250 (right) ensembles. Different bands correspond to fits with different choices of $t_{\rm min}$, as described in the main text. The opacity of each curve is determined by its AIC weight, given in \Cref{eq:AIC_weight}. The weights are also shown in the lower panel as a function of $\tmin$. The dashed horizontal lines represent the 16th and 84th percentiles.}
\end{figure}

In \Cref{fig:model_avg} we show the estimated energy shift for
the ground state in the $T_{1g}(0)$ irrep on both ensembles from ``Set 2''. The bands
represent the individual fits entering the weighted average. In the lower
panel of each plot, we report the AIC weights of the fits at the
corresponding values of $\tmin$.
For all correlators, $t_{\rm{max}}$ has been fixed to the last time
slice for which a reliable signal is observed.
Single-nucleon correlators have been analyzed using a variety of models
with $\nstates=1,\dots, 6$, varying $\tmin$ while keeping
$t_{\rm max}$ fixed.
The optimal model, i.e. a $(\tmin,\nstates)$ combination,
for single-nucleon correlators has been selected according to
the minimum value of the AIC; see \Cref{eq:AIC_weight}.
Subsequently, the conspiracy fits described in \Cref{sec:conspiracy} were applied
to two-nucleon correlators for different values of $\tmin$, at fixed $\nstates$,
ensuring that a sufficient number of data points were included in each
fit to avoid overfitting. Central values and statistical uncertainties 
were then computed as detailed in the main text.
The propagation of the uncertainty described in \Cref{sec:spectrum} aims
to incorporate into the bootstrap distribution
any variation arising from the arbitrary choice of 
$\tmin$. Since there is no a priori reason to expect
the corresponding systematic uncertainty
to be normally distributed, this procedure can,
in some cases, lead to asymmetric bootstrap distributions,
particularly for energy levels with worse signal-to-noise ratios.
Nevertheless, the difference between the uncertainties obtained by assuming normally distributed energy levels and those estimated from the corresponding quantiles is, in general, not larger than 20\%. We therefore assume normally distributed energy levels in the subsequent finite-volume analysis, as this provides the more conservative estimate of the uncertainty.

\subsection{Bootstrap analysis and prior dependence}
\label{sec:bootstrap_prior_details}

We provide further details on how uncertainties are estimated in the spectrum analysis presented in \Cref{sec:conspiracy}.
First, a standard non-linear regression is performed, using linear error propagation, to extract values of the model parameters from the averaged correlation-function data and our priors given in \Cref{eq:priors}.  We refer to this analysis as \texttt{boot0}.
Subsequently, we use bootstrap resampling to cross check the \texttt{boot0} uncertainty estimate and to provide correlated results that are used in the amplitude analysis.
In Bayesian inference, it is common to use the ``parametric bootstrap''~\cite{efron_parametric_bootstrap} to approximate marginalizing over the priors to obtain the final uncertainty estimate.
For parametric bootstrap, the Gaussian prior distributions are treated as follows: for each bootstrap resampling $i$, a new central value $\mu_i$ is drawn from $\mathcal{N}(\mu_{\texttt{boot0}}, \tilde{\sigma})$, where $\mu_{\texttt{boot0}}$ is the \texttt{boot0} posterior mean and $\tilde\sigma$ is the original prior width. For fitting that bootstrap resampling, the original prior is replaced by the resampled prior $\mathcal{N}(\mu_i, \tilde{\sigma})$ to determine the bootstrap posterior.
The resulting distribution of bootstrap posterior means then provides a more thorough uncertainty estimate for each model parameter.

In our analysis, following this procedure, we observed a large number of outliers in the bootstrap distribution, which we believe are not true statistical outliers.
In the present work, to mitigate this issue, we deviated from the parametric bootstrap in the following way.  First, we used the original prior mean, rather than \texttt{boot0} posterior when drawing new prior means, as we observed that it did not have a significant influence on the posteriors of the ground-state parameters we are interested in.
Second, instead of drawing the new prior means with variance according to the original prior, we used the posterior variance, multiplied by a ``widening factor'', $w=\{1,2,3,4,5\}$, which we observe reduces the outlier results.
This procedure could lead to underestimated uncertainties.
However, if the priors are non-informative (having much larger variance than the posteriors, such that the data is dictating the posterior distribution), then this ad-hoc procedure can be shown to have a negligible effect on the resulting predicted variance.

\begin{figure}
    \includegraphics[width=\textwidth]{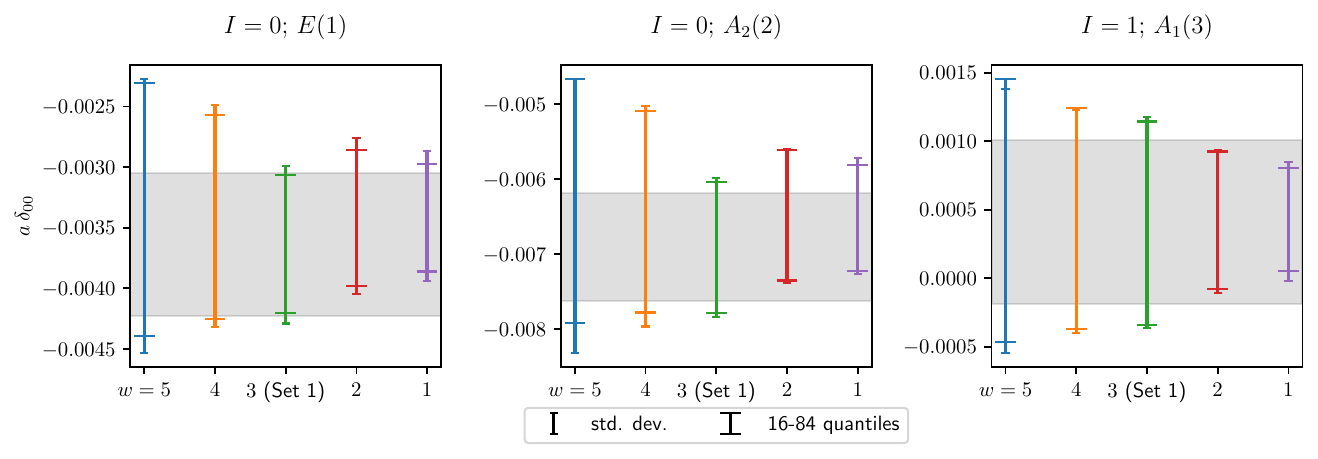}
    \caption{\label{fig:widening_factor_comparison}Effect of varying $w$ on the extracted finite-volume energy shifts $a\,\delta_{00}$ for three irrep levels on the J250 ensemble corresponding to Set 1: the isosinglet $E(1)$ and $A_2(2)$ ground states and the isovector $A_1(3)$ ground state. The gray band represents the \texttt{boot0} result.
    }
\end{figure}

For our case, the ground-state priors are all non-informative and determined by the numerical data, while the excited-state priors lie in-between non-informative and informative, particularly when 3 states are used in the single-nucleon correlator model.
Therefore, we perform a detailed comparison of this ad-hoc resampling strategy and its impact on the final uncertainties we quote here.
In Figure~\ref{fig:widening_factor_comparison}, we show the result of this study on three representative levels on the J250 ensemble, showing the resulting $a\delta_{00}$ parameter, which is the most important quantity in the finite-volume analysis.  While we do observe a trend that the uncertainty estimate increases with larger $w$ factors, indicating that the full uncertainty may not yet be saturated, with our current dataset, it is not clear if this is a meaningful trend or resulting from instabilities due to our relatively small statistical sample size.
Therefore, we also compare the results to that from \texttt{boot0}, which is obtained with linear error propagation.  Most often in LQCD correlator analysis, these uncertainties are consistent with bootstrap uncertainty estimates using the parametric bootstrap.  In future work, when we increase our statistical sample size, and move to make more robust and precise comparisons with experimental results, the estimate of uncertainties will be revisited.
For the purpose of the present work, we believe the consistency of the \texttt{boot0} uncertainty estimate and that from the chosen $w=3$ in Set 1 and $w=1$ in Set 2 is reasonable. We also note that the finite-volume analysis uses linear error propagation, making the use of uncertainties consistent with \texttt{boot0} a compatible approximation.

\subsection{Description of ancillary data}

The results from both analyses, for the D251 and J250 ensembles and for those energy levels used in the amplitude analysis, are provided as ancillary material to the arXiv submission using the \texttt{hdf5} format. All results are stored in a single file, \texttt{energylevels.h5}, that contains a series of hdf5 groups of the form
\begin{equation*}
    \textit{ENS}\texttt{/Set}X\texttt{/NN\_I}Y\texttt{/Pref}P_x P_y P_z\texttt{/}\textit{irrep}\texttt{/level\_}Z\,,
\end{equation*}
where $\textit{ENS}=\texttt{D251},\texttt{J250}$ labels the ensemble, $X=\texttt{1},\texttt{2}$ the analysis strategy, $Y=\texttt{0},\texttt{1}$ denotes the isospin channel, $P_x,P_y,P_z$ are the reference total momentum, in units of $2\pi/L$, \textit{irrep} is the corresponding irrep using the notation from \rcite{Morningstar:2013bda} and $Z$ denotes the energy level, starting from zero. 

Each of these groups contains three datasets: \texttt{E\_N0} and \texttt{E\_N1}, which contain the single-particle ground-state energies from the combined analysis, and \texttt{dE\_lab}, which contains the results for the energy shift. All data are in lattice units. The reference three-momenta squared of the single-particle states associated to the level, $\boldsymbol{p}_1$ and $\boldsymbol{p}_2$, are indicated as an attribute named \texttt{N\_momenta} of the data group, in units of $2\pi/L$. Datasets are one-dimensional with different sizes depending on the analysis strategy and ensemble.  The first entry corresponds to the analysis of the mean, or \texttt{boot0}, and the remaining elements are bootstrap samples.

To ensure reproducibility, we remind that, as indicted in the main text, fits to the spectrum are performed using the total energy in the lattice frame, reconstructed as $E=E_N^\text{cont}(\boldsymbol{p}_1^2)+E_N^\text{cont}(\boldsymbol{p}_2^2)+\delta E_{NN}$, normalized with the nucleon mass, where $E_N^\text{cont}(\boldsymbol{p}^2)=\sqrt{M_N^2+\boldsymbol{p}^2}$. For the nucleon mass, we use the single-particle results from the fit to the isospin-zero data in the rest-frame $T_{1\text{g}}$ irrep, whose key in the file is
\begin{equation*}
\textit{ENS}\texttt{/Set}X\texttt{/NN\_I0}\texttt{/Pref000}\texttt{/}\texttt{T1g}\texttt{/level\_0}\texttt{/E\_N0}\,.
\end{equation*}
Fits are performed correlated, where we consider the covariance matrix of the energy shift between all levels fitted, which we determine from the bootstrap samples. Finally, we indicate that two different valid choices were made for the amplitude analysis. The LHCQC analysis used the \texttt{boot0} results as the mean of the energy levels, while the PWQC analysis used the mean of the bootstrap samples.
The difference between the two choices is related to the bias of the bootstrap analysis, and corresponds to less than 40\% of the error. As such, it is negligible in the final result.

\section{Analysis based on the standard finite-volume formalism}
\label{app:standardFV}

To complement the results presented in the main text, we here present the fit results obtained from using the $S$-wave truncated standard QC. In order to suppress the $S$--$D$ mixing in the isospin-0 (deuteron) channel, helicity averaging is applied, i.e.\ we fit to linear combinations of the spectra belonging to different irreps.
As shown in \rcite{Briceno:2013bda}, the following linear combinations remove the leading dependence on $S$--$D$ mixing,
\begin{align}
    \boldsymbol n = (0, 0, n): E &= \frac{1}{3} E_{A_2} + \frac{2}{3} E_E , \\
    \boldsymbol n = (0, n, n): E &= \frac{1}{3} E_{A_2} + \frac{1}{3} E_{B_1} + \frac{1}{3} E_{B_2} ,
\end{align}
with integer $n$.  In the isospin-1 (dineutron) channel, no such averaging is applied, since we only fit to $A_1$-like irreps.

For isospin 0, we use the two- and three-parameter effective range expansions as defined in \Cref{eq:ERE}, denoted as ERE2 and ERE3, respectively.
For isospin 1, we also perform fits to the three-parameter rational model defined in \Cref{eq:rational3}, which we denote as rational3. The results from the fits are summarized in \Cref{tab:LuscherTablesCombined}. We note that in the case of the C103 ensemble, these correspond to a reanalysis of the spectra defining the $\chi^2$ as in \Cref{eq:chi2}, i.e., from the lab-frame energy shifts, rather than from $q_{\rm cm}^2$ as in \rcite{BaryonScattering:2025ziz}, and using the single-nucleon mass value from the simultaneous fit to the $T_{1g}(0)$ irrep.
Thus, even though the results are consistent with those in \rcite{BaryonScattering:2025ziz}, their numerical values are slightly different.

These results can be directly compared to those obtained using the finite-volume formalisms that include the effects of left-hand cuts.  Although the Lüscher formalism may appear sufficient to describe the lattice results, its omission of one-pion-exchange effects restricts the region in which it can be reliably applied. Even in this region, better descriptions are typically obtained when including the effect of left-hand cuts, as it makes it possible to describe higher partial waves without the need for additional parameters, include energy levels that lie on top of the cut, and describe both channels simultaneously. What is more, the use of physics-informed fit models leads to more reliable extrapolations to the subthreshold region, which are vital to reliably determining the properties of bound and virtual states.

\begin{figure*}
    \centering
    \includegraphics[width=\textwidth]{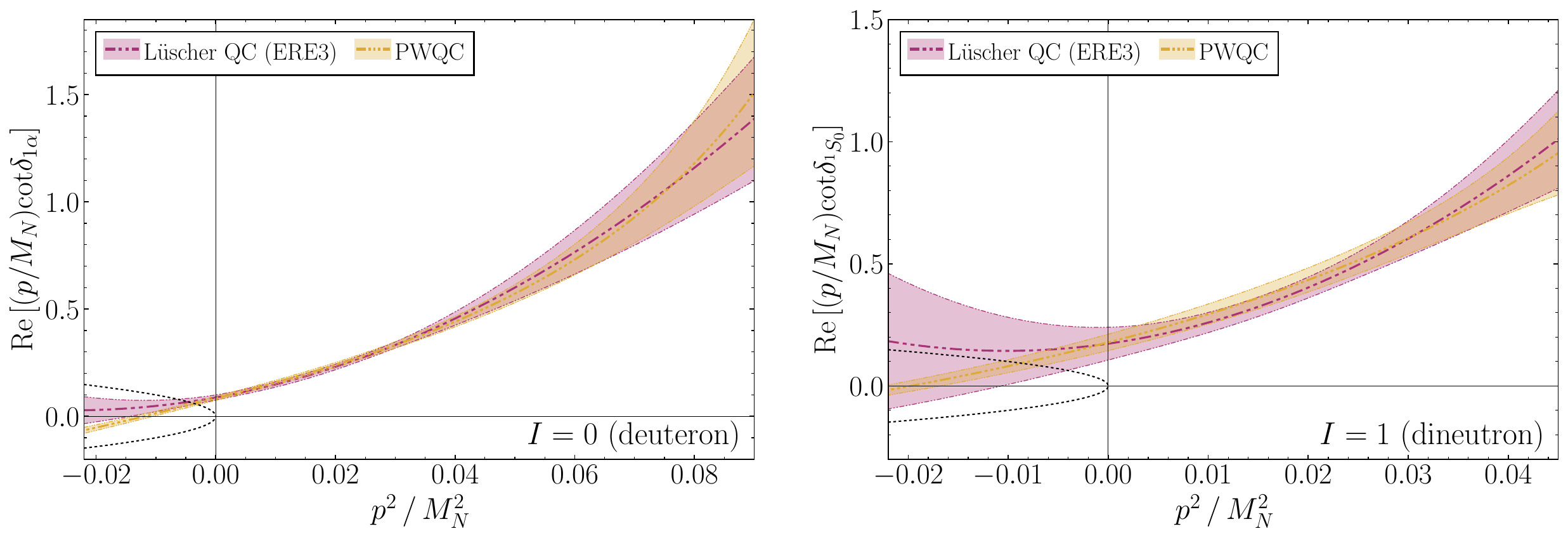}
    \caption{\label{fig:kcotI0I1C103Luscher} Comparison of the results of the lowest partial wave in the C103 ensemble ($M_\pi\approx 710$ MeV) for the $I=0$ ($\alpha$ wave, left) and $I=1$ ($S$ wave, right) channels, for both the PWQC and the standard Lüscher QC. Results from Lüscher's QC correspond to the ERE3 fits.}
\end{figure*}

An example is presented in \Cref{fig:kcotI0I1C103Luscher}, where we compare the results obtained with the standard formalism using ERE3 fits to the averaged results from the PWQC, which are summarized in \Cref{tab:PWQCtableC103,tab:PWQCtableC103scattering_isospin0,tab:PWQCtableC103scattering_isospin1}. In this ensemble, with a pion mass of $M_\pi\approx 710$ MeV, the data can be safely described using Lüscher's formalism, without any risk of it failing due to the presence of a left-hand cut. However, the data can also be described using the PWQC, and good agreement is observed between the two methods in the region where the lattice data lie. However, the PWQC captures higher partial waves without requiring additional parameters and describes both channels simultaneously, thus leading to a lower $\chi^2$ per degree of freedom overall. Moreover, PWQC results lead to a more reliable extrapolation: while mean predictions from the standard formalism do not predict a virtual state in the dineutron channel, the PWQC results necessarily contain one, as required by the presence of long-range interactions.

\begin{table*}[t]
    \centering
    \begin{tabular}{c c c c c c c c c c}
        \toprule
        Ensemble
        & $I$
        & Fit
        & Set
        & $\chi^2/\mathrm{dof}$
        & $1/a_0$ [1/fm]
        & $r_0$ [fm]
        & $P_0$ [fm${}^3$]
        & $(C_0,C_1,C_2)$ [fm, fm${}^3$, fm${}^2$]
        & $\kappa$ [MeV] \\
        \midrule

        \multirow{6}{*}{D251}
        & \multirow{4}{*}{$0$}
        & \multirow{2}{*}{ERE2}
        & 1 & 7.70 / 6
        & $-0.239(59)$ & 2.75(17) & -- & -- & $-37.3(7.4)$ \\
        &
        &
        & 2 & 7.23 / 6
        & $-0.325(97)$ & 3.11(26) & -- & -- & $-47(11)$ \\

        &
        &
        \multirow{2}{*}{ERE3}
        & 1 & 0.84 / 5
        & $-0.200(65)$ & 2.14(28) & 0.26(11) & -- & $-33.5(9.5)$ \\
        &
        &
        & 2 & 2.19 / 5
        & $-0.181(85)$ & 2.19(46) & 0.36(18) & -- & $-31(12)$ \\

        \cmidrule(lr){2-10}

        &
        \multirow{2}{*}{$1$}
        & \multirow{2}{*}{rational3}
        & 1 & 12.13 / 7
        & -- & -- & --
        & $\bigl(0.54(17),-0.259(82),0.86(53)\bigr)$
        & $-148(37)$ \\
        &
        &
        & 2 & 5.98 / 7
        & -- & -- & --
        & $\bigl(0.74(27),-0.30(10),1.16(68)\bigr)$
        & $-124(24)$ \\

        \midrule

        \multirow{6}{*}{J250}
        & \multirow{4}{*}{$0$}
        & \multirow{2}{*}{ERE2}
        & 1 & 21.22 / 6
        & $-0.41(12)$ & 3.14(29) & -- & -- & $-56(12)$ \\
        &
        &
        & 2 & 16.06 / 6
        & $-0.27(10)$ & 2.94(27) & -- & -- & $-41(12)$ \\

        &
        &
        \multirow{2}{*}{ERE3}
        & 1 & 2.63 / 5
        & $-0.12(10)$ & 1.25(40) & 0.89(22) & -- & $-23(18)$ \\
        &
        &
        & 2 & 8.36 / 5
        & $-0.099(99)$ & 1.90(39) & 0.49(18) & -- & $-18(17)$ \\

        \cmidrule(lr){2-10}

        &
        \multirow{2}{*}{$1$}
        & \multirow{2}{*}{rational3}
        & 1 & 3.05 / 7
        & -- & -- & --
        & $\bigl(1.69(61),-0.64(23),1.67(91)\bigr)$
        & $-80(24)$ \\
        &
        &
        & 2 & 3.67 / 7
        & -- & -- & --
        & $\bigl(1.25(66),-0.49(25),0.79(73)\bigr)$
        & $-108(45)$ \\
        \midrule

        \multirow{5}{*}{C103}
        & \multirow{2}{*}{$0$}
        & ERE2
        & \multirow{2}{*}{Ref.~\cite{BaryonScattering:2025ziz}}
        & 12.16 / 7
        & $-0.722(90)$ & 1.96(13) & -- & -- & $-96.4(9.9)$ \\
        &
        &
        ERE3
        &
        & 8.24 / 6
        & $-0.714(96)$ & 1.23(39) & 0.196(96) & -- & $-108(17)$ \\

        \cmidrule(lr){2-10}

        &
        \multirow{3}{*}{$1$}
        & ERE2
        & \multirow{3}{*}{Ref.~\cite{BaryonScattering:2025ziz}}
        & 4.40 / 6
        & $-1.14(34)$ & $3.34(39)$ & -- & -- & $-115(24)$ \\
        &
        & ERE3
        &
        & 1.02 / 5
        & $-1.41(54)$ & $1.4(1.5)$ & $0.53(36)$ & -- & -- \\
        &
        & rational3
        &
        & 1.49 / 5
        & -- & -- & -- & $\bigl(0.75(27), -0.138(54), 0.65(63)\bigr)$ & $-149(55)$ \\
        \bottomrule
    \end{tabular}
    \caption{
        Best-fit results for the C103, D251, and J250 ensembles in the
        isospin-$0$ and isospin-$1$ channels using the standard Lüscher
        quantization condition.
        A dash indicates that the corresponding coefficient is fixed
        to zero or is not applicable to the given parametrization.
        On C103, the ERE3 model predicts a resonance rather than a virtual state in $I = 1$.
    }
    \label{tab:LuscherTablesCombined}
\end{table*}

\section{Summary of individual amplitude fits}
\label{app:tables}

This appendix compiles the results from individual fits to the energy levels, as well as the associated predictions for scattering observables, for both ensembles and both finite-volume formalisms. A comparison of the spectra for different models and cutoff choices is also depicted.

\subsection{Fit results from the left-hand-cut formalisms}

We present the results from each individual fit to the energy levels from both extraction strategies (labeled as Set 1 and Set 2, as discussed in the main text) and both ensembles. Results obtained using the LHCQC formalism are presented in \Cref{tab:RFTlhcTableD251Part1,tab:RFTlhcTableD251Part2} for the D251 ensemble, and in \Cref{tab:RFTlhcTableJ250Part1,tab:RFTlhcTableJ250Part2} for J250.  Note that quantities presented in these tables with a hat correspond to the parameters of the $\hat{\cK}$ matrix. Physical observables obtained from solving the integral equations, \Cref{eq:Mfull,eq:M0_def,eq:ME_int_eq}, are those \emph{without} a hat, and will be presented later.

\begin{figure*}
     \centering
     \includegraphics[width=\textwidth]{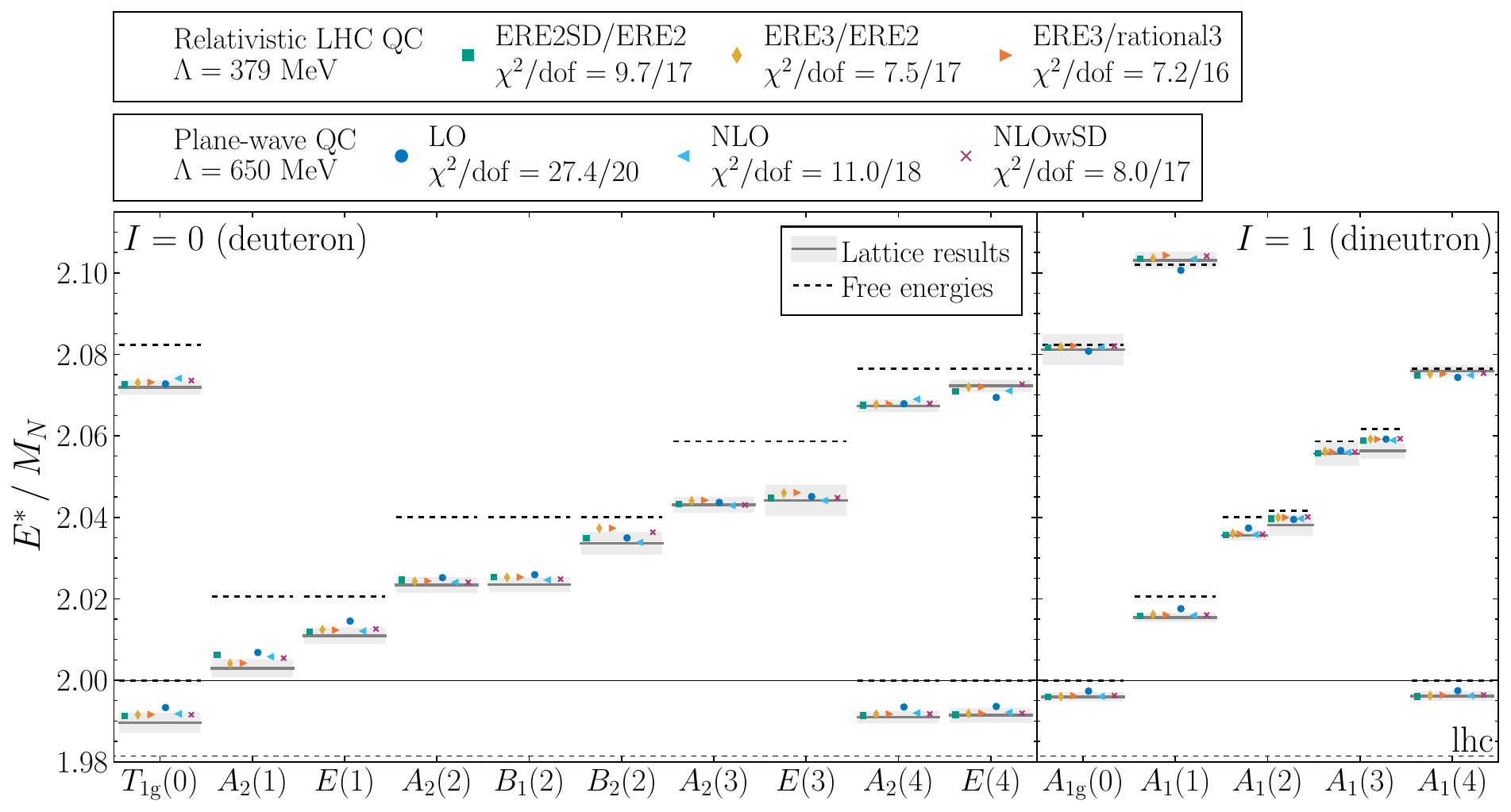}\vspace{0.8cm}
     \includegraphics[width=\textwidth]{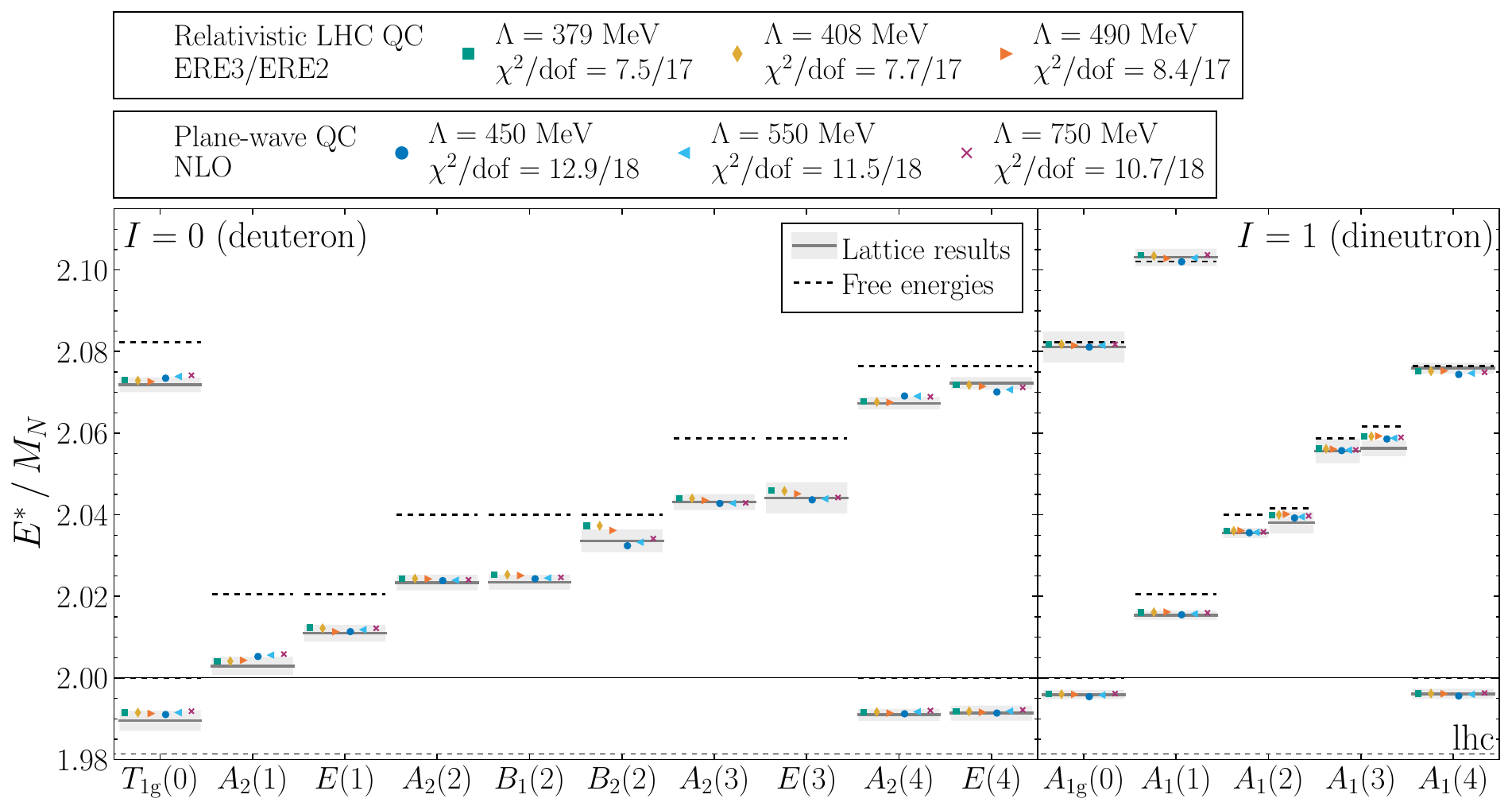}
     
     \caption{\label{fig:D251benergyfit} Same as \Cref{fig:D251energyfit} for the Set 2 energy data of ensemble D251. The best-fit results of each fit are presented in \Cref{tab:RFTlhcTableD251Part1,tab:RFTlhcTableD251Part2,tab:PWQCtableD251}. As in \Cref{fig:D251energyfit}, the predictions from the PWQC fit with the NLOwSD model are presented in the top panel for illustrative purposes; results for this model and cutoff value are not used in the final analysis. The values of the fit parameters are similar to those for $\Lambda=650$ MeV. }

 \end{figure*}

\begin{figure*}
     \centering
     \includegraphics[width=\textwidth]{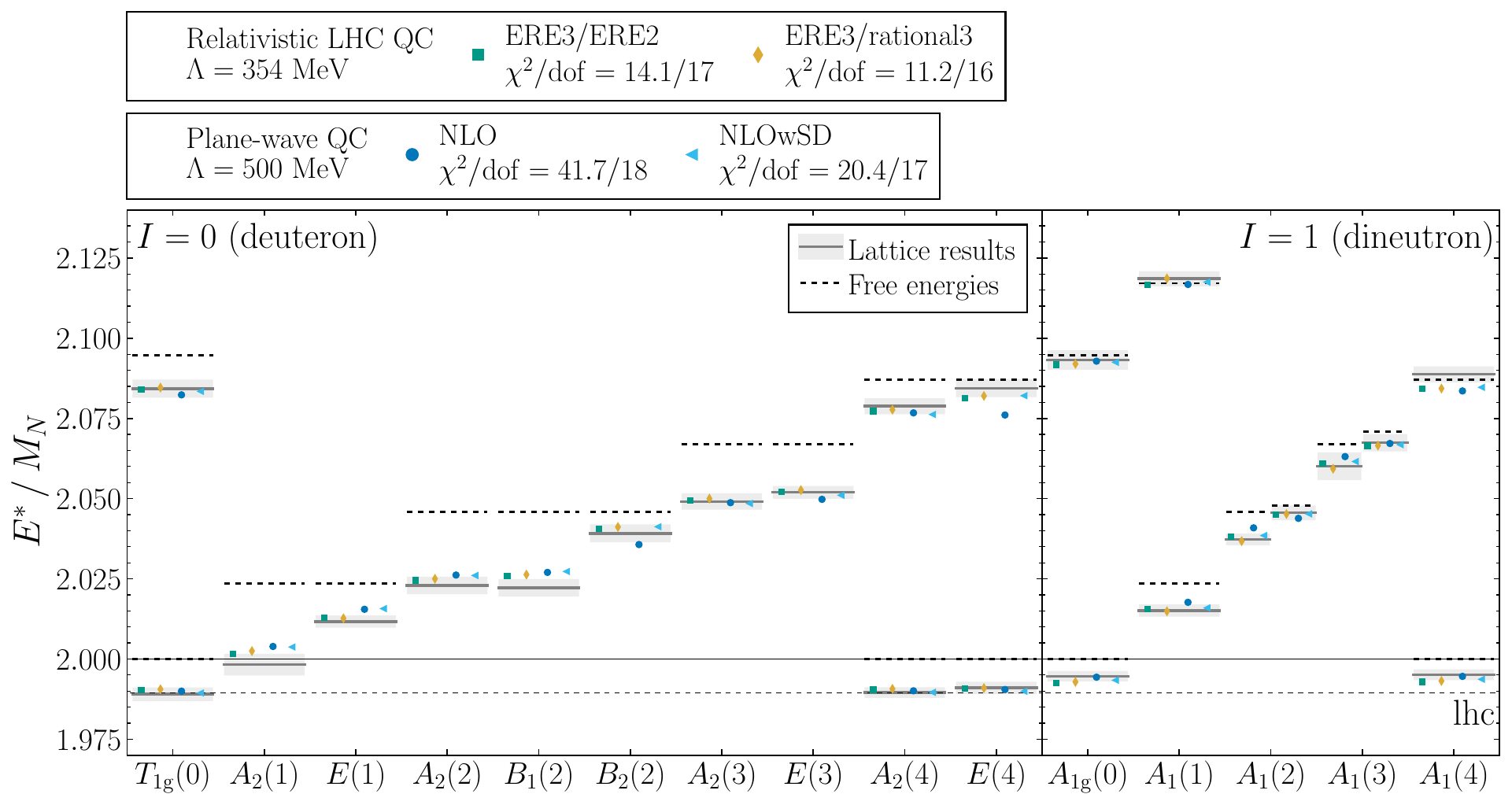}\vspace{0.8cm}
     \includegraphics[width=\textwidth]{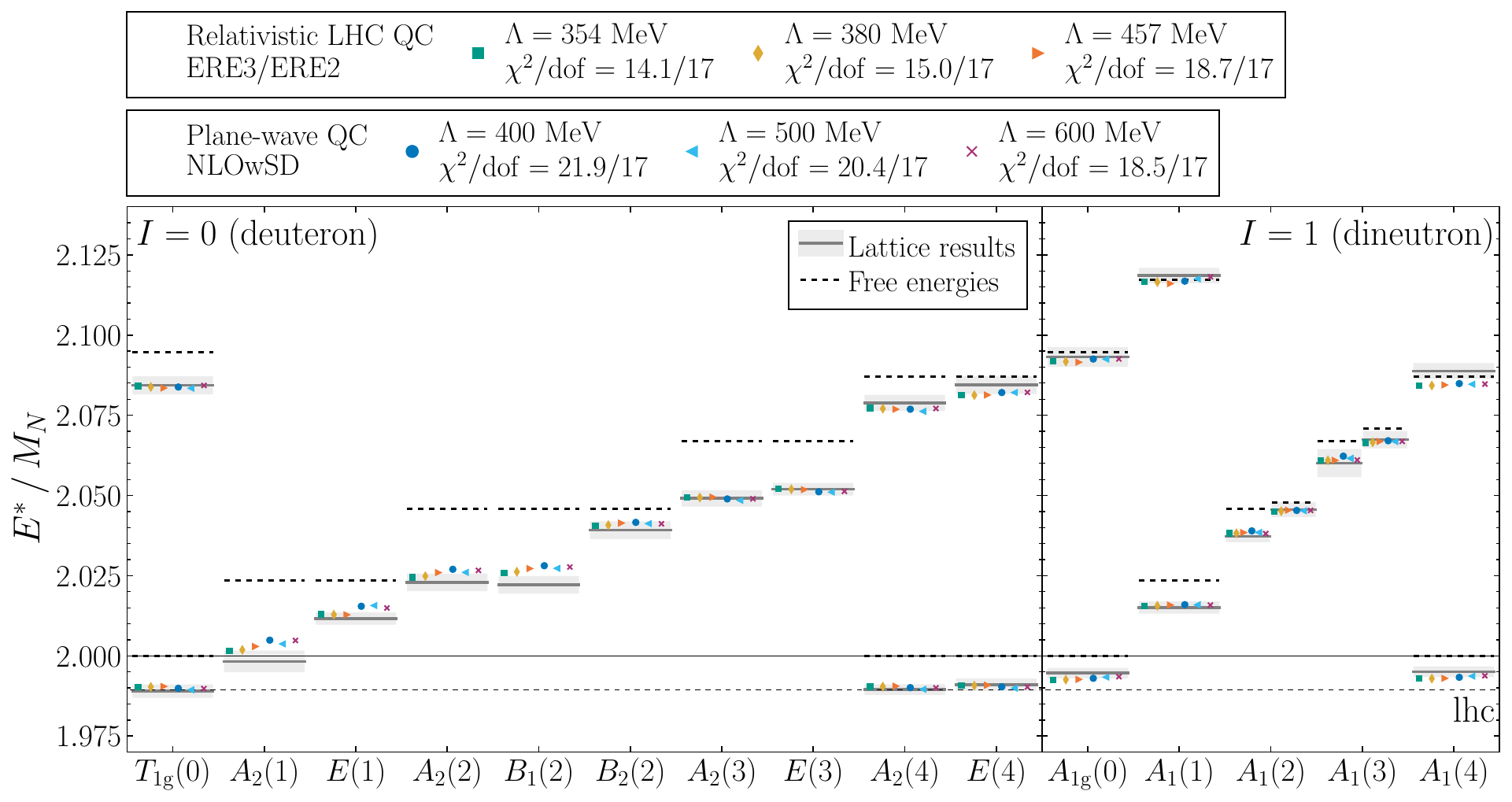}
     
     \caption{\label{fig:J250energyfit} Same as \Cref{fig:D251energyfit} for the Set 1 energy data of the J250 ensemble. The best-fit results of each fit are presented in \Cref{tab:RFTlhcTableJ250Part1,tab:RFTlhcTableJ250Part2,tab:PWQCtableJ250}.}

 \end{figure*}

\begin{figure*}
     \centering
     \includegraphics[width=\textwidth]{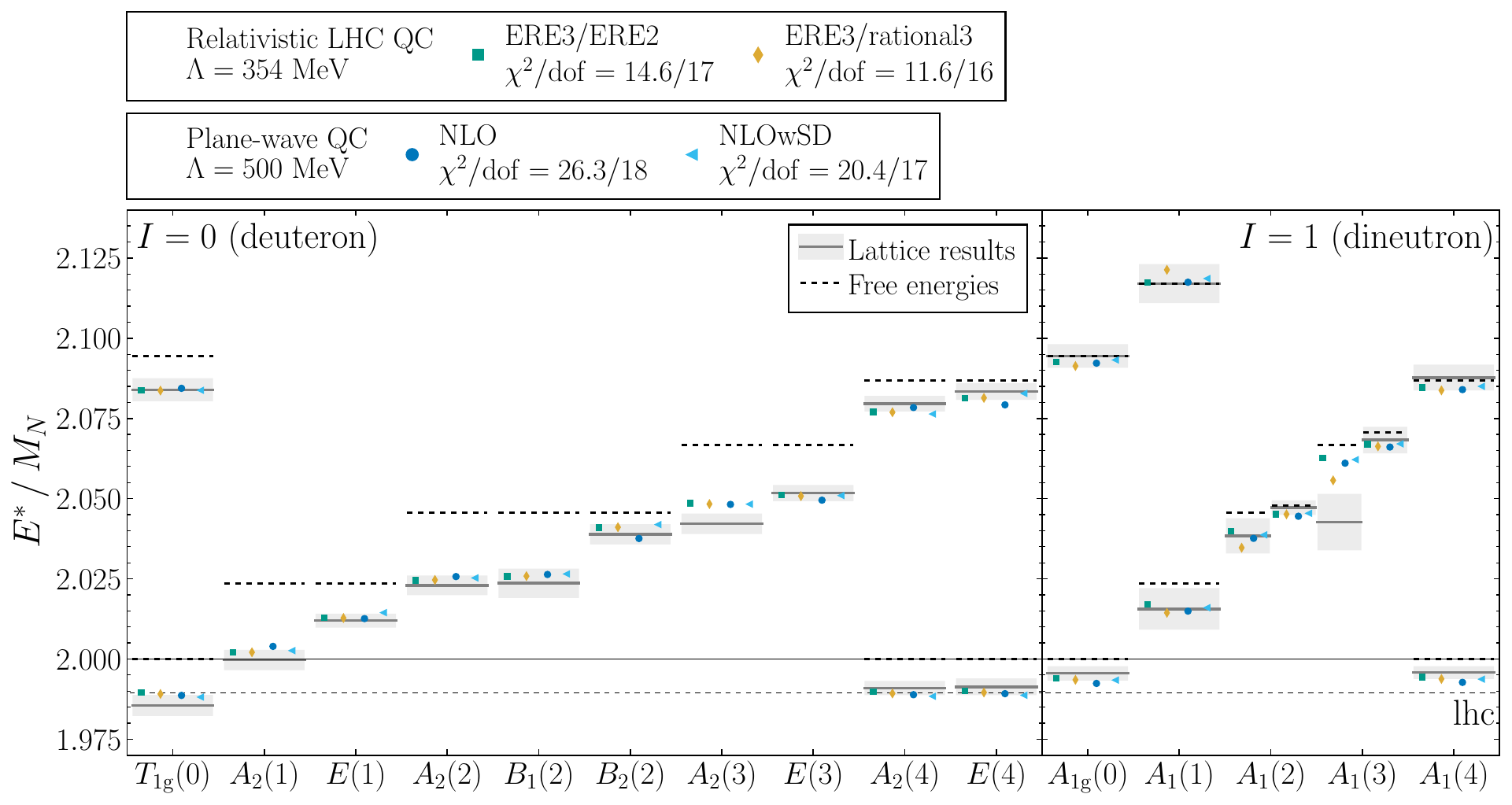}\vspace{0.8cm}
     \includegraphics[width=\textwidth]{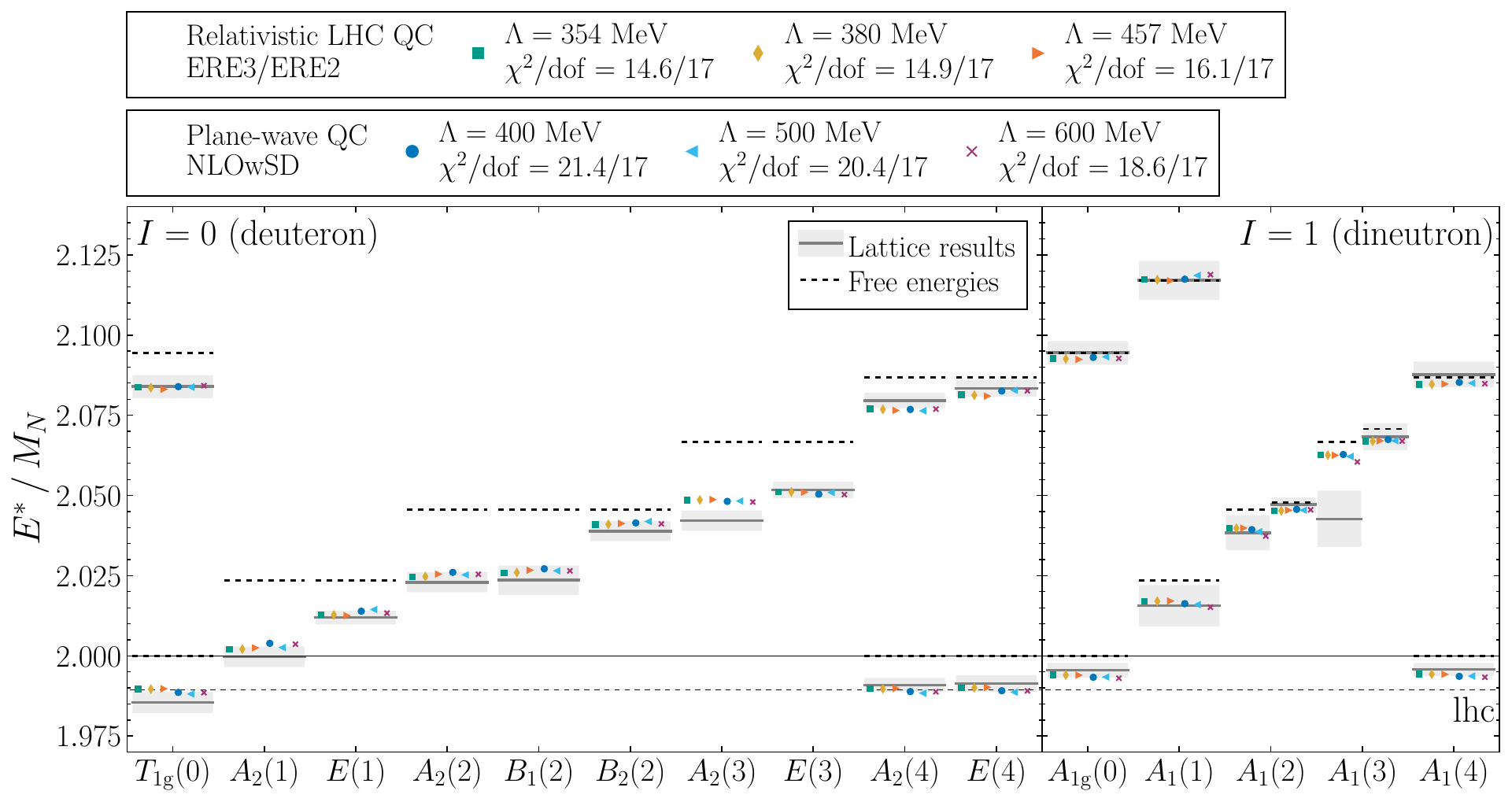}
     
     \caption{\label{fig:J250benergyfit} Same as \Cref{fig:D251energyfit} for the Set 2 energy data of the J250 ensemble. The best-fit results of each fit are presented in \Cref{tab:RFTlhcTableJ250Part1,tab:RFTlhcTableJ250Part2,tab:PWQCtableJ250}.}

 \end{figure*}

Analogous results for the EFT potential parameters obtained using the PWQC approach are presented in \Cref{tab:PWQCtableD251,tab:PWQCtableJ250} for the D251 and J250 ensembles, respectively. We also present in \Cref{tab:PWQCtableC103} the result from reanalyzing the energy levels of the C103 ensemble from~\rcite{BaryonScattering:2025ziz}. Finally, we present graphically the predicted energy levels of a selection of fits in \Cref{fig:D251benergyfit,fig:J250energyfit,fig:J250benergyfit}. These figures depict the dependence of the fit predictions on the finite-volume formalism, the fit model and the value of the cutoff for Set 2 data of D251 and Set 1 and 2 data of J250, respectively, and complement \Cref{fig:D251energyfit} in the main text, which presents analogous results for the fit to Set 1 data in D251.

\begin{table*}[h]
    \centering
    \begin{tabular}{c c c c c c c c c c c c c}
        Fit & $M_N^2 \alpha$ & $\Lambda$\,/\,MeV & Set & $\chi^2$\,/\,dof & $g_{\pi NN}^2$ & $1\,/\,(M_N \hat{a}_{0, {}^3S_1})$ & $M_N \hat{r}_{0, {}^3S_1}$ & $M_N^3 \hat{P}_{0, {}^3S_1}$ & $M_N^2 \hat\theta$ & $M_N \hat{C}_{0, {}^1S_0}$ & $M_N^3 \hat{C}_{1, {}^1S_0}$ & $M_N^2 \hat{C}_{2, {}^1S_0}$ \\
        \toprule
        \multirow{8}{*}{\rotatebox{90}{ERE3/rational3}} & \multirow{2}{*}{7.5} & \multirow{2}{*}{379} & 1 & 10.5/16 & $102(12)$ & $-0.039(16)$ & $6.9(2.5)$ & $83(32)$ & $-$ & $4.6(1.5)$ & $-11(15)$ & $27(21)$ \\
        & & & 2 & 7.2/16 & $122(21)$ & $-0.026(23)$ & $8.9(3.3)$ & $167(114)$ & $-$ & $6.7(2.9)$ & $-13(21)$ & $32(26)$ \\
        \cmidrule{2-13}
        & \multirow{2}{*}{6.5} & \multirow{2}{*}{408} & 1 & 10.7/16 & $100(12)$ & $-0.042(18)$ & $6.8(2.7)$ & $95(38)$ & $-$ & $4.8(1.7)$ & $-13(16)$ & $27(22)$ \\
        & & & 2 & 7.3/16 & $115(20)$ & $-0.027(24)$ & $9.0(3.4)$ & $179(127)$ & $-$ & $7.1(3.1)$ & $-16(21)$ & $34(27)$ \\
        \cmidrule{2-13}
        & \multirow{2}{*}{5.5} & \multirow{2}{*}{443} & 1 & 11.1/16 & $98(12)$ & $-0.047(20)$ & $6.8(3.2)$ & $115(49)$ & $-$ & $5.1(1.9)$ & $-15(17)$ & $28(23)$ \\
        & & & 2 & 7.3/16 & $101(17)$ & $-0.024(25)$ & $9.3(3.0)$ & $153(98)$ & $-$ & $7.5(3.5)$ & $-22(22)$ & $37(30)$ \\
        \cmidrule{2-13}
        & \multirow{2}{*}{4.5} & \multirow{2}{*}{490} & 1 & 11.8/16 & $95(11)$ & $-0.056(24)$ & $6.8(3.8)$ & $150(71)$ & $-$ & $5.5(2.2)$ & $-19(18)$ & $30(25)$ \\
        & & & 2 & 7.4/16 & $96(14)$ & $-0.025(27)$ & $9.4(3.4)$ & $191(122)$ & $-$ & $8.4(4.2)$ & $-29(24)$ & $41(33)$ \\
        \bottomrule
    \end{tabular}
    \caption{Best-fit results to the D251 energies using the LHCQC with the ERE3 model for $I=0$ and the rational3 model for $I = 1$, for a selection of cutoff values. In this formalism, we set the cutoff in terms of $M_N^2 \alpha$. The column labeled $\Lambda$/MeV is an approximate conversion to the convention used by the PWQC. Those coefficients marked with a hyphen are set to zero in that particular fit.
    }
    \label{tab:RFTlhcTableD251Part1}
\end{table*}

\begin{table*}
    \centering
    \begin{tabular}{c c c c c c c c c c c c}
        Fit & $M_N^2 \alpha$ & $\Lambda$\,/\,MeV & Set & $\chi^2$\,/\,dof & $g_{\pi NN}^2$ & $1\,/\,(M_N \hat{a}_{0, {}^3S_1})$ & $M_N \hat{r}_{0, {}^3S_1}$ & $M_N^3 \hat{P}_{0, {}^3S_1}$ & $M_N^2 \hat\theta$ & $1/(M_N \hat{a}_{0, {}^1S_0})$ & $M_N \hat{r}_{0, {}^1S_0}$ \\
        \toprule
        \multirow{8}{*}{\rotatebox{90}{ERE3/ERE2}} & \multirow{2}{*}{7.5} & \multirow{2}{*}{379} & 1 & 11.0/17 & $104(12)$ & $-0.041(17)$ & $6.7(2.6)$ & $86(33)$ & $-$ & $-0.212(68)$ & $15.4(3.4)$ \\
        & & & 2 & 7.5/17 & $125(20)$ & $-0.027(23)$ & $9.1(3.3)$ & $171(118)$ & $-$ & $-0.139(56)$ & $11.5(2.8)$ \\
        \cmidrule{2-12}
        & \multirow{2}{*}{6.5} & \multirow{2}{*}{408} & 1 & 11.3/17 & $102(12)$ & $-0.044(18)$ & $6.7(2.8)$ & $99(40)$ & $-$ & $-0.203(68)$ & $15.4(3.5)$ \\
        & & & 2 & 7.7/17 & $119(19)$ & $-0.028(25)$ & $9.2(3.5)$ & $190(136)$ & $-$ & $-0.130(56)$ & $11.8(2.8)$ \\
        \cmidrule{2-12}
        & \multirow{2}{*}{5.5} & \multirow{2}{*}{443} & 1 & 11.8/17 & $100(11)$ & $-0.050(21)$ & $6.6(3.3)$ & $121(51)$ & $-$ & $-0.191(67)$ & $15.5(3.6)$ \\
        & & & 2 & 8.0/17 & $106(17)$ & $-0.027(26)$ & $9.6(3.2)$ & $168(119)$ & $-$ & $-0.122(55)$ & $12.8(2.8)$ \\
        \cmidrule{2-12}
        & \multirow{2}{*}{4.5} & \multirow{2}{*}{490} & 1 & 12.7/17 & $96(11)$ & $-0.061(25)$ & $6.5(4.1)$ & $161(76)$ & $-$ & $-0.176(67)$ & $15.9(3.7)$ \\
        & & & 2 & 8.4/17 & $99(14)$ & $-0.027(27)$ & $9.9(3.4)$ & $196(132)$ & $-$ & $-0.109(55)$ & $13.2(2.7)$ \\
        \midrule
        \multirow{8}{*}{\rotatebox{90}{ERE2SD/ERE2}} & \multirow{2}{*}{7.5} & \multirow{2}{*}{379} & 1 & 11.0/17 & $171(40)$ & $0.066(65)$ & $5.0(3.8)$ & $-$ & $5.1(2.3)$ & $-0.159(56)$ & $7.9(2.8)$ \\
        & & & 2 & 9.7/17 & $162(32)$ & $0.076(50)$ & $5.9(2.9)$ & $-$ & $5.3(1.7)$ & $-0.113(52)$ & $8.4(2.2)$ \\
        \cmidrule{2-12}
        & \multirow{2}{*}{6.5} & \multirow{2}{*}{408} & 1 & 12.6/17 & $158(37)$ & $0.083(78)$ & $5.1(3.6)$ & $-$ & $4.8(2.2)$ & $-0.152(56)$ & $8.6(3.2)$ \\
        & & & 2 & 10.5/17 & $149(33)$ & $0.091(64)$ & $5.8(3.4)$ & $-$ & $4.9(1.8)$ & $-0.108(54)$ & $9.0(2.7)$ \\
        \cmidrule{2-12}
        & \multirow{2}{*}{5.5} & \multirow{2}{*}{443} & 1 & 15.1/17 & $143(33)$ & $0.104(90)$ & $5.2(4.0)$ & $-$ & $4.4(2.0)$ & $-0.144(58)$ & $9.6(3.5)$ \\
        & & & 2 & 11.9/17 & $134(34)$ & $0.109(83)$ & $5.8(4.3)$ & $-$ & $4.5(1.9)$ & $-0.101(56)$ & $9.8(3.1)$ \\
        \cmidrule{2-12}
        & \multirow{2}{*}{4.5} & \multirow{2}{*}{490} & 1 & 19.0/17 & $126(33)$ & $0.13(13)$ & $5.4(5.6)$ & $-$ & $3.9(2.2)$ & $-0.136(61)$ & $11.1(4.3)$ \\
        & & & 2 & 14.1/17 & $119(32)$ & $0.13(11)$ & $5.8(5.3)$ & $-$ & $3.9(1.9)$ & $-0.094(59)$ & $10.8(3.6)$ \\
        \bottomrule
    \end{tabular}
    \caption{Best-fit results to the D251 energies using the LHCQC with the ERE2SD or ERE3 model for $I=0$ and the ERE2 model for $I = 1$, for a selection of cutoff values. In this formalism, we set the cutoff in terms of $M_N^2 \alpha$. The column labeled $\Lambda$/MeV is an approximate conversion to the convention used by the PWQC. Those coefficients marked with a hyphen are set to zero in that particular fit.}
    \label{tab:RFTlhcTableD251Part2}
\end{table*}

\begin{table*}[h]
    \centering
    \begin{tabular}{c c c c c c c c c c c c c}
        Fit & $M_N^2 \alpha$ & $\Lambda$\,/\,MeV & Set & $\chi^2$\,/\,dof & $g_{\pi NN}^2$ & $1\,/\,(M_N \hat{a}_{0, {}^3S_1})$ & $M_N \hat{r}_{0, {}^3S_1}$ & $M_N^3 \hat{P}_{0, {}^3S_1}$ & $M_N^2 \hat\theta$ & $M_N \hat{C}_{0, {}^1S_0}$ & $M_N^3 \hat{C}_{1, {}^1S_0}$ & $M_N^2 \hat{C}_{2, {}^1S_0}$ \\
        \toprule
        \multirow{8}{*}{\rotatebox{90}{ERE3/rational3}} & \multirow{2}{*}{7.5} & \multirow{2}{*}{354} & 1 & 11.2/16 & $112(27)$ & $-0.017(21)$ & $3.8(2.9)$ & $161(90)$ & $-$ & $23(15)$ & $-63(40)$ & $87(63)$ \\
        & & & 2 & 11.6/16 & $110(29)$ & $-0.007(20)$ & $7.4(2.4)$ & $94(61)$ & $-$ & $19(17)$ & $-96(81)$ & $45(49)$ \\
        \cmidrule{2-13}
        & \multirow{2}{*}{6.5} & \multirow{2}{*}{380} & 1 & 11.8/16 & $105(26)$ & $-0.020(22)$ & $3.9(3.0)$ & $169(102)$ & $-$ & $27(21)$ & $-81(55)$ & $101(84)$ \\
        & & & 2 & 11.8/16 & $107(28)$ & $-0.008(21)$ & $7.6(2.6)$ & $105(77)$ & $-$ & $23(24)$ & $-120(119)$ & $52(65)$ \\
        \cmidrule{2-13}
        & \multirow{2}{*}{5.5} & \multirow{2}{*}{413} & 1 & 12.7/16 & $96(24)$ & $-0.022(24)$ & $4.0(3.2)$ & $173(113)$ & $-$ & $33(32)$ & $-112(91)$ & $122(123)$ \\
        & & & 2 & 12.1/16 & $103(27)$ & $-0.012(23)$ & $7.9(2.9)$ & $122(102)$ & $-$ & $30(41)$ & $-166(214)$ & $67(103)$ \\
        \cmidrule{2-13}
        & \multirow{2}{*}{4.5} & \multirow{2}{*}{457} & 1 & 13.9/16 & $81(18)$ & $-0.020(26)$ & $4.2(2.9)$ & $159(82)$ & $-$ & $39(41)$ & $-160(153)$ & $144(164)$ \\
        & & & 2 & 12.5/16 & $95(25)$ & $-0.016(27)$ & $8.4(3.2)$ & $140(135)$ & $-$ & $49(102)$ & $-285(581)$ & $107(242)$ \\
        \bottomrule
    \end{tabular}
    \caption{Same as \Cref{tab:RFTlhcTableD251Part1} for the J250 ensemble.}
    \label{tab:RFTlhcTableJ250Part1}
\end{table*}

\begin{table*}
    \centering
    \begin{tabular}{c c c c c c c c c c c c}
        Fit & $M_N^2 \alpha$ & $\Lambda$\,/\,MeV & Set & $\chi^2$\,/\,dof & $g_{\pi NN}^2$ & $1\,/\,(M_N \hat{a}_{0, {}^3S_1})$ & $M_N \hat{r}_{0, {}^3S_1}$ & $M_N^3 \hat{P}_{0, {}^3S_1}$ & $M_N^2 \hat\theta$ & $1/(M_N \hat{a}_{0, {}^1S_0})$ & $M_N \hat{r}_{0, {}^1S_0}$ \\
        \toprule
        \multirow{8}{*}{\rotatebox{90}{ERE3/ERE2}} & \multirow{2}{*}{7.5} & \multirow{2}{*}{354} & 1 & 14.1/17 & $126(25)$ & $-0.019(22)$ & $3.7(3.2)$ & $189(115)$ & $-$ & $-0.039(25)$ & $9.2(1.7)$ \\
        & & & 2 & 14.6/17 & $114(29)$ & $-0.012(22)$ & $6.4(2.9)$ & $114(74)$ & $-$ & $-0.084(55)$ & $9.9(3.2)$ \\
        \cmidrule{2-12}
        & \multirow{2}{*}{6.5} & \multirow{2}{*}{380} & 1 & 15.0/17 & $119(23)$ & $-0.023(24)$ & $3.8(3.5)$ & $205(135)$ & $-$ & $-0.033(25)$ & $9.2(1.7)$ \\
        & & & 2 & 14.9/17 & $111(28)$ & $-0.014(24)$ & $6.4(3.2)$ & $131(96)$ & $-$ & $-0.074(55)$ & $9.7(3.2)$ \\
        \cmidrule{2-12}
        & \multirow{2}{*}{5.5} & \multirow{2}{*}{413} & 1 & 16.4/17 & $111(21)$ & $-0.029(26)$ & $4.1(3.7)$ & $223(157)$ & $-$ & $-0.026(26)$ & $9.3(1.7)$ \\
        & & & 2 & 15.3/17 & $107(26)$ & $-0.019(26)$ & $6.6(3.6)$ & $155(131)$ & $-$ & $-0.063(54)$ & $9.7(3.2)$ \\
        \cmidrule{2-12}
        & \multirow{2}{*}{4.5} & \multirow{2}{*}{457} & 1 & 18.7/17 & $100(18)$ & $-0.037(30)$ & $4.5(4.0)$ & $239(181)$ & $-$ & $-0.018(26)$ & $9.6(1.8)$ \\
        & & & 2 & 16.1/17 & $99(23)$ & $-0.025(31)$ & $6.9(4.2)$ & $188(181)$ & $-$ & $-0.051(53)$ & $9.9(3.3)$ \\
        \midrule
        \multirow{8}{*}{\rotatebox{90}{ERE2SD/ERE2}} & \multirow{2}{*}{7.5} & \multirow{2}{*}{354} & 1 & 23.7/17 & $159(24)$ & $0.083(31)$ & $3.8(1.3)$ & $-$ & $5.53(72)$ & $-0.024(24)$ & $7.7(1.1)$ \\
        & & & 2 & 19.9/17 & $144(46)$ & $0.060(55)$ & $5.8(3.1)$ & $-$ & $4.3(1.8)$ & $-0.050(47)$ & $8.2(2.8)$ \\
        \cmidrule{2-12}
        & \multirow{2}{*}{6.5} & \multirow{2}{*}{380} & 1 & 24.4/17 & $147(24)$ & $0.100(39)$ & $3.8(1.6)$ & $-$ & $5.13(83)$ & $-0.019(24)$ & $7.8(1.3)$ \\
        & & & 2 & 20.2/17 & $139(48)$ & $0.083(74)$ & $5.3(3.2)$ & $-$ & $4.3(1.9)$ & $-0.040(49)$ & $8.0(3.0)$ \\
        \cmidrule{2-12}
        & \multirow{2}{*}{5.5} & \multirow{2}{*}{413} & 1 & 26.0/17 & $134(20)$ & $0.120(44)$ & $3.8(1.7)$ & $-$ & $4.66(79)$& $-0.013(25)$ & $8.0(1.3)$ \\
        & & & 2 & 21.0/17 & $131(48)$ & $0.108(99)$ & $4.9(3.8)$ & $-$ & $4.1(2.0)$ & $-0.031(52)$ & $8.0(3.2)$ \\
        \cmidrule{2-12}
        & \multirow{2}{*}{4.5} & \multirow{2}{*}{457} & 1 & 28.8/17 & $119(19)$ & $0.145(58)$ & $3.9(2.2)$ & $-$ & $4.13(87)$ & $-0.006(26)$ & $8.4(1.4)$ \\
        & & & 2 & 22.6/17 & $118(44)$ & $0.14(13)$ & $4.8(4.8)$ & $-$ & $3.8(2.1)$ & $-0.023(55)$ & $8.3(3.5)$ \\
        \bottomrule
    \end{tabular}
    \caption{Same as \Cref{tab:RFTlhcTableD251Part2} for the J250 ensemble.}
    \label{tab:RFTlhcTableJ250Part2}
\end{table*}

\begin{table*}
  \centering
  \tabcolsep=0.15cm
      \begin{tabular}{c c c c c c c c c c}
    \toprule
     Fit & $\Lambda\,/\,\mathrm{MeV}$ & Set & $\chi^2\,/\,\mathrm{dof}$ & $g_{\pi NN}^2$ & $\tilde{C}_{{}^3S_1}\,/\,10^3$ &  $\tilde{C}_{{}^1S_0}\,/\,10^3$ & $C_{{}^3S_1}\,/\, \text{MeV}^{-2}$ & $C_{{}^3 S_1-{}^3D_1}\,/\, \text{MeV}^{-2}$ &  $C_{{}^1S_0}\,/\, \text{MeV}^{-2}$\\
    \midrule
    \multirow{8}{*}{\rotatebox{90}{LO}} & \multirow{2}{*}{450} & 1 & 27.3/20  & $290(39)$  & $-0.83(33)$  & $0.28(23)$  & $-$ & $-$ & $-$\\
 &  & 2 & 25.2/20  & $281(47)$  & $-0.91(40)$  & $0.08(25)$  & $-$ & $-$ & $-$\\\cmidrule{2-10}
 & \multirow{2}{*}{550} & 1 & 25.4/20  & $247(25)$  & $0.17(35)$  & $0.45(18)$  & $-$ & $-$ & $-$\\
 &  & 2 & 27.3/20  & $231(32)$  & $0.09(44)$  & $0.29(19)$  & $-$ & $-$ & $-$\\\cmidrule{2-10}
 & \multirow{2}{*}{650} & 1 & 24.0/20  & $223(20)$  & $1.10(46)$  & $0.55(16)$  & $-$ & $-$ & $-$\\
 &  & 2 & 27.4/20  & $205(26)$  & $0.90(55)$  & $0.39(17)$  & $-$ & $-$ & $-$\\\cmidrule{2-10}
 & \multirow{2}{*}{750} & 1 & 22.6/20  & $208(18)$  & $2.09(66)$  & $0.61(15)$  & $-$ & $-$ & $-$\\
 &  & 2 & 26.9/20  & $190(23)$  & $1.73(76)$  & $0.45(16)$  & $-$ & $-$ & $-$\\\midrule
\multirow{8}{*}{\rotatebox{90}{NLO}} & \multirow{2}{*}{450} & 1 & 15.1/18  & $279(48)$  & $-1.16(55)$  & $-0.46(34)$  & $0.0014(20)$ & $-$ & $0.0065(21)$\\
 &  & 2 & 12.9/18  & $237(46)$  & $-2.03(59)$  & $-0.90(38)$  & $0.0055(26)$ & $-$ & $0.0067(23)$\\\cmidrule{2-10}
 & \multirow{2}{*}{550} & 1 & 11.2/18  & $218(35)$  & $-0.67(62)$  & $-0.25(30)$  & $0.0036(20)$ & $-$ & $0.0057(20)$\\
 &  & 2 & 11.5/18  & $185(34)$  & $-1.46(62)$  & $-0.66(33)$  & $0.0076(28)$ & $-$ & $0.0065(22)$\\\cmidrule{2-10}
 & \multirow{2}{*}{650} & 1 & 9.8/18  & $189(30)$  & $0.01(73)$  & $0.00(25)$  & $0.0049(24)$ & $-$ & $0.0056(22)$\\
 &  & 2 & 11.0/18  & $161(29)$  & $-0.56(72)$  & $-0.37(26)$  & $0.0098(38)$ & $-$ & $0.0067(25)$\\\cmidrule{2-10}
 & \multirow{2}{*}{750} & 1 & 9.2/18  & $172(26)$  & $1.3(1.1)$  & $0.44(38)$  & $0.0068(36)$ & $-$ & $0.0061(28)$\\
 &  & 2 & 10.7/18  & $148(26)$  & $3.0(4.0)$  & $0.25(54)$  & $0.0152(87)$ & $-$ & $0.0076(35)$\\\midrule
\multirow{6}{*}{\rotatebox{90}{NLOwSD}} & \multirow{2}{*}{400} & 1 & 10.1/17  & $215(99)$  & $-1.77(70)$  & $-0.94(45)$  & $0.0050(43)$ & $0.0143(60)$ & $0.0083(24)$\\
 &  & 2 & 8.7/17  & $140(100)$  & $-2.67(77)$  & $-1.47(51)$  & $0.0116(64)$ & $0.0176(83)$ & $0.0085(26)$\\\cmidrule{2-10}
 & \multirow{2}{*}{500} & 1 & 9.0/17  & $203(43)$  & $-0.2(1.0)$  & $-0.53(32)$  & $0.0090(53)$ & $0.0114(63)$ & $0.0068(20)$\\
 &  & 2 & 9.0/17  & $141(63)$  & $1.3(3.8)$  & $-1.04(44)$  & $0.027(19)$ & $0.024(14)$ & $0.0079(25)$\\\cmidrule{2-10}
 & \multirow{2}{*}{600} & 1 & 7.9/17  & $161(56)$  & $0.6(1.3)$  & $-0.34(37)$  & $0.0105(76)$ & $0.0076(68)$ & $0.0065(23)$\\
 &  & 2 & 7.9/17  & $108(61)$  & $4(10)$  & $-0.85(42)$  & $0.029(26)$ & $0.017(16)$ & $0.0083(28)$\\
    \bottomrule
    \end{tabular}
    
  \caption{Best-fit results to the D251 energies using the PWQC, for a selection of cutoff values. Those coefficients marked with a hyphen are set to zero in that particular fit.}
  \label{tab:PWQCtableD251}
\end{table*}

\begin{table*}
  \centering
  \tabcolsep=0.15cm
      \begin{tabular}{c c c c c c c c c c}
    \toprule
     Fit & $\Lambda\,/\,\mathrm{MeV}$ & Set & $\chi^2\,/\,\mathrm{dof}$ & $g_{\pi NN}^2$ & $\tilde{C}_{{}^3S_1}\,/\,10^3$ &  $\tilde{C}_{{}^1S_0}\,/\,10^3$ & $C_{{}^3S_1}\,/\, \text{MeV}^{-2}$ & $C_{{}^3 S_1-{}^3D_1}\,/\, \text{MeV}^{-2}$ &  $C_{{}^1S_0}\,/\, \text{MeV}^{-2}$\\
    \midrule
    \multirow{10}{*}{\rotatebox{90}{NLO}} & \multirow{2}{*}{300} & 1 & 47.1/18  & $316(77)$  & $-1.57(60)$  & $-1.49(52)$  & $-0.0139(30)$ & $-$ & $0.0062(44)$\\
 &  & 2 & 40.2/18  & $340(100)$  & $-1.89(77)$  & $-1.02(78)$  & $-0.0141(39)$ & $-$ & $0.0097(62)$\\\cmidrule{2-10}
 & \multirow{2}{*}{400} & 1 & 32.7/18  & $196(45)$  & $-1.61(62)$  & $-1.66(43)$  & $0.0011(27)$ & $-$ & $0.0100(28)$\\
 &  & 2 & 31.4/18  & $202(54)$  & $-2.06(74)$  & $-1.52(67)$  & $0.0031(33)$ & $-$ & $0.0104(45)$\\\cmidrule{2-10}
 & \multirow{2}{*}{500} & 1 & 41.7/18  & $220(25)$  & $0.34(60)$  & $-0.34(38)$  & $-0.0022(24)$ & $-$ & $0.0046(24)$\\
 &  & 2 & 26.3/18  & $156(38)$  & $-1.79(74)$  & $-1.62(60)$  & $0.0074(33)$ & $-$ & $0.0108(43)$\\\cmidrule{2-10}
 & \multirow{2}{*}{600} & 1 & 22.9/18  & $136(29)$  & $-0.80(74)$  & $-1.13(28)$  & $0.0071(33)$ & $-$ & $0.0101(29)$\\
 &  & 2 & 23.7/18  & $135(31)$  & $-1.19(77)$  & $-1.31(49)$  & $0.0102(43)$ & $-$ & $0.0121(53)$\\\cmidrule{2-10}
 & \multirow{2}{*}{700} & 1 & 21.6/18  & $126(27)$  & $0.48(96)$  & $-0.20(71)$  & $0.0099(52)$ & $-$ & $0.0120(44)$\\
 &  & 2 & 22.4/18  & $124(28)$  & $1.2(2.4)$  & $0.3(2.3)$  & $0.0153(84)$ & $-$ & $0.0156(93)$\\\midrule
\multirow{8}{*}{\rotatebox{90}{NLOwSD}} & \multirow{2}{*}{300} & 1 & 26.3/17  & $116(99)$  & $-2.41(72)$  & $-2.04(52)$  & $-0.0048(48)$ & $0.0341(80)$ & $0.0049(46)$\\
 &  & 2 & 22.8/17  & $10(130)$  & $-3.60(92)$  & $-2.30(76)$  & $0.0010(71)$ & $0.041(11)$ & $0.0032(82)$\\\cmidrule{2-10}
 & \multirow{2}{*}{400} & 1 & 21.9/17  & $114(52)$  & $-1.81(71)$  & $-1.96(44)$  & $0.0103(54)$ & $0.0193(71)$ & $0.0103(28)$\\
 &  & 2 & 21.4/17  & $85(82)$  & $-2.66(97)$  & $-2.05(75)$  & $0.0157(80)$ & $0.0215(86)$ & $0.0113(47)$\\\cmidrule{2-10}
 & \multirow{2}{*}{500} & 1 & 20.4/17  & $114(37)$  & $1.3(2.8)$  & $-1.59(37)$  & $0.022(12)$ & $0.022(11)$ & $0.0097(25)$\\
 &  & 2 & 20.4/17  & $97(45)$  & $2.1(5.5)$  & $-1.72(57)$  & $0.034(23)$ & $0.027(16)$ & $0.0113(43)$\\\cmidrule{2-10}
 & \multirow{2}{*}{600} & 1 & 18.5/17  & $104(33)$  & $5(10)$  & $-1.30(30)$  & $0.024(21)$ & $0.016(14)$ & $0.0106(30)$\\
 &  & 2 & 18.6/17  & $86(50)$  & $12(30)$  & $-1.48(54)$  & $0.042(47)$ & $0.023(25)$ & $0.0135(58)$\\
    \bottomrule
    \end{tabular}
    
  \caption{Same as \Cref{tab:PWQCtableD251} for the J250 ensemble.}
  \label{tab:PWQCtableJ250}
\end{table*}

\begin{table*}
  \centering
  \tabcolsep=0.15cm
      \begin{tabular}{c c c c c c c c c}
    \toprule
     Fit & $\Lambda\,/\,\mathrm{MeV}$ & $\chi^2\,/\,\mathrm{dof}$ & $g_{\pi NN}^2$ & $\tilde{C}_{{}^3S_1}/10^3$ &  $\tilde{C}_{{}^1S_0}/10^3$ & $C_{{}^3S_1}/ \text{MeV}^{-2}$ &  $C_{{}^1S_0}/ \text{MeV}^{-2}$\\
    \midrule
    \multirow{4}{*}{\rotatebox{90}{LO}} & \multirow{1}{*}{650}  & 15.2/20  & $1.22(17) \times 10^{3}$  & $0.10(52)$  & $-0.16(22)$  & $-$ & $-$\\
 & \multirow{1}{*}{750}  & 14.3/20  & $780(110)$  & $0.61(58)$  & $0.00(21)$  & $-$ & $-$\\
 & \multirow{1}{*}{850}  & 15.5/20  & $656(80)$  & $1.83(80)$  & $0.31(21)$  & $-$ & $-$\\
 & \multirow{1}{*}{950}  & 16.1/20  & $573(67)$  & $3.2(1.2)$  & $0.55(22)$  & $-$ & $-$\\\midrule
\multirow{4}{*}{\rotatebox{90}{NLO}} & \multirow{1}{*}{650}  & 15.3/18  & $1.16(32) \times 10^{3}$  & $-0.0(1.2)$  & $-0.00(60)$  & $-0.0005(17)$ & $-0.0008(18)$\\
 & \multirow{1}{*}{750}  & 14.5/18  & $700(210)$  & $-0.0(1.3)$  & $-0.00(60)$  & $0.0011(16)$ & $-0.0002(17)$\\
 & \multirow{1}{*}{850}  & 10.6/18  & $460(150)$  & $-0.1(1.3)$  & $-0.57(51)$  & $0.0023(16)$ & $0.0031(21)$\\
 & \multirow{1}{*}{950}  & 10.0/18  & $290(110)$  & $-0.5(1.2)$  & $-0.53(44)$  & $0.0040(16)$ & $0.0044(27)$\\
    \bottomrule
    \end{tabular}
    
  \caption{Same as \Cref{tab:PWQCtableD251} for the C103 ensemble.}
  \label{tab:PWQCtableC103}
\end{table*}

\subsection{Scattering observables}
\label{app:tablesscattering}

In this section, we collect the results for the scattering observables obtained after solving the relevant integral equation with the parameters presented before. Results from the LHCQC formalism are presented in \Cref{tab:LHCQCtableD251scattering_isospin0,tab:LHCQCtableD251scattering_isospin1} for the D251 ensemble and in 
\Cref{tab:LHCQCtableJ250scattering_isospin0,tab:LHCQCtableJ250scattering_isospin1} for the J250 ensemble, while results from the PWQC can be found in \Cref{tab:PWQCtableD251scattering_isospin0,tab:PWQCtableD251scattering_isospin1,tab:PWQCtableJ250scattering_isospin0,tab:PWQCtableJ250scattering_isospin1}. We also present results for C103 obtained using the PWQC in \Cref{tab:PWQCtableC103scattering_isospin0,tab:PWQCtableC103scattering_isospin1}. In all cases, the last row of these tables represents the average result for both energy-determination strategies, and includes both a statistical and a systematic error defined following \Cref{eq:statavg,eq:systavg}, where each dataset is assigned an equal 1/2 weight for this average.

\begin{table*}
  \centering
  \tabcolsep=0.15cm
      \begin{tabular}{c c c c c c c c c c}
    \toprule
     Fit & $M_N^2\alpha$ & $\Lambda\,/\,{\rm MeV}$ & Set & $\chi^2\,/\,\mathrm{dof}$ & $a_0^{-1}$ / fm${}^{-1}$ & $a_0$ / fm &  $r_0$ / fm & $P_0$ / fm${}^3$ & $\kappa$ / MeV \\
    \midrule
    \multirow{8}{*}{\rotatebox{90}{ERE2SD/ERE2}} & \multirow{2}{*}{7.5} & \multirow{2}{*}{379}  & 1 & 11.0/17  & $-0.241(64)$  & $-4.2(1.1)$  & $2.35(21)$  & $0.019(99)$  & $-38.6(8.1)$ \\ &  & & 2 & 9.7/17  & $-0.206(88)$  & $-4.8(2.1)$  & $2.53(31)$  & $0.153(87)$  & $-34(12)$ \\\cmidrule{2-10}
 & \multirow{2}{*}{6.5} & \multirow{2}{*}{408}  & 1 & 12.6/17  & $-0.246(64)$  & $-4.1(1.1)$  & $2.40(17)$  & $0.06(11)$  & $-39.2(9.0)$ \\ &  & & 2 & 10.5/17  & $-0.218(92)$  & $-4.6(1.9)$  & $2.58(32)$  & $0.094(81)$  & $-35(13)$ \\\cmidrule{2-10}
 & \multirow{2}{*}{5.5} & \multirow{2}{*}{443}  & 1 & 15.1/17  & $-0.256(64)$  & $-3.90(98)$  & $2.47(19)$  & $0.04(13)$  & $-40.4(8.7)$ \\ &  & & 2 & 11.9/17  & $-0.234(90)$  & $-4.3(1.6)$  & $2.65(35)$  & $0.062(68)$  & $-37(11)$ \\\cmidrule{2-10}
 & \multirow{2}{*}{4.5} & \multirow{2}{*}{490}  & 1 & 19.0/17  & $-0.276(66)$  & $-3.62(86)$  & $2.59(23)$  & $0.026(90)$  & $-42.6(8.5)$ \\ &  & & 2 & 14.1/17  & $-0.258(93)$  & $-3.9(1.4)$  & $2.76(38)$  & $0.05(10)$  & $-40(11)$ \\\midrule
\multirow{8}{*}{\rotatebox{90}{ERE3/ERE2}} & \multirow{2}{*}{7.5} & \multirow{2}{*}{379}  & 1 & 11.0/17  & $-0.243(74)$  & $-4.1(1.3)$  & $1.80(36)$  & $0.29(11)$  & $-41(11)$ \\ &  & & 2 & 7.5/17  & $-0.15(10)$  & $-6.8(4.6)$  & $2.06(48)$  & $0.78(46)$  & $-26(15)$ \\\cmidrule{2-10}
 & \multirow{2}{*}{6.5} & \multirow{2}{*}{408}  & 1 & 11.3/17  & $-0.243(75)$  & $-4.1(1.3)$  & $1.77(38)$  & $0.32(12)$  & $-41(11)$ \\ &  & & 2 & 7.7/17  & $-0.14(10)$  & $-7.0(4.9)$  & $2.04(49)$  & $0.83(50)$  & $-25(16)$ \\\cmidrule{2-10}
 & \multirow{2}{*}{5.5} & \multirow{2}{*}{443}  & 1 & 11.8/17  & $-0.243(76)$  & $-4.1(1.3)$  & $1.74(40)$  & $0.38(14)$  & $-41(11)$ \\ &  & & 2 & 8.0/17  & $-0.138(95)$  & $-7.2(5.0)$  & $2.05(45)$  & $0.74(42)$  & $-24(15)$ \\\cmidrule{2-10}
 & \multirow{2}{*}{4.5} & \multirow{2}{*}{490}  & 1 & 12.7/17  & $-0.244(77)$  & $-4.1(1.3)$  & $1.69(44)$  & $0.47(18)$  & $-41(12)$ \\ &  & & 2 & 8.4/17  & $-0.124(95)$  & $-8.1(6.2)$  & $2.01(44)$  & $0.82(44)$  & $-22(15)$ \\\midrule
\multirow{8}{*}{\rotatebox{90}{ERE3/rational3}} & \multirow{2}{*}{7.5} & \multirow{2}{*}{379}  & 1 & 10.5/16  & $-0.239(73)$  & $-4.2(1.3)$  & $1.81(35)$  & $0.29(11)$  & $-40(11)$ \\ &  & & 2 & 7.2/16  & $-0.15(10)$  & $-6.6(4.4)$  & $2.04(48)$  & $0.77(45)$  & $-26(15)$ \\\cmidrule{2-10}
 & \multirow{2}{*}{6.5} & \multirow{2}{*}{408}  & 1 & 10.7/16  & $-0.238(74)$  & $-4.2(1.3)$  & $1.79(37)$  & $0.33(12)$  & $-40(11)$ \\ &  & & 2 & 7.3/16  & $-0.15(10)$  & $-6.8(4.6)$  & $2.01(49)$  & $0.80(47)$  & $-26(16)$ \\\cmidrule{2-10}
 & \multirow{2}{*}{5.5} & \multirow{2}{*}{443}  & 1 & 11.1/16  & $-0.238(74)$  & $-4.2(1.3)$  & $1.76(39)$  & $0.38(14)$  & $-40(11)$ \\ &  & & 2 & 7.3/16  & $-0.142(95)$  & $-7.1(4.7)$  & $2.02(43)$  & $0.71(36)$  & $-25(15)$ \\\cmidrule{2-10}
 & \multirow{2}{*}{4.5} & \multirow{2}{*}{490}  & 1 & 11.8/16  & $-0.236(76)$  & $-4.2(1.4)$  & $1.72(43)$  & $0.47(18)$  & $-40(12)$ \\ &  & & 2 & 7.4/16  & $-0.127(96)$  & $-7.9(5.9)$  & $1.96(45)$  & $0.82(42)$  & $-23(15)$ \\\midrule
 \multicolumn{5}{c}{AIC average}    & $-0.196(85)(50)$  & $-5.5(3.5)(1.5)$  & $2.01(40)(25)$  & $0.48(31)(29)$  & $-33(13)(8)$ \\    \bottomrule
    \end{tabular}
    
  \caption{Results for the scattering length, effective range, shape parameter and binding momentum in the isospin 0 channel, obtained from the analysis using the LHCQC on D251 levels with different $\hat{\mathcal{K}}$ parametrizations and cutoffs. The final line shows an AIC average of the results in the table.}
  \label{tab:LHCQCtableD251scattering_isospin0}
\end{table*}

\begin{table*}
  \centering
  \tabcolsep=0.15cm
      \begin{tabular}{c c c c c c c c c c}
    \toprule
     Fit & $M_N^2\alpha$ & $\Lambda/{\rm MeV}$ & Set & $\chi^2\,/\,\mathrm{dof}$ & $a_0^{-1}$ / fm${}^{-1}$ & $a_0$ / fm &  $r_0$ / fm & $P_0$ / fm${}^3$ & $\kappa$ / MeV \\
    \midrule
    \multirow{8}{*}{\rotatebox{90}{ERE2SD/ERE2}} & \multirow{2}{*}{7.5} & \multirow{2}{*}{379}  & 1 & 11.0/17  & $-1.39(32)$  & $-0.72(17)$  & $6.4(1.2)$  & $0.90(31)$  & $-104.6(8.3)$ \\ &  & & 2 & 9.7/17  & $-1.09(31)$  & $-0.92(26)$  & $5.13(93)$  & $0.57(22)$  & $-97(12)$ \\\cmidrule{2-10}
 & \multirow{2}{*}{6.5} & \multirow{2}{*}{408}  & 1 & 12.6/17  & $-1.40(33)$  & $-0.72(17)$  & $6.3(1.2)$  & $0.90(31)$  & $-105.5(8.6)$ \\ &  & & 2 & 10.5/17  & $-1.10(32)$  & $-0.91(26)$  & $5.11(91)$  & $0.60(25)$  & $-98(12)$ \\\cmidrule{2-10}
 & \multirow{2}{*}{5.5} & \multirow{2}{*}{443}  & 1 & 15.1/17  & $-1.41(34)$  & $-0.71(17)$  & $6.3(1.1)$  & $0.95(34)$  & $-106.5(9.2)$ \\ &  & & 2 & 11.9/17  & $-1.11(33)$  & $-0.90(27)$  & $5.13(90)$  & $0.64(27)$  & $-98(13)$ \\\cmidrule{2-10}
 & \multirow{2}{*}{4.5} & \multirow{2}{*}{490}  & 1 & 19.0/17  & $-1.44(36)$  & $-0.70(17)$  & $6.4(1.1)$  & $1.04(40)$  & $-108(10)$ \\ &  & & 2 & 14.1/17  & $-1.13(35)$  & $-0.88(27)$  & $5.22(91)$  & $0.70(29)$  & $-99(14)$ \\\midrule
\multirow{8}{*}{\rotatebox{90}{ERE3/ERE2}} & \multirow{2}{*}{7.5} & \multirow{2}{*}{379}  & 1 & 11.0/17  & $-1.55(41)$  & $-0.64(17)$  & $6.8(1.2)$  & $1.21(43)$  & $-111(11)$ \\ &  & & 2 & 7.5/17  & $-1.16(35)$  & $-0.86(26)$  & $5.33(93)$  & $0.73(27)$  & $-100(13)$ \\\cmidrule{2-10}
 & \multirow{2}{*}{6.5} & \multirow{2}{*}{408}  & 1 & 11.3/17  & $-1.56(41)$  & $-0.64(17)$  & $6.8(1.2)$  & $1.24(45)$  & $-111(11)$ \\ &  & & 2 & 7.7/17  & $-1.16(35)$  & $-0.86(26)$  & $5.34(91)$  & $0.75(28)$  & $-100(14)$ \\\cmidrule{2-10}
 & \multirow{2}{*}{5.5} & \multirow{2}{*}{443}  & 1 & 11.8/17  & $-1.56(41)$  & $-0.64(17)$  & $6.9(1.2)$  & $1.28(48)$  & $-111(12)$ \\ &  & & 2 & 8.0/17  & $-1.15(35)$  & $-0.87(27)$  & $5.36(89)$  & $0.78(29)$  & $-100(15)$ \\\cmidrule{2-10}
 & \multirow{2}{*}{4.5} & \multirow{2}{*}{490}  & 1 & 12.7/17  & $-1.57(42)$  & $-0.64(17)$  & $7.0(1.2)$  & $1.33(51)$  & $-111(12)$ \\ &  & & 2 & 8.4/17  & $-1.15(35)$  & $-0.87(27)$  & $5.40(89)$  & $0.80(29)$  & $-99(15)$ \\\midrule
\multirow{8}{*}{\rotatebox{90}{ERE3/rational3}} & \multirow{2}{*}{7.5} & \multirow{2}{*}{379}  & 1 & 10.5/16  & $-1.58(44)$  & $-0.63(17)$  & $6.3(1.3)$  & $0.87(73)$  & $-114(12)$ \\ &  & & 2 & 7.2/16  & $-1.22(39)$  & $-0.82(26)$  & $5.1(1.0)$  & $0.54(47)$  & $-104(15)$ \\\cmidrule{2-10}
 & \multirow{2}{*}{6.5} & \multirow{2}{*}{408}  & 1 & 10.7/16  & $-1.59(44)$  & $-0.63(17)$  & $6.2(1.3)$  & $0.87(73)$  & $-114(12)$ \\ &  & & 2 & 7.3/16  & $-1.22(39)$  & $-0.82(26)$  & $5.03(99)$  & $0.54(46)$  & $-105(15)$ \\\cmidrule{2-10}
 & \multirow{2}{*}{5.5} & \multirow{2}{*}{443}  & 1 & 11.1/16  & $-1.59(44)$  & $-0.63(17)$  & $6.2(1.3)$  & $0.87(73)$  & $-115(12)$ \\ &  & & 2 & 7.3/16  & $-1.20(39)$  & $-0.83(27)$  & $4.88(95)$  & $0.56(43)$  & $-105(16)$ \\\cmidrule{2-10}
 & \multirow{2}{*}{4.5} & \multirow{2}{*}{490}  & 1 & 11.8/16  & $-1.59(44)$  & $-0.63(17)$  & $6.2(1.3)$  & $0.88(72)$  & $-115(12)$ \\ &  & & 2 & 7.4/16  & $-1.20(39)$  & $-0.83(27)$  & $4.80(93)$  & $0.55(42)$  & $-106(17)$ \\\midrule
 \multicolumn{5}{c}{AIC average}    & $-1.34(38)(19)$  & $-0.76(22)(11)$  & $5.9(1.1)(0.7)$  & $0.87(43)(25)$  & $-106(13)(6)$ \\    \bottomrule
    \end{tabular}
    
  \caption{Same as \Cref{tab:LHCQCtableD251scattering_isospin0} for the isospin 1 channel data in D251.}
  \label{tab:LHCQCtableD251scattering_isospin1}
\end{table*}

\begin{table*}
  \centering
  \tabcolsep=0.15cm
      \begin{tabular}{c c c c c c c c c c}
    \toprule
     Fit & $M_N^2\alpha$ & $\Lambda/{\rm MeV}$ & Set & $\chi^2\,/\,\mathrm{dof}$ & $a_0^{-1}$ / fm${}^{-1}$ & $a_0$ / fm &  $r_0$ / fm & $P_0$ / fm${}^3$ & $\kappa$ / MeV \\
    \midrule
    \multirow{8}{*}{\rotatebox{90}{ERE2SD/ERE2}} & \multirow{2}{*}{7.5} & \multirow{2}{*}{354}  & 1 & 23.7/17  & $-0.199(86)$  & $-5.0(2.2)$  & $2.61(24)$  & $-0.36(28)$  & $-32(11)$ \\ &  & & 2 & 19.9/17  & $-0.150(90)$  & $-6.7(4.0)$  & $2.59(30)$  & $-0.27(22)$  & $-25(13)$ \\\cmidrule{2-10}
 & \multirow{2}{*}{6.5} & \multirow{2}{*}{380}  & 1 & 24.4/17  & $-0.210(77)$  & $-4.8(1.8)$  & $2.63(26)$  & $-0.46(26)$  & $-34(10)$ \\ &  & & 2 & 20.2/17  & $-0.154(96)$  & $-6.5(4.0)$  & $2.59(29)$  & $-0.24(20)$  & $-26(13)$ \\\cmidrule{2-10}
 & \multirow{2}{*}{5.5} & \multirow{2}{*}{413}  & 1 & 26.0/17  & $-0.229(89)$  & $-4.4(1.7)$  & $2.68(28)$  & $-0.36(25)$  & $-36(11)$ \\ &  & & 2 & 21.0/17  & $-0.171(98)$  & $-5.8(3.4)$  & $2.63(32)$  & $-0.29(20)$  & $-28(14)$ \\\cmidrule{2-10}
 & \multirow{2}{*}{4.5} & \multirow{2}{*}{457}  & 1 & 28.8/17  & $-0.261(93)$  & $-3.8(1.4)$  & $2.78(31)$  & $-0.36(24)$  & $-40(12)$ \\ &  & & 2 & 22.6/17  & $-0.200(98)$  & $-5.0(2.5)$  & $2.72(34)$  & $-0.25(19)$  & $-32(12)$ \\\midrule
\multirow{8}{*}{\rotatebox{90}{ERE3/ERE2}} & \multirow{2}{*}{7.5} & \multirow{2}{*}{354}  & 1 & 14.1/17  & $-0.085(96)$  & $-12(13)$  & $1.57(39)$  & $0.42(47)$  & $-16(17)$ \\ &  & & 2 & 14.6/17  & $-0.08(10)$  & $-12(15)$  & $1.88(41)$  & $0.32(36)$  & $-15(17)$ \\\cmidrule{2-10}
 & \multirow{2}{*}{6.5} & \multirow{2}{*}{380}  & 1 & 15.0/17  & $-0.089(98)$  & $-11(12)$  & $1.59(40)$  & $0.48(50)$  & $-16(17)$ \\ &  & & 2 & 14.9/17  & $-0.08(10)$  & $-13(16)$  & $1.88(43)$  & $0.39(41)$  & $-15(18)$ \\\cmidrule{2-10}
 & \multirow{2}{*}{5.5} & \multirow{2}{*}{413}  & 1 & 16.4/17  & $-0.098(99)$  & $-10(10)$  & $1.61(41)$  & $0.53(53)$  & $-18(17)$ \\ &  & & 2 & 15.3/17  & $-0.08(10)$  & $-12(16)$  & $1.87(45)$  & $0.47(49)$  & $-15(18)$ \\\cmidrule{2-10}
 & \multirow{2}{*}{4.5} & \multirow{2}{*}{457}  & 1 & 18.7/17  & $-0.11(10)$  & $-9.0(8.1)$  & $1.64(42)$  & $0.59(56)$  & $-20(17)$ \\ &  & & 2 & 16.1/17  & $-0.08(11)$  & $-12(15)$  & $1.85(48)$  & $0.57(60)$  & $-15(18)$ \\\midrule
\multirow{8}{*}{\rotatebox{90}{ERE3/rational3}} & \multirow{2}{*}{7.5} & \multirow{2}{*}{354}  & 1 & 11.2/16  & $-0.105(98)$  & $-9.5(8.9)$  & $1.54(37)$  & $0.38(39)$  & $-19(17)$ \\ &  & & 2 & 11.6/16  & $-0.070(92)$  & $-14(19)$  & $2.02(38)$  & $0.33(30)$  & $-13(16)$ \\\cmidrule{2-10}
 & \multirow{2}{*}{6.5} & \multirow{2}{*}{380}  & 1 & 11.8/16  & $-0.11(10)$  & $-9.0(8.1)$  & $1.55(38)$  & $0.42(40)$  & $-20(17)$ \\ &  & & 2 & 11.8/16  & $-0.068(93)$  & $-15(20)$  & $2.02(39)$  & $0.37(34)$  & $-13(16)$ \\\cmidrule{2-10}
 & \multirow{2}{*}{5.5} & \multirow{2}{*}{413}  & 1 & 12.7/16  & $-0.12(10)$  & $-8.3(6.9)$  & $1.54(39)$  & $0.46(40)$  & $-22(17)$ \\ &  & & 2 & 12.1/16  & $-0.068(94)$  & $-15(20)$  & $2.02(40)$  & $0.44(41)$  & $-13(16)$ \\\cmidrule{2-10}
 & \multirow{2}{*}{4.5} & \multirow{2}{*}{457}  & 1 & 13.9/16  & $-0.131(92)$  & $-7.6(5.4)$  & $1.51(36)$  & $0.48(28)$  & $-24(15)$ \\ &  & & 2 & 12.5/16  & $-0.069(93)$  & $-14(19)$  & $2.02(41)$  & $0.51(48)$  & $-13(16)$ \\\midrule
 \multicolumn{5}{c}{AIC average}    & $-0.090(97)(21)$  & $-12(15)(2)$  & $1.77(39)(23)$  & $0.41(41)(10)$  & $-17(17)(4)$ \\    \bottomrule
    \end{tabular}
    
  \caption{Same as \Cref{tab:LHCQCtableD251scattering_isospin0} for the isospin 0 channel data in J250.}
  \label{tab:LHCQCtableJ250scattering_isospin0}
\end{table*}

\begin{table*}
  \centering
  \tabcolsep=0.15cm
      \begin{tabular}{c c c c c c c c c c}
    \toprule
     Fit & $M_N^2\alpha$ & $\Lambda/{\rm MeV}$ & Set & $\chi^2\,/\,\mathrm{dof}$ & $a_0^{-1}$ / fm${}^{-1}$ & $a_0$ / fm &  $r_0$ / fm & $P_0$ / fm${}^3$ & $\kappa$ / MeV \\
    \midrule
    \multirow{8}{*}{\rotatebox{90}{ERE2SD/ERE2}} & \multirow{2}{*}{7.5} & \multirow{2}{*}{354}  & 1 & 23.7/17  & $-0.54(14)$  & $-1.86(47)$  & $4.42(56)$  & $-0.19(26)$  & $-62.0(8.7)$ \\ &  & & 2 & 19.9/17  & $-0.67(26)$  & $-1.48(57)$  & $5.0(1.2)$  & $-0.25(55)$  & $-70(13)$ \\\cmidrule{2-10}
 & \multirow{2}{*}{6.5} & \multirow{2}{*}{380}  & 1 & 24.4/17  & $-0.55(15)$  & $-1.82(49)$  & $4.40(58)$  & $-0.17(31)$  & $-63.0(9.2)$ \\ &  & & 2 & 20.2/17  & $-0.67(26)$  & $-1.49(58)$  & $4.9(1.1)$  & $-0.25(58)$  & $-70(13)$ \\\cmidrule{2-10}
 & \multirow{2}{*}{5.5} & \multirow{2}{*}{413}  & 1 & 26.0/17  & $-0.56(14)$  & $-1.78(45)$  & $4.39(53)$  & $-0.12(29)$  & $-64.2(9.0)$ \\ &  & & 2 & 21.0/17  & $-0.68(27)$  & $-1.47(58)$  & $4.8(1.1)$  & $-0.23(61)$  & $-70(13)$ \\\cmidrule{2-10}
 & \multirow{2}{*}{4.5} & \multirow{2}{*}{457}  & 1 & 28.8/17  & $-0.58(15)$  & $-1.72(44)$  & $4.42(53)$  & $-0.06(30)$  & $-65.6(9.2)$ \\ &  & & 2 & 22.6/17  & $-0.70(28)$  & $-1.43(57)$  & $4.8(1.2)$  & $-0.18(64)$  & $-72(14)$ \\\midrule
\multirow{8}{*}{\rotatebox{90}{ERE3/ERE2}} & \multirow{2}{*}{7.5} & \multirow{2}{*}{354}  & 1 & 14.1/17  & $-0.56(14)$  & $-1.79(44)$  & $4.38(50)$  & $-0.06(30)$  & $-64.0(8.9)$ \\ &  & & 2 & 14.6/17  & $-0.81(32)$  & $-1.23(48)$  & $5.3(1.3)$  & $-0.19(69)$  & $-77(12)$ \\\cmidrule{2-10}
 & \multirow{2}{*}{6.5} & \multirow{2}{*}{380}  & 1 & 15.0/17  & $-0.56(14)$  & $-1.78(44)$  & $4.36(50)$  & $-0.04(30)$  & $-64.4(9.0)$ \\ &  & & 2 & 14.9/17  & $-0.80(32)$  & $-1.24(49)$  & $5.3(1.3)$  & $-0.18(69)$  & $-76(12)$ \\\cmidrule{2-10}
 & \multirow{2}{*}{5.5} & \multirow{2}{*}{413}  & 1 & 16.4/17  & $-0.57(14)$  & $-1.76(43)$  & $4.36(50)$  & $0.00(29)$  & $-65.0(9.1)$ \\ &  & & 2 & 15.3/17  & $-0.80(31)$  & $-1.25(49)$  & $5.2(1.3)$  & $-0.14(69)$  & $-77(13)$ \\\cmidrule{2-10}
 & \multirow{2}{*}{4.5} & \multirow{2}{*}{457}  & 1 & 18.7/17  & $-0.58(14)$  & $-1.72(42)$  & $4.39(51)$  & $0.05(28)$  & $-65.9(9.1)$ \\ &  & & 2 & 16.1/17  & $-0.80(31)$  & $-1.25(48)$  & $5.2(1.3)$  & $-0.07(69)$  & $-77(13)$ \\\midrule
\multirow{8}{*}{\rotatebox{90}{ERE3/rational3}} & \multirow{2}{*}{7.5} & \multirow{2}{*}{354}  & 1 & 11.2/16  & $-0.55(15)$  & $-1.83(50)$  & $3.70(53)$  & $-0.38(45)$  & $-66(11)$ \\ &  & & 2 & 11.6/16  & $-0.60(27)$  & $-1.66(74)$  & $3.23(99)$  & $-0.99(74)$  & $-71(16)$ \\\cmidrule{2-10}
 & \multirow{2}{*}{6.5} & \multirow{2}{*}{380}  & 1 & 11.8/16  & $-0.54(15)$  & $-1.84(51)$  & $3.63(52)$  & $-0.34(43)$  & $-66(11)$ \\ &  & & 2 & 11.8/16  & $-0.59(27)$  & $-1.68(75)$  & $3.14(96)$  & $-0.97(73)$  & $-71(16)$ \\\cmidrule{2-10}
 & \multirow{2}{*}{5.5} & \multirow{2}{*}{413}  & 1 & 12.7/16  & $-0.54(15)$  & $-1.84(51)$  & $3.53(51)$  & $-0.29(41)$  & $-66(11)$ \\ &  & & 2 & 12.1/16  & $-0.59(26)$  & $-1.70(76)$  & $3.03(92)$  & $-0.94(71)$  & $-71(16)$ \\\cmidrule{2-10}
 & \multirow{2}{*}{4.5} & \multirow{2}{*}{457}  & 1 & 13.9/16  & $-0.55(15)$  & $-1.83(51)$  & $3.37(49)$  & $-0.22(37)$  & $-67(12)$ \\ &  & & 2 & 12.5/16  & $-0.58(26)$  & $-1.74(78)$  & $2.88(87)$  & $-0.88(69)$  & $-71(17)$ \\\midrule
 \multicolumn{5}{c}{AIC average}    & $-0.61(23)(9)$  & $-1.67(59)(21)$  & $3.88(85)(79)$  & $-0.45(57)(36)$  & $-69(13)(4)$ \\    \bottomrule
    \end{tabular}
    
  \caption{Same as \Cref{tab:LHCQCtableD251scattering_isospin0} for the isospin 1 channel data in J250.}
  \label{tab:LHCQCtableJ250scattering_isospin1}
\end{table*}

\begin{table*}
  \centering
  \tabcolsep=0.15cm
      \begin{tabular}{c c c c c c c c c}
    \toprule
     Fit & $\Lambda\,/\,\mathrm{MeV}$ & Set & $\chi^2\,/\,\mathrm{dof}$ & $a_0^{-1}$ / fm${}^{-1}$ & $a_0$ / fm &  $r_0$ / fm &  $P_0$ / fm${}^3$ & $\kappa$ / MeV \\
    \midrule
    \multirow{8}{*}{\rotatebox{90}{LO}} & \multirow{2}{*}{450} & 1 & 27.3/20  & $-0.292(62)$  & $-3.42(73)$  & $2.73(16)$  & $-0.191(51)$  & $-44.1(7.3)$ \\ &  & 2 & 25.2/20  & $-0.296(76)$  & $-3.38(87)$  & $2.71(20)$  & $-0.193(61)$  & $-44.6(8.9)$ \\\cmidrule{2-9}
 & \multirow{2}{*}{550} & 1 & 25.4/20  & $-0.316(66)$  & $-3.17(67)$  & $2.68(15)$  & $-0.234(64)$  & $-47.1(7.5)$ \\ &  & 2 & 27.3/20  & $-0.384(92)$  & $-2.61(62)$  & $2.81(21)$  & $-0.300(90)$  & $-54.3(9.4)$ \\\cmidrule{2-9}
 & \multirow{2}{*}{650} & 1 & 24.0/20  & $-0.319(67)$  & $-3.14(66)$  & $2.63(15)$  & $-0.220(67)$  & $-47.6(7.6)$ \\ &  & 2 & 27.4/20  & $-0.418(99)$  & $-2.39(56)$  & $2.82(22)$  & $-0.32(10)$  & $-58.0(9.6)$ \\\cmidrule{2-9}
 & \multirow{2}{*}{750} & 1 & 22.6/20  & $-0.317(67)$  & $-3.15(67)$  & $2.60(14)$  & $-0.197(65)$  & $-47.6(7.7)$ \\ &  & 2 & 26.9/20  & $-0.43(10)$  & $-2.33(55)$  & $2.81(22)$  & $-0.30(10)$  & $-59.2(9.7)$ \\\midrule
\multirow{8}{*}{\rotatebox{90}{NLO}} & \multirow{2}{*}{450} & 1 & 15.1/18  & $-0.248(59)$  & $-4.03(96)$  & $2.61(17)$  & $-0.108(85)$  & $-38.9(7.4)$ \\ &  & 2 & 12.9/18  & $-0.182(75)$  & $-5.5(2.3)$  & $2.42(18)$  & $0.042(79)$  & $-30(11)$ \\\cmidrule{2-9}
 & \multirow{2}{*}{550} & 1 & 11.2/18  & $-0.254(60)$  & $-3.94(93)$  & $2.49(15)$  & $-0.079(73)$  & $-40.0(7.6)$ \\ &  & 2 & 11.5/18  & $-0.215(81)$  & $-4.6(1.8)$  & $2.38(17)$  & $0.039(77)$  & $-35(11)$ \\\cmidrule{2-9}
 & \multirow{2}{*}{650} & 1 & 9.8/18  & $-0.259(61)$  & $-3.86(91)$  & $2.42(14)$  & $-0.061(68)$  & $-40.8(7.7)$ \\ &  & 2 & 11.0/18  & $-0.232(85)$  & $-4.3(1.6)$  & $2.36(17)$  & $0.044(76)$  & $-37(11)$ \\\cmidrule{2-9}
 & \multirow{2}{*}{750} & 1 & 9.2/18  & $-0.262(62)$  & $-3.82(90)$  & $2.39(14)$  & $-0.047(65)$  & $-41.3(7.8)$ \\ &  & 2 & 10.7/18  & $-0.239(87)$  & $-4.2(1.5)$  & $2.34(17)$  & $0.055(76)$  & $-38(11)$ \\\midrule
\multirow{6}{*}{\rotatebox{90}{NLOwSD}} & \multirow{2}{*}{400} & 1 & 10.1/17  & $-0.245(60)$  & $-4.1(1.0)$  & $2.54(19)$  & $-0.04(13)$  & $-38.7(7.7)$ \\ &  & 2 & 8.7/17  & $-0.188(80)$  & $-5.3(2.2)$  & $2.43(18)$  & $0.15(12)$  & $-31(11)$ \\\cmidrule{2-9}
 & \multirow{2}{*}{500} & 1 & 9.0/17  & $-0.245(59)$  & $-4.08(99)$  & $2.53(15)$  & $-0.034(87)$  & $-38.7(7.5)$ \\ &  & 2 & 9.0/17  & $-0.224(80)$  & $-4.5(1.6)$  & $2.62(23)$  & $0.19(14)$  & $-36(10)$ \\\cmidrule{2-9}
 & \multirow{2}{*}{600} & 1 & 7.9/17  & $-0.250(61)$  & $-4.00(97)$  & $2.40(16)$  & $-0.00(10)$  & $-39.7(7.8)$ \\ &  & 2 & 7.9/17  & $-0.216(85)$  & $-4.6(1.8)$  & $2.38(18)$  & $0.17(12)$  & $-35(11)$ \\\midrule
 \multicolumn{4}{c}{AIC average}    & $-0.236(73)(24)$  & $-4.3(1.4)(0.5)$  & $2.43(17)(8)$  & $0.034(96)(93)$  & $-37.8(9.6)(3.2)$ \\    \bottomrule
    \end{tabular}
    
  \caption{Results for the scattering length, effective range, shape parameter and binding momentum for the isospin 0 channel, determined from the predictions obtained using the results from PWQC fits to the D251 energy levels. The last row presents averaged results obtained using the AIC on all the results in the table.}
  \label{tab:PWQCtableD251scattering_isospin0}
\end{table*}

\begin{table*}
  \centering
  \tabcolsep=0.15cm
      \begin{tabular}{c c c c c c c c c}
    \toprule
     Fit & $\Lambda\,/\,\mathrm{MeV}$ & Set & $\chi^2\,/\,\mathrm{dof}$ & $a_0^{-1}$ / fm${}^{-1}$ & $a_0$ / fm &  $r_0$ / fm &  $P_0$ / fm${}^3$ & $\kappa$ / MeV \\
    \midrule
    \multirow{8}{*}{\rotatebox{90}{LO}} & \multirow{2}{*}{450} & 1 & 27.3/20  & $-1.92(39)$  & $-0.52(11)$  & $10.9(2.6)$  & $2.6(1.6)$  & $-106.1(5.3)$ \\ &  & 2 & 25.2/20  & $-1.69(39)$  & $-0.59(14)$  & $9.1(2.2)$  & $1.5(1.1)$  & $-104.1(6.8)$ \\\cmidrule{2-9}
 & \multirow{2}{*}{550} & 1 & 25.4/20  & $-2.02(36)$  & $-0.496(90)$  & $10.8(2.3)$  & $2.5(1.4)$  & $-109.2(4.5)$ \\ &  & 2 & 27.3/20  & $-1.91(42)$  & $-0.52(12)$  & $9.6(2.2)$  & $1.6(1.1)$  & $-109.5(6.0)$ \\\cmidrule{2-9}
 & \multirow{2}{*}{650} & 1 & 24.0/20  & $-2.05(34)$  & $-0.487(80)$  & $10.5(2.0)$  & $2.3(1.2)$  & $-111.0(4.1)$ \\ &  & 2 & 27.4/20  & $-2.03(42)$  & $-0.49(10)$  & $9.7(2.2)$  & $1.5(1.0)$  & $-112.4(5.5)$ \\\cmidrule{2-9}
 & \multirow{2}{*}{750} & 1 & 22.6/20  & $-2.09(33)$  & $-0.479(75)$  & $10.4(1.9)$  & $2.1(1.1)$  & $-112.5(3.9)$ \\ &  & 2 & 26.9/20  & $-2.09(42)$  & $-0.478(97)$  & $9.7(2.1)$  & $1.44(97)$  & $-114.2(5.2)$ \\\midrule
\multirow{8}{*}{\rotatebox{90}{NLO}} & \multirow{2}{*}{450} & 1 & 15.1/18  & $-1.12(23)$  & $-0.89(18)$  & $6.6(1.2)$  & $1.36(50)$  & $-91.6(7.2)$ \\ &  & 2 & 12.9/18  & $-0.96(23)$  & $-1.04(25)$  & $5.35(98)$  & $0.76(30)$  & $-88.6(9.3)$ \\\cmidrule{2-9}
 & \multirow{2}{*}{550} & 1 & 11.2/18  & $-1.24(26)$  & $-0.80(17)$  & $6.4(1.2)$  & $1.05(41)$  & $-98.1(7.2)$ \\ &  & 2 & 11.5/18  & $-1.08(26)$  & $-0.92(22)$  & $5.28(99)$  & $0.61(25)$  & $-95.7(9.3)$ \\\cmidrule{2-9}
 & \multirow{2}{*}{650} & 1 & 9.8/18  & $-1.33(28)$  & $-0.75(16)$  & $6.3(1.2)$  & $0.89(37)$  & $-102.3(7.1)$ \\ &  & 2 & 11.0/18  & $-1.17(28)$  & $-0.86(21)$  & $5.3(1.0)$  & $0.53(23)$  & $-100.1(9.2)$ \\\cmidrule{2-9}
 & \multirow{2}{*}{750} & 1 & 9.2/18  & $-1.39(29)$  & $-0.72(15)$  & $6.3(1.2)$  & $0.80(34)$  & $-104.9(7.1)$ \\ &  & 2 & 10.7/18  & $-1.21(29)$  & $-0.82(20)$  & $5.2(1.0)$  & $0.48(21)$  & $-102.7(9.2)$ \\\midrule
\multirow{6}{*}{\rotatebox{90}{NLOwSD}} & \multirow{2}{*}{400} & 1 & 10.1/17  & $-1.20(27)$  & $-0.83(19)$  & $6.6(1.3)$  & $1.40(54)$  & $-95.8(9.9)$ \\ &  & 2 & 8.7/17  & $-1.03(27)$  & $-0.97(26)$  & $5.1(1.1)$  & $0.82(27)$  & $-94(13)$ \\\cmidrule{2-9}
 & \multirow{2}{*}{500} & 1 & 9.0/17  & $-1.24(26)$  & $-0.80(17)$  & $6.4(1.1)$  & $1.09(39)$  & $-98.6(7.9)$ \\ &  & 2 & 9.0/17  & $-1.13(29)$  & $-0.88(22)$  & $5.2(1.0)$  & $0.69(24)$  & $-99(11)$ \\\cmidrule{2-9}
 & \multirow{2}{*}{600} & 1 & 7.9/17  & $-1.37(30)$  & $-0.73(16)$  & $6.2(1.2)$  & $0.87(36)$  & $-104.9(9.2)$ \\ &  & 2 & 7.9/17  & $-1.20(30)$  & $-0.83(21)$  & $4.9(1.0)$  & $0.54(19)$  & $-105(12)$ \\\midrule
 \multicolumn{4}{c}{AIC average}    & $-1.23(28)(13)$  & $-0.82(19)(8)$  & $5.7(1.1)(0.7)$  & $0.78(32)(23)$  & $-100.9(9.5)(4.2)$ \\    \bottomrule
    \end{tabular}
    
  \caption{Same as \Cref{tab:PWQCtableD251scattering_isospin0} for the isospin 1 channel of ensemble D251.}
  \label{tab:PWQCtableD251scattering_isospin1}
\end{table*}

\begin{table*}
  \centering
  \tabcolsep=0.15cm
      \begin{tabular}{c c c c c c c c c}
    \toprule
     Fit & $\Lambda\,/\,\mathrm{MeV}$ & Set & $\chi^2\,/\,\mathrm{dof}$ & $a_0^{-1}$ / fm${}^{-1}$ & $a_0$ / fm &  $r_0$ / fm &  $P_0$ / fm${}^3$ & $\kappa$ / MeV \\
    \midrule
    \multirow{10}{*}{\rotatebox{90}{NLO}} & \multirow{2}{*}{300} & 1 & 47.1/18  & $-0.244(84)$  & $-4.1(1.4)$  & $3.40(40)$  & $-1.12(45)$  & $-36.5(9.4)$ \\ &  & 2 & 40.2/18  & $-0.152(81)$  & $-6.6(3.5)$  & $3.06(38)$  & $-0.73(39)$  & $-25(11)$ \\\cmidrule{2-9}
 & \multirow{2}{*}{400} & 1 & 32.7/18  & $-0.214(77)$  & $-4.7(1.7)$  & $2.96(28)$  & $-0.57(26)$  & $-33.6(9.6)$ \\ &  & 2 & 31.4/18  & $-0.125(74)$  & $-8.0(4.8)$  & $2.67(26)$  & $-0.31(21)$  & $-21(11)$ \\\cmidrule{2-9}
 & \multirow{2}{*}{500} & 1 & 41.7/18  & $-0.214(74)$  & $-4.7(1.6)$  & $3.06(29)$  & $-0.78(27)$  & $-33.5(9.1)$ \\ &  & 2 & 26.3/18  & $-0.124(75)$  & $-8.1(4.9)$  & $2.50(22)$  & $-0.23(17)$  & $-22(11)$ \\\cmidrule{2-9}
 & \multirow{2}{*}{600} & 1 & 22.9/18  & $-0.219(77)$  & $-4.6(1.6)$  & $2.70(23)$  & $-0.41(20)$  & $-34.8(9.8)$ \\ &  & 2 & 23.7/18  & $-0.127(76)$  & $-7.9(4.7)$  & $2.42(20)$  & $-0.20(16)$  & $-22(12)$ \\\cmidrule{2-9}
 & \multirow{2}{*}{700} & 1 & 21.6/18  & $-0.221(77)$  & $-4.5(1.6)$  & $2.65(23)$  & $-0.38(19)$  & $-35.2(9.9)$ \\ &  & 2 & 22.4/18  & $-0.129(77)$  & $-7.7(4.6)$  & $2.39(20)$  & $-0.17(16)$  & $-22(12)$ \\\midrule
\multirow{8}{*}{\rotatebox{90}{NLOwSD}} & \multirow{2}{*}{300} & 1 & 26.3/17  & $-0.247(89)$  & $-4.0(1.5)$  & $2.74(36)$  & $-0.45(53)$  & $-38(11)$ \\ &  & 2 & 22.8/17  & $-0.154(88)$  & $-6.5(3.7)$  & $2.36(28)$  & $0.18(47)$  & $-26(13)$ \\\cmidrule{2-9}
 & \multirow{2}{*}{400} & 1 & 21.9/17  & $-0.210(77)$  & $-4.8(1.7)$  & $2.72(24)$  & $-0.25(26)$  & $-34(10)$ \\ &  & 2 & 21.4/17  & $-0.127(77)$  & $-7.9(4.8)$  & $2.47(24)$  & $0.04(29)$  & $-22(12)$ \\\cmidrule{2-9}
 & \multirow{2}{*}{500} & 1 & 20.4/17  & $-0.180(70)$  & $-5.5(2.1)$  & $2.67(20)$  & $-0.18(19)$  & $-29.6(9.5)$ \\ &  & 2 & 20.4/17  & $-0.108(71)$  & $-9.3(6.1)$  & $2.50(18)$  & $0.06(17)$  & $-19(11)$ \\\cmidrule{2-9}
 & \multirow{2}{*}{600} & 1 & 18.5/17  & $-0.198(74)$  & $-5.1(1.9)$  & $2.59(20)$  & $-0.19(20)$  & $-32.2(9.9)$ \\ &  & 2 & 18.6/17  & $-0.116(74)$  & $-8.6(5.5)$  & $2.37(20)$  & $0.03(20)$  & $-20(12)$ \\\midrule
 \multicolumn{4}{c}{AIC average}    & $-0.163(75)(44)$  & $-6.6(3.9)(1.8)$  & $2.53(21)(13)$  & $-0.14(21)(16)$  & $-27(11)(6)$ \\    \bottomrule
    \end{tabular}
    
  \caption{Same as \Cref{tab:PWQCtableD251scattering_isospin0} for the isospin 0 channel of the J250 ensemble.}
  \label{tab:PWQCtableJ250scattering_isospin0}
\end{table*}

\begin{table*}
  \centering
  \tabcolsep=0.15cm
      \begin{tabular}{c c c c c c c c c}
    \toprule
     Fit & $\Lambda\,/\,\mathrm{MeV}$ & Set & $\chi^2\,/\,\mathrm{dof}$ & $a_0^{-1}$ / fm${}^{-1}$ & $a_0$ / fm &  $r_0$ / fm &  $P_0$ / fm${}^3$ & $\kappa$ / MeV \\
    \midrule
    \multirow{10}{*}{\rotatebox{90}{NLO}} & \multirow{2}{*}{300} & 1 & 47.1/18  & $-0.58(14)$  & $-1.72(40)$  & $6.1(1.2)$  & $0.37(57)$  & $-59.7(6.6)$ \\ &  & 2 & 40.2/18  & $-0.73(26)$  & $-1.36(48)$  & $7.9(2.6)$  & $1.6(1.8)$  & $-64.3(8.4)$ \\\cmidrule{2-9}
 & \multirow{2}{*}{400} & 1 & 32.7/18  & $-0.54(12)$  & $-1.84(42)$  & $4.81(76)$  & $-0.07(19)$  & $-61.0(7.3)$ \\ &  & 2 & 31.4/18  & $-0.58(22)$  & $-1.71(64)$  & $5.2(1.4)$  & $0.02(33)$  & $-63(11)$ \\\cmidrule{2-9}
 & \multirow{2}{*}{500} & 1 & 41.7/18  & $-0.94(22)$  & $-1.07(25)$  & $7.6(1.8)$  & $0.01(56)$  & $-74.7(5.0)$ \\ &  & 2 & 26.3/18  & $-0.52(21)$  & $-1.92(77)$  & $4.2(1.1)$  & $-0.26(27)$  & $-61(13)$ \\\cmidrule{2-9}
 & \multirow{2}{*}{600} & 1 & 22.9/18  & $-0.63(14)$  & $-1.58(36)$  & $4.37(69)$  & $-0.46(20)$  & $-69.0(7.5)$ \\ &  & 2 & 23.7/18  & $-0.51(21)$  & $-1.95(80)$  & $3.85(94)$  & $-0.33(29)$  & $-62(14)$ \\\cmidrule{2-9}
 & \multirow{2}{*}{700} & 1 & 21.6/18  & $-0.66(15)$  & $-1.52(34)$  & $4.31(69)$  & $-0.53(22)$  & $-71.0(7.5)$ \\ &  & 2 & 22.4/18  & $-0.52(21)$  & $-1.93(80)$  & $3.72(89)$  & $-0.36(31)$  & $-63(14)$ \\\midrule
\multirow{8}{*}{\rotatebox{90}{NLOwSD}} & \multirow{2}{*}{300} & 1 & 26.3/17  & $-0.69(18)$  & $-1.45(38)$  & $5.0(1.2)$  & $0.08(47)$  & $-71(11)$ \\ &  & 2 & 22.8/17  & $-0.75(33)$  & $-1.33(57)$  & $3.9(1.9)$  & $0.6(1.4)$  & $-83(27)$ \\\cmidrule{2-9}
 & \multirow{2}{*}{400} & 1 & 21.9/17  & $-0.61(15)$  & $-1.64(40)$  & $4.38(76)$  & $-0.03(24)$  & $-67.8(8.8)$ \\ &  & 2 & 21.4/17  & $-0.65(27)$  & $-1.54(64)$  & $4.3(1.4)$  & $0.11(48)$  & $-71(15)$ \\\cmidrule{2-9}
 & \multirow{2}{*}{500} & 1 & 20.4/17  & $-0.66(15)$  & $-1.52(36)$  & $4.33(73)$  & $-0.32(23)$  & $-71.1(8.2)$ \\ &  & 2 & 20.4/17  & $-0.65(27)$  & $-1.53(63)$  & $4.2(1.1)$  & $-0.16(35)$  & $-72(15)$ \\\cmidrule{2-9}
 & \multirow{2}{*}{600} & 1 & 18.5/17  & $-0.67(15)$  & $-1.50(35)$  & $4.13(69)$  & $-0.40(25)$  & $-72.7(8.3)$ \\ &  & 2 & 18.6/17  & $-0.57(24)$  & $-1.75(74)$  & $3.62(96)$  & $-0.18(38)$  & $-68(16)$ \\\midrule
 \multicolumn{4}{c}{AIC average}    & $-0.62(20)(6)$  & $-1.63(57)(16)$  & $4.05(92)(29)$  & $-0.27(38)(21)$  & $-70(13)(4)$ \\    \bottomrule
    \end{tabular}
    
  \caption{Same as \Cref{tab:PWQCtableD251scattering_isospin0} for the isospin 1 channel of the J250 ensemble.}
  \label{tab:PWQCtableJ250scattering_isospin1}
\end{table*}

\begin{table*}
  \centering
  \tabcolsep=0.15cm
      \begin{tabular}{c c c c c c c c}
    \toprule
     Fit & $\Lambda\,/\,\mathrm{MeV}$ & $\chi^2\,/\,\mathrm{dof}$ & $a_0^{-1}$ / fm${}^{-1}$ & $a_0$ / fm  &  $r_0$ / fm &  $P_0$ / fm${}^3$ & $\kappa$ / MeV \\
    \midrule
    \multirow{4}{*}{\rotatebox{90}{LO}} & \multirow{1}{*}{650} & 15.2/20  & $-0.628(62)$  & $-1.59(16)$  & $1.902(77)$  & $0.0252(42)$  & $-87.4(6.6)$ \\
 & \multirow{1}{*}{750} & 14.3/20  & $-0.704(70)$  & $-1.42(14)$  & $1.811(76)$  & $0.0109(59)$  & $-96.4(7.3)$ \\
 & \multirow{1}{*}{850} & 15.5/20  & $-0.714(73)$  & $-1.40(14)$  & $1.805(75)$  & $0.0121(89)$  & $-97.5(7.5)$ \\
 & \multirow{1}{*}{950} & 16.1/20  & $-0.714(74)$  & $-1.40(14)$  & $1.798(74)$  & $0.018(11)$  & $-97.6(7.6)$ \\\midrule
\multirow{4}{*}{\rotatebox{90}{NLO}} & \multirow{1}{*}{650} & 15.3/18  & $-0.661(75)$  & $-1.51(17)$  & $1.86(10)$  & $0.010(43)$  & $-91.2(7.6)$ \\
 & \multirow{1}{*}{750} & 14.5/18  & $-0.658(74)$  & $-1.52(17)$  & $1.781(94)$  & $0.036(34)$  & $-92.0(7.7)$ \\
 & \multirow{1}{*}{850} & 10.6/18  & $-0.684(77)$  & $-1.46(16)$  & $1.715(88)$  & $0.051(27)$  & $-95.7(8.1)$ \\
 & \multirow{1}{*}{950} & 10.0/18  & $-0.679(77)$  & $-1.47(17)$  & $1.643(75)$  & $0.069(21)$  & $-96.2(8.3)$ \\\midrule
 \multicolumn{3}{c}{AIC average}    & $-0.685(73)(27)$  & $-1.46(16)(6)$  & $1.763(79)(86)$  & $0.035(18)(23)$  & $-95.1(7.7)(3.2)$ \\    \bottomrule
    \end{tabular}
    
  \caption{Same as \Cref{tab:PWQCtableD251scattering_isospin0} for the isospin 0 channel of the C103 ensemble.}
  \label{tab:PWQCtableC103scattering_isospin0}
\end{table*}

\begin{table*}
  \centering
  \tabcolsep=0.15cm
      \begin{tabular}{c c c c c c c c}
    \toprule
     Fit & $\Lambda\,/\,\mathrm{MeV}$ & $\chi^2\,/\,\mathrm{dof}$ & $a_0^{-1}$ / fm${}^{-1}$ & $a_0$ / fm  &  $r_0$ / fm &  $P_0$ / fm${}^3$ &  $\kappa$ / MeV \\
    \midrule
    \multirow{4}{*}{\rotatebox{90}{LO}} & \multirow{1}{*}{650} & 15.2/20  & $-1.38(18)$  & $-0.724(95)$  & $2.82(23)$  & $0.180(38)$  & $-140(10)$ \\
 & \multirow{1}{*}{750} & 14.3/20  & $-1.55(19)$  & $-0.646(81)$  & $2.71(23)$  & $0.146(34)$  & $-154(10)$ \\
 & \multirow{1}{*}{850} & 15.5/20  & $-1.59(19)$  & $-0.629(76)$  & $2.67(22)$  & $0.146(34)$  & $-157(10)$ \\
 & \multirow{1}{*}{950} & 16.1/20  & $-1.62(19)$  & $-0.616(74)$  & $2.64(22)$  & $0.145(34)$  & $-160(10)$ \\\midrule
\multirow{4}{*}{\rotatebox{90}{NLO}} & \multirow{1}{*}{650} & 15.3/18  & $-1.64(38)$  & $-0.61(14)$  & $2.99(34)$  & $0.155(73)$  & $-154(19)$ \\
 & \multirow{1}{*}{750} & 14.5/18  & $-1.80(42)$  & $-0.56(13)$  & $2.87(33)$  & $0.144(72)$  & $-167(21)$ \\
 & \multirow{1}{*}{850} & 10.6/18  & $-1.31(27)$  & $-0.76(16)$  & $2.45(20)$  & $0.189(41)$  & $-142(18)$ \\
 & \multirow{1}{*}{950} & 10.0/18  & $-1.36(28)$  & $-0.73(15)$  & $2.37(18)$  & $0.189(40)$  & $-148(20)$ \\\midrule
 \multicolumn{3}{c}{AIC average}    & $-1.46(24)(13)$  & $-0.69(12)(6)$  & $2.59(22)(17)$  & $0.169(40)(20)$  & $-150(15)(7)$ \\    \bottomrule
    \end{tabular}
    
  \caption{Same as \Cref{tab:PWQCtableD251scattering_isospin0} for the isospin 1 channel of the C103 ensemble.}
  \label{tab:PWQCtableC103scattering_isospin1}
\end{table*}

\clearpage
\bibliography{cited_refs}   

\end{document}